# STRUCTURAL COMPOSITES FOR MULTIFUNCTIONAL APPLICATIONS: CURRENT CHALLENGES AND FUTURE TRENDS

C. González[1,2], J. J. Vilatela[1], J. M. Molina-Aldareguía[1], C. S. Lopes[1], J. LLorca[1,2,*]

[1] IMDEA Materials Institute
C/ Eric Kandel 2, 28906 Getafe, Madrid, Spain

[2] Department of Materials Science, Polytechnic University of Madrid
E. T. S. de Ingenieros de Caminos, 28040 Madrid, Spain

**Abstract**

This review paper summarizes the current state-of-art and challenges for the future developments of fiber-reinforced composites for structural applications with multifunctional capabilities. After a brief analysis of the reasons of the successful incorporation of fiber-reinforced composites in many different industrial sectors, the review analyzes three critical factors that will define the future of composites. The first one is the application of novel fiber-deposition and preforming techniques together with innovative liquid moulding strategies, which will be combined by optimization tools based on novel multiscale modelling approaches, so fiber-reinforced composites with optimized properties can be designed and manufactured for each application. In addition, composite applications will be enhanced by the incorporation of multifunctional capabilities. Among them, electrical conductivity, energy storage (structural supercapacitors and batteries) and energy harvesting (piezoelectric and solar energy) seem to be the most promising ones.

1. Introduction

In the current situation of technological development, the strongest materials available for structural applications are in the form of small diameter fibers (in the range 5 to 40 µm). Strength of fibrous materials is enhanced by the reduced defect size, while the orientation of crystalline domains comprised of stiff and strong conjugated C-C or C-N bonds in graphitic planes or closed-packed polymer chains parallel to the fiber axis improves dramatically the stiffness and strength of polymer and carbon fibers [1-2]. However, the application of fibrous materials (either in the form of woven or nonwoven textiles) in practical applications is hindered by a number of issues: negligible compressive strength of fibers due to buckling, rapid strength degradation due to the generation of surface defects by fretting and/or environmental attack, permeability, etc. These problems can be overcome by infiltrating the fiber preform with a polymeric matrix, leading to the development of composite materials.

The application of fiber-reinforced composites (FRC) (and, in particular, of polymer-matrix composites) as structural materials has grown continuously during the last 50 years owing to their unique combination of low density, high stiffness and strength as well as toughness. The success of FRC as structural materials can be traced to their simple –but efficient– hierarchical structure with the three levels of organization: ply, laminate and structural element (Fig. 1) with different dominant length scales. Owing to this hierarchical structure, FRC are manufactured following a bottom-up approach, which begins with the individual plies that are stacked to form multidirectional laminates, which, in turn, are assembled intro structural elements. The properties at each level can be carefully controlled from the matrix and fiber volume fraction and spatial distribution at the ply level, the stacking sequence and the fiber architecture in each ply at the laminate level and, finally, the spatial disposition of the laminates to form the structural element. As a result, FRC with optimum mechanical properties can be easily designed for each application.

In addition, the hierarchical structure provides another critical property for structural applications: toughness. While stiffness and strength are mainly provided by the fibers, it is worth noting that FRC can dissipate large amounts of energy before fracture (in the range 10 to 50 kJ/m$^2$) and are, thus, flaw insensitive. In the absence of plasticity, the high toughness of structural composites is also a result of their hierarchical structure [3-5] that leads to the energy dissipation at different length scales during deformation and produces tough materials from brittle constituents. This behavior has also been reported in structural biomaterials, which also present high toughness due to their complex hierarchical structure [6-8].

As a result, FRC nowadays have widespread use in structural applications in transportation, sports, construction, civil infrastructures, energy generation, and many other industrial sectors. From the viewpoint of the development of the next generation of composite materials for structural applications, it is important to analyze the current challenges to extend the application of FRC as well as the goals that drive their evolution. These challenges are summarized in three main areas in this review, related to processing, multiscale design and optimization, and development of structural composites with multifunctional capabilities.

Processing of FRC is more expensive than that of other structural materials because of the high cost of the fibers (particularly carbon) and of the manufacturing strategy, that produces the components one by one in a batch process. Moreover, manufacturing of FRC has been hindered by the lack of automated processing techniques as well as the need of expensive equipment (autoclaves) to obtain high-performance, defect-free composites. These limitations are being overcome by the application of novel fiber-deposition and preforming techniques together with innovative liquid moulding strategies. They will be used in combination with advanced simulation techniques to take advantage of the large design space provided by the hierarchical structure of FRC. The final objective is that composite materials and structures can be designed and optimized *in silico*, before they are actually manufactured, leading to the implementation of a right-first-time approach to composite manufacturing.

FRC show excellent mechanical properties, but it is obvious that the full capabilities of these materials have not been attained. The stiffness and the strength of FRC are mainly controlled by the fibers but other properties (such as toughness, ductility, impact resistance, etc.) depend very much on the actual features of their hierarchical microstructure. These properties have been optimized so far by means costly trial and error strategies and they have been mainly applied to the laminate and structural element level, because experimental results and simulation techniques are readily available at this length scales. However, large improvements on the properties at the laminate and structural element level may be attained by controlling the microstructure and the properties at the µm-nm level. This is becoming feasible due to the new approaches based on multiscale modeling and characterization techniques available nowadays [3-4] which enable assessing the effects of microstructure on large structures, and will lead to the design of composites with optimized hierarchical microstructures.

Finally, the last challenge from the viewpoint of structural applications is to develop FRC with the properties that metals (their main competitors) and FRC lack, namely high electrical and thermal conductivity, fire retardancy, impermeability for gases and liquids, etc. These challenges have been addressed through modification of the matrix properties by means of nanoreinforcements or by adding layers with suitable properties within the hierarchical structure. Moreover, future developments in this area will be directed to achieve functionalities that cannot be obtained with metallic materials while maintaining the structural properties. They include structural FRC that can harvest energy and store energy and, thus, that can encompass two functions (structural and energy management) into one component.

**2. Novel manufacturing strategies**

*2.1. Advanced deposition techniques*
The application of FRC has been hampered by the limitations of the traditional hand-laying manufacturing process, which lead to poor quality, large variability and high labor costs. In addition, the insufficient knowledge about the structural behavior of composites, especially in terms of damage and failure due to their hierarchical structure, has limited the design envelope. Up to now, most composite structures were manufactured until recently with a very limited combination of laminate lay-ups (quasi-isotropic, unidirectional, etc.). This panorama is changing rapidly with the widespread application of automated systems capable of manufacturing large composite parts with high quality, using arbitrary lay-up configurations. For instance, Automated Fiber Placement (AFP) systems are in use for the automated manufacturing of complex aeronautical composite laminate parts of the new Boeing 787 and Airbus A350 aircrafts.

The AFP technology allows the automated production of high-quality advanced composite structures with tailored properties. This capability expands tremendously the design space of FRC allowing for impressive improvements in terms of stiffness, buckling, strength, notch and cut-out insensitivity. Several examples are given in the following paragraphs.

2.1.1. Improved damage tolerance through dispersed-ply laminates

The stacking sequence of laminates in industrial practice is often limited to combinations of 0º, 90º, ±45º fiber angle plies which is in line with the limitations of traditional layup processes to ensure a precise placement of fibers. This practice, in spite of being advantageous due to its simplicity and readiness, can be inefficient in terms of structural behavior. For instance, a laminate might present good specific stiffness and strength properties, but may show a poor response to impact loads. However, it is possible to design and build alternative laminates that keep the same in-plane and bending stiffness and much better impact resistance, by means of the combination of advanced optimization tools and AFP. The ply orientations are dispersed in the [-90º, 90˚] range (e.g. at intervals of 5˚) in these non-conventional laminates, which can be easily manufactured with AFP technology. By using the whole range of possible ply angles, it is possible to control at which interfaces the largest delaminations might occur and, in this way, improve the damage tolerance of a given laminate without sacrificing its stiffness.

Advanced optimization tools, together with high-fidelity numerical models based on the finite element method coupled with continuum damage mechanics at the mesoscale level [9, 10], were employed to design such non-conventional laminates and improve the impact and the compression after impact responses of composites, just by tailoring the stacking sequence. The designed laminates maintained target in-plane and bending stiffness properties, hence ruling out the effect of stiffness variation on the impact resistance and damage tolerance responses. This procedure is possible since multiple optima exist, i.e. there is more than one stacking sequence that satisfies a given design criterion in a sufficiently large design space, such as the one provided by AFP manufacturing. Different strategies, either by using Genetic Algorithms [11] or Ant Colony Optimization (ACO) algorithms [12,13], were explored to optimize the low velocity impact behavior of dispersed laminates with quasi-isotropic [12,13] and orthotropic stiffness properties [11]. Regarding impact resistance, the key to delaying the initiation of the most influencing failure mechanism - interply delamination – seems to be designing for the most uniform through-the-thickness profiles of interlaminar shear stresses as the laminate is deflected by the nearly quasi-static out-of-plane load. Regarding damage tolerance, which is controlled by the final extension and through-the-thickness location of delaminations generated with the impact, the optimal results are more difficult to obtain and rationalize because of the intricacy of the mechanisms involved in the impact and the compression after impact. Some of the ply interfaces in the optimized laminates have low ply orientation mismatch, or even allow the forming of ply clusters, while other interfaces have large mismatch angles. It was observed that sublaminate sequences with interfaces with low mismatch orientations behave similarly to ply clusters within which no (or small) delaminations are formed. These sublaminates, if majorly oriented in the direction of the applied loading, confer the laminate a relatively high stability and compression-after-impact strength. Delaminations sufficiently large to divide the laminate in sublaminates will only occur at the interfaces with large ply orientation mismatches, the vicinity of ply clusters, and at the backface of the laminate (due to ply splitting) [14-17]. Experimental compression-after-impact results of optimized dispersed stacking sequence laminates revealed damage tolerance improvements in the order of 20% with respect to conventional laminates with equivalent stiffness properties [14-17]. Recently developed optimization techniques point to even better designs, but these results have not yet been confirmed experimentally.

2.1.2. Fiber-Steering for improved composite performance

The fiber-steering capability of typical AFP machines allows the production of composite panels with non-conventional, dispersed-ply lay-ups whose orientation varies continuously from point to point. A continuously varying lay-up is achieved by steering the fibers within each ply, as shown in Figure 2. In such cases, the laminate stiffness also varies with the in-plane coordinates of the laminate; hence these configurations are also termed Variable-Stiffness Panels (VSP) [18].

The in-plane stiffness variation allowed by fiber-steering demonstrated a great capacity for load redistribution from the central sections of the panels to their supported edge sections. As a result, the buckling loads can be improved by a factor of two, and even higher, as compared with the most efficient straight-fiber designs with the same mass [18-20]. The higher the freedom to steer, i.e. the smaller the allowable radius of curvature of the fibres, the higher the potential for increasing the structural performance. Furthermore, the residual thermal stresses generated during the curing of these configurations are beneficial under compression loads, hence increasing the buckling resistance of VSP [19,15]. The advantages in terms of strength performance are also remarkable, with demonstrated improvements well above 50% [19-22]. Furthermore, it is possible to design VSP that do not develop critical stress concentrations around central holes due to the potential for the load redistribution far away from the hole. Thus, these VSP can actually be much less insensitive to the presence of holes in terms of buckling (compressive and shear) and failure loads [23, 24]. This is demonstrated in Figure 2c) that shows the finite element predictions of compressive failure loads of conventional and non-conventional, simply supported composite panels with holes of different diameters. The composite layups are the optimal ones that can be achieved given the fiber orientation possibilities in each case. On the one side, Figure 2c) shows that the optimal VSP has a much higher failure load than the straight-fiber panel for any hole size up to at least 2/3 of the panel size. And, on the other side, that the higher failure load is kept constant up to a hole size of ½ the panel size for the VSP whilst a linear decrease in compressive failure load with hole radius is associated with the conventional panel. This is a remarkable achievement and a great promise regarding weight savings in comparison to conventional solutions where the thickness is typically increased to mitigate the effects of stress concentrations around cut-outs, e.g. windows and doors. The fiber-steered configurations with the highest critical buckling loads are very similar to the optimum variable-stiffness designs in terms of failure. This means that a VSP can be designed simultaneously for the optimum buckling and strength responses [23, 24]. This is not possible with (straight-fibers) conventional laminates, where there is always a trade-off between buckling and strength.

Practical manufacturing conditions impose some constraints to ideal designs and result in some manufacturing defects. Typical AFP heads have the capability of delivering a small number of fiber tows along a reference path and within a limited course width. In order to lay-up an entire ply, the machine has to perform several passes, shifting the reference course from pass to pass. The shifting is, therefore, made discretely at multiples of the course width. As a result of the curvature, the boundaries of neighboring courses do not perfectly match. This leads to irregular regions between courses that can be accounted for by prescribing the shift distance so that no gaps occur. However, this practice produces overlap regions for curved reference paths. The AFP machine can be programmed to cut tows individually so that no thickness builds up to prevent the overlap regions. This practice results in laminates that contain what can be regarded as manufacturing defects, in essence small wedge-like areas free of fibers due to the dropping of the individual finite-thickness tows. Examples of VSP with these two types of heterogeneities are shown in Figure 3 [20]. Experimental and numerical studies concluded that the overlaps have an influence on the buckling and failure behaviors of VSP [20, 21] while gaps only influence their strength [25,26].

During AFP manufacturing of VSP, ply staggering is advised, i.e. the shifting of the fiber placement courses of one layer, in one of the planar directions, in relation to the adjacent layer with the same fiber orientation distribution. This is done to avoid the formation of course edges and of resin rich areas, which would occur at the same places through-the-thickness of a laminate in clustered plies with the same fiber angle distribution. Experimental and numerical analyses revealed a positive influence of ply staggering on the panel strength [25, 26], while the buckling load was not affected [20-21]. Another strategy to mitigate the effects of the manufacturing defects on VSP is the placement of ±45º layers on the top and bottom layers [18, 19].

Experimental and numerical analyses were carried to exactly quantify the effects of tow-drop defects found in a realistic AFP manufacturing environment on laminate in-plane strength, toughness [27] and damage tolerance [28]. Plain strength, open-hole tensile (OHT) strength and compression-after-impact (CAI) tests were performed. Test specimens were designed to represent the ply discontinuities at the edges of adjacent tow-placement courses (up to 32 tows can be placed in a single course) and compared with equivalent baseline laminates without defects. The results have shown that un-notched VSP are more influenced by the defects than notched ones, which are practically unaffected. This leads to the realization that defects have a more pronounced effect on laminate resistance than on toughness. Moreover, the staggering of plies can largely reduce the effect of defects in un-notched VSP composites [27]. The influence of defects on the impact damage tolerance of steered laminates was found to be relevant only for very low impact energies [28].

2.1.3. Dry Fiber Placement and Combination with other Manufacturing Processes

AFP is normally used to place narrow strips of tacky thermoset or thermoplastic prepreg material onto a mould to form a laminate, that in the case of thermoset resins, must be cured afterwards. Although the narrowness of the material strips offers more flexibility in terms of part geometry than hand lay-up of plies, components made with these fiber placement techniques for pre-impregnated materials are limited with respect to the fiber trajectories and steering that can be accomplished by the machine. In addition, part integration within a component is often limited by the complicated processing conditions.

As discussed in the next section, a new AFP capability is currently under development that combines the advantages of Liquid Moulding (LM) technology to form highly complicated and integrated parts, with the versatility of an AFP machine to manufacture complex preforms in an automated way. Furthermore, the use of the AFP technology for preform manufacturing within the framework of LM will reduce the amount of waste, fabric structure variability and uncertainties concerning the local fiber volume fraction and thus increase final mechanical properties. The final composite part can be obtained by injecting the produced preform with resin using Resin Transfer Moulding (RTM) or resin infusion processes [29].

*2.2 Innovative liquid moulding*

Manufacturing techniques for thermoset-based FRCs are traditionally classified depending on the way the resin is incorporated during processing. Prepreg sheets are semi-finished products in which the initial dry fabric (unidirectional or woven) was pre-impregnated with the resin and maintained in a semicured condition (B-stage) at low temperature prior to the final stacking, assembly and consolidation into a laminate. Autoclave consolidation of angle-ply laminates using thermoset prepreg sheets was developed to obtain composite parts with high fiber volume fraction (≈65%) and low void content (<2%). However, the large capital investments required by autoclave manufacturing, in combination with high recurring operational costs (e.g. energy consumption, $N_2$ fillings, long cure cycles) has been a powerful driving force to seek other manufacturing routes that can provide the same properties at lower cost. This is the case for the development of a new generation of out-of-autoclave (OoA) prepregs that can be consolidated in a vacuum bag [30-31]. These prepregs include partially impregnated tows that provide the appropriate evacuation channels for the migration of voids and volatiles during the consolidation and curing process. As a result, FRC parts with low porosity and high fiber volume fraction can be manufactured without the application of autoclave pressure.

Nevertheless, the majority of the OoA processing techniques are currently related to liquid moulding (LM) processes in which the resin is forced to infiltrate into a dry preform by means of a pressure gradient [32, 33]. The main objective of this technology is to fully impregnate the fabric as a result of the resin propagation between fibers and fiber bundles, while reducing as much as possible the voids entrapped and the dry spots. The impregnation driving force comes from the pressure differential between the inlet and outlet gates and the resistance to resin flow is controlled by the permeability of the porous fiber bed and the rheological properties of the fluid. Different LM processes are nowadays well-established at the industrial level. They include closed molding techniques, such as resin transfer molding (RTM), and open molding ones, such as vacuum-assisted resin infusion (VARI), among many others.

In RTM, the dry preform is placed in a mould cavity of controlled thickness and impregnation is driven by the pressure difference between the inlet and outlet. The main advantages of RTM are the ability to provide near net-shape composite parts with very good dimensional control as well as the elimination of volatiles generated during curing that are effectively transported to the venting ports of the system. However, large pressure gradients between inlet and outlet have to be applied to fully impregnate the fiber fabric for high fiber volume fractions, because the fabric permeability decreases as the fiber volume fraction increases. High-pressure RTM (HP-RTM) [34, 35] combines mechanical presses with advanced injection systems to apply pressures of 70-100 bars. This technique, in combination with fast curing epoxy resins, is amenable for high production rates (as the ones required by the automotive industry). In compression RTM (C-RTM), the mould gap thickness is initially relaxed to increase the fabric permeability, and then reduced, once the part is filled, to achieve the required fiber volume fraction while applying a uniform hydrostatic pressure field [33]. Finally, SQRTM (Same Qualified RTM) uses previously qualified prepregs that are placed in a standard resin transfer mould. Resin is then injected into the mould and the uniform hydrostatic pressure exerted by the resin mimics the pressure conditions during autoclave consolidation. Similar principles are used in the QuickStep process [36] in which a heat transfer fluid replaces the pressurized autoclave gases for consolidation. It should be finally noted that RTM technologies are limited to small to medium parts due to recurrent costs associated with the closed mould.

The size limitation is overcome in open mould strategies, in which one part of the mould is substituted by a flexible bag and the impregnation is driven by the atmospheric pressure as the bag is connected to a vacuum pump [37]. Fluid flow in VARI is mainly controlled by the competition between fabric compaction and fluid permeability. The atmospheric pressure over the vacuum bag is directly transferred to the dry fiber network before impregnation, leading to an initial compaction of the laminate. As the fluid infiltrates the fiber preform, the atmospheric pressure is shared between the fiber network and the fluid, resulting in the well-known elastic spring-back effect. The stress transfer mechanisms between the fluid and the fiber network strongly influence the mesoscopic resin flow, because the preform compaction and the permeability are no longer homogeneous in the infused part, leading to important changes in the pressure distribution, fluid velocity and filling times [38, 39]. One of the mayor drawbacks of VARI, particularly for very large components, is the long infiltration time because of the limited driving force (vacuum pressure). To speed up the process, a resin distribution medium with high permeability (as compared to the fiber preform) can be placed on top of the fiber preform (e.g. SCRIMP or Seemann Composites Resin Infusion Molding Process), so infiltration takes place simultaneously in-plane and through-the-thickness and the filling time is significantly reduced.

A major drawback of liquid moulding is the presence of voids and air entrapments that reduce the mechanical properties. In order to manufacture high quality composite parts with reduced porosity is critical to understand the physical mechanisms that control the interaction between the liquid and the fabric during infiltration. Macrodefects or scraps result from dry or poorly impregnated regions that appear when the resin flow reaches the outlet gate prior to complete filling. They are created by processing disturbances, such as undesired flow channels running along the mould cavity/inserts edges (race-tracking [40]), differences in the reinforcement volume fraction due to differences in clamping pressure, changes in the textile architecture by shearing and premature resin gelification. As a result, the resin reaches the venting port before the whole part is totally filled, leading to dry or poorly impregnated regions. These disturbances are very sensitive to the way the preform is cut, stacked and placed on the mould by the user, resulting in a hardly predictable behavior.

Even if the part has been completely filled with resin, voids can be generated in the material due to the inhomogeneous fluid flow propagation caused by the dual-scale porosity of the textile preform [41]. Standard reinforcements used in composite manufacturing are produced by weaving tows containing thousands of fibers arranged in different fabric architectures (woven, non-crimp, stitched fabrics, etc.), leading to intratow and intertow porosity, also known as dual-scale porosity. This dual-scale porosity produces an inhomogeneous resin flow through the fabric. Macroflow occurs between adjacent tows and is essentially controlled by the external pressure gradient, while microflow within the fiber tows is driven by capillary forces. This dual-scale flow (micro-macro) is responsible for the generation of voids by the direct competition between viscous and capillary forces. Experimental results have demonstrated that void formation in engineering composites depends on the ratio between viscous and capillary forces through the non-dimensional modified capillary number [42], $Ca^*=\mu v/(\gamma \cos\theta)$, where $\mu$ and $v$ stand for the resin viscosity and the average resin velocity, respectively, while γ and θ are the fluid surface tension and the contact angle. Viscous forces are predominant over capillary ones for high capillary numbers ($Ca^* > 10^{-2}$), and voids are generated by the rapid flow propagation along the yarn-to-yarn free gaps while void entrapments are generated at the intratow level. On the contrary, the fluid velocity is small for low capillary numbers ($Ca^* < 10^{-3}$) and the wicking effects caused by the intratow capillary forces become predominant. In this case, entrapments are generated at the tow interfaces rather than at the intratow spaces. The optimum filling rate (as dictated by the capillary number) will lead to a uniform progress of both the microflow and macroflow and to a minimum void volume fraction. Optimum capillary numbers that lead to a minimum void content have been reported in specific material systems [43]. Visualizing and understanding these resin flow mechanisms at the micro/meso level during liquid moulding processing is essential to manufacture high quality composite parts with low volume fraction of voids and defects [43-47].

The progress of the macroscopic resin flow can be studied by direct optical monitoring using transparent acrylic moulds in RTM [48, 49] or over the surface of the vacuum bag in VARI [50]. Resin flow and curing can also be tracked with embedded optical Bragg sensors due to the change of the refractive index of the surrounding media as a result of resin infiltration [51, 52] or by means of sensors that detect the changes in electrical resistivity [53]. Very recently, carbon nanotube (CNT) yarns embedded in the fabric were used to monitor the resin infiltration and curing during the fabrication of a structural composite by VARI [54]. The resin molecules penetrate the porous CNT yarn structure, leading to large changes in the electrical conductivity that can be used to track the flow front (Figure 4). The CNT yarns were placed in the mid-plane of the E-glass woven laminate (Figure 4a, 4b and 4c) and the laminate was compacted inside the vacuum bag. The evolution of the flow front position was also measured with CCD cameras and compared with the predictions from the calibrated CNT yarns sensors. Both measurements were in good agreement (Fig. 4d) and showed that the fluid propagation was controlled by Darcy's law (Fig. 4e).

Despite the individual complexity of these techniques, they provide valuable information about the macroscopic flow propagation in a composite part during the infusion/injection process, albeit with different advantages and disadvantages. The main drawback is related to their modest spatial resolution, which can only track the macroscopic averaged resin flow, neglecting the local distribution at the intra and inter-tow level. To overcome these problems,

Magnetic Resonance Imaging methods (MRI) can be used to track the front flow by mapping the fluid concentration inside porous samples, including different fiber preforms [55-57]. For instance, MRI measurements were used to study capillary driven transversal flow in bundles of aligned fibers using blends of water and corn syrup containing protonated liquids [55]. As a result, the evolution of the wet portion of the fiber bundle with respect to time was obtained and compared with analytical models of fluid propagation.

Very recently [58], in-situ vacuum assisted infiltration experiments were carried out using Synchrotron X-ray Computed tomography (SXCT) to evaluate the mechanisms of microfluid flow in a single tow of E-glass fiber. The high resolution of the SXCT images with an effective pixel size of ≈1.25μm$^2$ allowed a detailed reconstruction of individual glass fibers of ≈16μm in diameter within the tow, while the contrast between the different phases (air, fluid and fibers) was enough to track the fluid front position and shape as well as the void transport during infiltration (Figure 5). The ability of this technique to provide detailed information of the microfluid flow and the void transport in composite materials was clearly established. The fluid propagation at the microscopic level as well as the mechanisms of void transport within the tow were related to the wetting mechanisms between the fluid and the fibers, the rheological properties of the fluid, and the local microstructural details (fiber volume fraction, fiber orientation) of the fiber tow. For instance, Figure 5a) shows a cross section perpendicular to the fiber tow of the partially impregnated volume where the different phases (fibers, fluid and voids) are clearly distinguished. The fiber volume fraction was only locally disturbed when voids were dragged along the fiber direction. The local distribution of fibers within the tow is inhomogeneous (fiber clusters and empty spaces) and fibers are not parallel between them and exhibit convergent/divergent trajectories with significant twisting and meandering. As a result, flow progress along the bundle is inhomogeneous, leading to the formation of preferential channels in the fiber direction, Figure 5c).

2.2.1. Prediction of manufacturing defects and on-line control strategies

During resin injection, dry spots can be formed due to the generation of unexpected flow patterns. Race-tracking and variations in the fiber volume content are the most important factors responsible for these processing disturbances. Race-tracking occurs due to a non-perfect fit of the fiber preform to the mould or insert edges creating a channel or gap where resin flows more easily towards the outlet gate. Race-tracking may strongly modify the flow patterns diverting the resin to the venting ports and its influence depends on the channel size which, in addition, is controlled by preform cutting and handling operations. For instance, Figure 6a shows a lab-scale induced race tracking on a rectangular acrylic mould. The resin flows easily in the upper edge from the inlet to the outlet of the mould. A dry spot is formed when the resin in the race tracking channel arrives to the outlet gate because the pressure gradient is exhausted and the resin flow is slowed down in the dry region. Simulation of flow process can be used to study the severity of race tracking and the effects on the dry spot formation. Figure 6b) represents the flow pattern modification in the presence of the race tracking showing how the resin is diverted directly to the outlet gate of the mould.

Other variations of preform permeability can be produced due to non homogeneous clamping pressure distribution or severe fabric shearing generated during draping. Therefore, realistic simulation of filling operations, injection or infusion, should include a range of possible scenarios to optimize the process and reduce as much as possible the defect formation in the composite part. The best design configuration will be attained by an appropriate combination of the different input parameters using mathematical optimization methods to minimize an objective cost function. Among them, the most important processing tunable parameters are the position of the inlet and outlet gates, the injection strategy (pressure or flow controlled), and several active and passive strategies, such as opening/closing the inlet and venting ports when required.

Heuristic or experience-based methods [59] have been used traditionally by industry to obtain reasonable good solutions of the optimization problem. However, this situation has changed with the availability of advanced simulation techniques that can provide a numerical solution to the optimization problem. This is the case, for instance, for the determination of the optimum location of the injection and venting ports, which can be obtained by means of algorithms based on Monte Carlo lotteries, neural networks, etc. [59-67].

Although simulation can *a priori* establish the best configuration of mould injection and venting ports for a given part geometry and material system, the uncertainties associated with the permeability variations are a major hurdle. As a result, on-the-fly control strategies for injection moulding become crucial if process automation and robustness is required. The key point in mould filling control relies on the prompt and proper response to permeability variations in real-time. This real-time feed-back and the corresponding closed-loop control during filling operations seems to be the most efficient approach to remove defects by dry spot formation during liquid moulding [68, 69]. The schematic of a simple rectangular RTM mould, where three independent injection ports are used to prescribe uniform flow paths overcoming permeability variations, is shown in Figure 7a. The flow front is captured by means of an optical system and this information is processed in real time to modify the fluid flow/pressure in the injection/venting ports. This on-the-fly simulation-controlled closed loop technique has been validated in the laboratory, but the complexity of the system for real parts (e.g. many different possible race-tracking runners, etc.) makes this technique currently prohibitively expensive for industrial environments. Moreover, on-the-fly control systems require on-line high performance computing to provide results in real time.

The use of sensor information [70] can address a wider range of unconventional flow disturbances than on-the-fly simulation control devices. The sensor reading will be affected by the process disturbances and, as a result, a number of specific actions can be taken to improve or to correct the resin flow. In this case, simulation cases run prior to the injection can help identifying possible disturbance scenarios from the sensor readings, allowing to take the appropriate actions without running on-the-fly simulations. Advani *et al.* [71-74] developed efficient control strategies based on predefined commands for the resin injection and venting ports for given sensor readings. Passive controls were only implemented in venting ports by opening/closing instructions when an appropriate sensor response was attained. Active controls, on the other hand, were implemented at injection ports. They controlled not only opening/closing actions, but the pressure and the flow rate at which the resin was introduced in the mould. The experimental and simulated flow fronts obtained with this approach are shown in Figure 7b) for a flat rectangular panel with a triangular insert. The simulation allowed identifying possible disturbance scenarios and taking specific corrective actions to fill completely the composite part, eliminating dry spots.

*2.3. Out-of-autoclave consolidation of prepregs*

Another strategy to reduce the large costs associated with the autoclave consolidation of prepregs, while maintaining the mechanical properties, is the development of a new generation of out-of-autoclave (OoA) prepregs which are specifically designed to be consolidated in a vacuum bag in an oven [75,76]. OoA prepregs make use of partially impregnated tows that are envisioned to serve as evacuation channels for voids and volatiles during consolidation, enabling the production of composite parts with high fiber volume fraction and low porosity. The resin system in OoA prepregs is carefully selected to maintain low viscosity level and a wider processing window during volatiles extraction to ensure that the evacuation channels remain open during OoA consolidation under atmospheric pressure.

Quality of OoA laminates is very sensitive to the processing conditions and, in particular, to the way debulking is performed in the laminate kit. Debulking is always necessary to extract as much as possible entrapped air between the adjacent layers during manual lay-up. However, excessive debulking in OoA prepregs may be detrimental as the intraply channels could be partially closed by excessive pressure, hampering the evacuation of the voids and volatiles during consolidation. In addition, OoA prepregs are more sensitive to environmental effects (such as water absorption from humid environments or prepreg defrosting [77]) that can reduce significantly the gel time of the resin. For instance, an increase of humidity from 50 to 90% decreases the gel time of MTM44-1 epoxy system (Advanced Composite Group) from 160 to 130 minutes [78] and shortens the processing window.

The capacity of an OoA prepreg to evacuate voids is measured using the gas permeability of the material [78], which is $\approx 10^{-14}$ m$^2$ in the fiber direction and considerably lower in the perpendicular one [79-81]. The gas permeability of the prepreg also changes during the consolidation: it decreases at the beginning due to the reduction of the resin viscosity and increases afterwards due to closure of the air channels and resin gelation [78]. XCT analysis has revealed itself as a powerful technique to visualize the void nature in OoA prepregs [82,83]. This is demonstrated in Fig. 8, which shows the porosity within the tows (also known as intraply voids) as well as the voids entrapped between adjacent layers during lay-up (interply porosity) in HexPly M56 [+45/0/-45/90]$_{3s}$ fresh and cured laminates manufactured using hand lay-up, Figures 8a) and 8b), respectively. To this end, 20 x 20 mm$^2$ samples were extracted from the center of 400 x 400 mm$^2$ panels and analyzed by XCT with a pixel size of 23 μm. This resolution was low enough to allow segmentation of most of the internal porosity. The results revealed a total porosity of ≈6.1% in the fresh laminate, divided between 1.9% and 4.2% for intraply and interply voids, respectively. The laminate was cured using the standard cycle recommended by the prepreg provider (one hour dwell at 110ºC followed by two hours at 180ºC) and the porosity decreased to 1.6%, which was mainly interply porosity generated during lay-up operations not removed during debulking. These results show that the evacuation channels are effective to reduce the intraply porosity during the cure cycle.

The use of OoA prepregs to alleviate the costs associated with autoclaves should be coupled with AFP or automated tape laying (ATL) techniques. During production trials with these techniques, the hot compaction roll pressure [85] removes most of the intraply porosity and the cured composite part was almost void free. This is shown in the tomograms in Figures 8c) and d), which show the intraply and interplay porosity of HexPly M56 [+45/0/-45/90]3s laminates deposited by AFP and cured out-of-autoclave with the same cure cycle indicated above. The final porosity after curing was very small (<0.1%) revealing that OoA consolidation of prepregs is able to produce void-free composites if the appropriate processing conditions are followed [77].

## 3. Multiscale virtual design and optimization strategies

FRC present a hierarchical structure and, thus, processing is carried out and properties develop following a bottom-up approach. Manufacturing begins by the arrangement of the fibers into textile preforms, which are infiltrated with the resin, stacked to form a laminate that has to adopt the component shape and finally cured. Similarly, individual plies with different fiber architecture and/or orientation are combined to form a laminate with particular mechanical properties, and different laminates are assembled into a structural component. So far, design and optimization of processing and properties of FRC have been carried out using costly and time-consuming trial and error approaches. Nevertheless, this scenario is changing rapidly due to the ever increasing power of digital computers, the maturation of numerical modeling tools and the development of multiscale modeling strategies that can deal with the inherent complexity of hierarchical materials, in which the structure and properties at small length scales determine the macroscopic behavior. In this context, there is an enormous to take advantage of the large design envelope of FRC and to carry out the optimization of the processing and properties *in silico*, before the materials are actually manufactured in the laboratory. An integrated multiscale modeling strategy for virtual processing and virtual testing of FRC is schematically depicted in Figure 9 and detailed below.

*3.1 Virtual processing*

The first step in the virtual processing of FRCs manufactured by LM is fabric draping and forming. The complex lay-ups required in LM can be obtained by draping and forming of flat and flexible textile fabrics, as well as hot forming of thermoset and thermoplastic prepregs. Rigid tools, holders and grippers are used to impose a prescribed mould geometry to the fabric and to provide the required stability to be handled before resin injection. Draping and forming simulation is a crucial operation for the optimization of composite manufacturing. It should guarantee that the process is free of defects such as wrinkles and yarn damages while predicting the final position and orientation of the individual yarns after forming. In addition, draping is necessary to produce near-net shape preforms avoiding the excessive production of scraps.

A fabric or a laminate prepreg should adapt to the tool surface and to the external constraints of holders and grippers by in-plane and out-of-plane deformations that combine axial deformation, rotation and shearing of the yarns, as well as relative sliding between the plies. It should be noted that the final deformation of the fabrics strongly influences other important variables such as the permeability (e.g. fiber volume fraction, fabric shearing, nesting, etc.) as well as wrinkle formation. The first methods developed to address the effect of deformation on the fabric drapability were based on mapping textiles onto a given surface as the pin-jointed net model or kinematic draping models [86-89]. In this approach, textile yarns were assumed to behave as inextensible bars and adjacent yarn rotation and sliding were the only possible mechanisms. Although pin jointed models were able to provide fast predictions of textile draping and forming, they could not deal neither with load boundary conditions nor with tool-fabric and intra-ply sliding.

Fortunately, other possibilities are available that allow to treat more accurately the draping problem. These methodologies, also known as mechanical models, solve the stress interactions between the plies and the mould, and can be classified into discrete, continuum and semi-discrete approaches. In discrete models [95, 99], yarns are modeled using structural finite elements such as beams, trusses or springs. The different deformation mechanisms, such as the yarn extension, yarn shearing, viscoelastic response, etc., are entrusted to these elements (Figure 10a). These models are simple to implement in standard finite element codes, but their size is too large for big parts as the discretization is linked to the repetitive unit cell of the fabric. To overcome these problems, a second class of continuum models can be used [90, 91]. Plies are discretized using shells or membranes and the mechanical behavior is introduced by means of appropriate constitutive equations developed under the framework of finite strains and rotations [92, 93]. In addition, interply sliding can be taken into account in these models by introducing relative displacements between adjacent plies controlled by contact algorithms. More efficient methods can be achieved by the combination of the previous approaches in a semi-discrete framework [94], based on standard finite elements which account internally for the deformation modes of a number of unit cells of the fabric through the appropriate strain energy formulations, Figure 10b).

Finally, more sophisticated models are based on the detailed finite element modeling of a unit cell at the mesoscale, which contains precise information of the fabric architecture [98]. For instance, Figure 11a shows a woven fabric structure reconstructed from images acquired using X-ray computed tomography. The dry yarns of the fabric are modeled as transversely isotropic rate-dependent hypoelastic solids. As a consequence, important effects as yarn cross section change due to yarn-to-yarn interactions as well as interyarn sliding can be modeled and compared with experimental results obtained with XCT, Figure 11b. More sophisticated models based in the interactions between individual fibers in dry bundles [100] can also be used, Figure 11c. All these models provide very fundamental information about the deformation processes at the fiber and fiber yarn level and can be used to tune the parameters used in simulations presented in the previous paragraphs to capture the macroscopic fabric deformation [98].

The mechanisms of interply sliding are especially critical to predict accurately forming and wrinkling of preforms containing multiples plies, including dry fabrics as well as thermoset and thermoplastics prepregs layups [97]. The use of multiple layers modeled through standard shell or membrane elements coupled by friction interactions becomes prohibitive for realistic composite parts and an intermediate approach, known as multi–layer material model, was developed to speed up computations [96]. This special formulation of the finite element shell contains a lay-up that accounts for ply level deformations as well as relative displacements between plies by minimization of the total strain energy using an implicit updated Lagrangian finite element scheme.

After forming, the fabric is placed into the mould cavity and is impregnated with resin. The simulation of the resin impregnation is carried out using fluid and solid computational mechanics, including the thermal interaction between the composite part and the mould as a result of the exothermal reactions during resin cure (Figure 12). The majority of the mould filling simulation strategies are based in Darcy's law for porous materials to relate the pressure gradient in the resin with the local average fluid velocity, through the fluid viscosity µ and the permeability tensor of the fiber preform **K**. Darcy's law assumes that viscous flow is dominant and that the Reynolds number is low. Both hypotheses are usually valid in LM of composites, Figure 12a). By invoking the mass conservation law, a boundary value problem for the pressure field can be obtained and solved under the appropriate boundary conditions (e.g. prescribed pressure, velocity or slip-free conditions at mould edges, etc.). The partial differential equation can be solved using any standard numerical approach although specific problems dealing with the propagation of moving boundaries during fluid flow should be addressed. The most popular technique to account for moving boundaries is the finite element method in combination with the control volume technique (FE/CV) [101], which is implemented in commercial software packages. The success of this technique to track the fluid front propagation is based on the use of a fixed discretization without the need of re-meshing (Figure 13), in contrast with the moving-mesh propagation algorithms.

Alternative methods to track the propagation of fluid flow through a porous media are the level-set method [103, 104], arbitrary Lagrangian Eulerian algorithms [105] and phase-field methods [106]. In general, all these techniques can describe accurately the flow front position from a mathematical viewpoint for a given set of material properties (fluid viscosity, permeability, etc.) [107-108]. However, there are still important open issues in LM simulation that will require the development of new constitutive equations and numerical methods. For instance, a complete constitutive model to account for fabric deformability during infusion process requires solving simultaneously the solid and fluid governing equations [109-112]. Most of the theories currently used to account for the fluid-solid interaction are based on Terzaghi's approximation, adapted to the specific case of composite manufacturing, which relates the hydrostatic pressure on the porous medium with the effective stress transmitted to the fiber bed skeleton [113, 114]. The influence of the fluid flow on the elasticity of the porous fabric is not fully understood due to the complexity of the physical phenomena, including yarn/fiber lubrication effects, ply nesting and accommodation, etc. [115]. The mechanical behavior of the porous material due to the presence of the fluid is of a crucial importance during resin infusion due to the stress transfer between the fabric and the fluid during the infiltration. As the fluid progresses, the stress is transferred to the resin, leading to fabric relaxation and thickness increment, which influences the permeability and, as a consequence, the pressure gradient between the inlet and outlet gates.

Future improvements of simulation of LM processes will come from multiscale strategies that take into account the hierarchical structure of FRC (Figure 12). In particular, fluid flow at different length scales should be analyzed. At the macroscopic scale, impregnation is modeled using the Darcy's law with a homogenized permeability of the fabric. However, flow is not homogeneous at lower scales, and preferential channels are formed as a consequence of the dual scale porosity of the fabric [114] (Figure 12b). Viscous flow is dominant in the yarn-to-yarn channels while wicking flow occurs inside the yarns triggered by capillary forces. The simulation of the latter phenomena at the microlevel (Figure 12c) is still under development due to the enormous difficulties to account for the surface tension effects on the moving flow front [115, 116].

Heat transfer between the composite part and the tool should also be accounted for in the numerical simulation of LM, especially in thick composites where exothermal reactions may modify significantly the actual cure cycle. The heat released by the resin can be included into the energy balance equation, taking into account the internal transmission and dispersion terms depending on the fluid velocity. Typically, the local temperature equilibrium is assumed to be equal in the fiber and the resin and this assumption is valid in the case of slow resin flow during impregnation. However, two different heat balance equations and temperature fields for fiber reinforcement and matrix should be used for the fiber and the resin in the case of fast reaction rates or fast flows [117]. Finally, prescribed boundary and initial conditions should be imposed to reflect the current manufacturing process (i.e. mould interaction, air convection, etc.).

Residual stresses are inherent to composite manufacturing and have a strong influence on the final quality of the component. They appear because of the mismatch in the coefficient of thermal expansions between resin and fibers and between part and mould [118, 119], as well as because of the volumetric changes in the resin during cure. Residual stresses are only partially relieved after demoulding leading to part distortion [119] as well as strength reductions [120-122]. Accurate predictions of the residual stress state at the micro and meso level require the use of specific models to account for the evolution of the viscoelastic response of the resin during manufacturing and, particularly, from the gelation point until the end of the process [124, 125]. However, current approaches use crude approximations for this behavior. They often neglect the viscoelastic relaxation of the resin, but account for the evolution of the elastic modulus of the resin as a function of the temperature and the degree of cure [126]. All these phenomena (together with the laminate stacking sequence, the resin cure kinetics, etc.) will influence the component distortion after demoulding [127]. It should be noted that component distortion may hamper the assembly operations when tight geometry tolerances are required, as in aircraft manufacturing, and accurate simulation tools capable of predicting these effects are critical [128-132].

*3.2 Virtual testing*

The design and certification of composites structures (mainly in the aerospace industry) has been very different to the approaches followed for other structural materials (e.g. metals) because of a number of reasons that include the hierarchical structure of FRCs, the complexity and variety of the failure mechanisms and the uncertainties introduced in the manufacturing process. As a result, certification of components was carried out by means of experimental campaigns following a bottom-up approach and advanced numerical simulations following a top-down strategy [133]. The experimental tests spanned from small coupons to panels, subcomponents up to the final global structure to obtain the mechanical properties at the different length scales. In parallel, design was carried out using a top-down approach (Figure 14), which began by the analysis of the whole structure [133]. The results of these analyses were used to provide the boundary conditions for more refined analyses at lower length scales, which used the material properties obtained from the corresponding mechanical tests. The modeling detail was increased as successive analyses 'zoom in' on structural regions, identified as being potentially strength critical. While this approach was successful from the design and certification viewpoint, it has several serious drawbacks from the viewpoint of cost and materials optimization. Firstly, the pyramid of mechanical tests (from coupons to structures) is very expensive. Secondly, any change in the properties of either matrix or fibers require to repeat the whole testing campaign, leading to a large inertia to introduce new materials. Finally, material properties are used as input at every length scale, but are not optimized for each situation, leading to sub-optimum designs.

These limitations are being overcome by the novel bottom-up, multiscale simulation strategies that follow the hierarchical structure of FRC [3, 4, 134]. This new strategy comes about as a result of recent advances in different areas, including micromechanical characterization of constituent properties, development of accurate modeling tools for composites at the micro and mesoscale, coherent strategies to pass information between length scales, and the use of parallelization techniques to increase the power of digital computers. As a result, "virtual tests" are starting to be used in engineering applications to limit the number of costly experimental tests to certify the safety of composite structures, and to reduce development time. The overall multiscale simulation scheme is depicted in Figure 9 and takes advantage of the natural separation of length scales between different entities (ply, laminate and structural elements) found in FRC. This allows multiscale modeling to be carried out by computing the properties of one entity (e.g. individual plies) at the relevant length scale, homogenizing the results into a constitutive model, and passing this information to the next length scale to determine the mechanical behavior of the larger entity (e.g. laminate). This procedure is repeated between laminates and structural elements. At each length scale, the main materials properties (stiffness, strength and toughness) are determined for different loading conditions and passed as input for the simulations at the next length scale. Thus, multiscale modeling is carried out through the transfer of information between the three different length scales rather than by coupling different simulation techniques. Obviously, this simulation strategy reduces enormously the number of mechanical tests at different length scales (because the behavior can be predicted with the models). In addition, the influence of the changes in the constituent properties on the structural behavior can be readily predicted, reducing the inertia to introduce new materials and paving the way for optimization of the material behavior at different length scales.

The four main ingredients of this multiscale modeling strategy are the determination of the constituent properties (matrix, fibers and interface) by means of experimental micromechanics, and the simulation of the mechanical properties at the ply, laminate and structural component level using the finite element method by means of computational micromechanics, mesomechanics and mechanics, respectively. The last three items have been recently reviewed [3, 4, 134] and will be briefly summarized in this review. However, the experimental characterization of the constituents will be presented in more detail.

3.2.1 *In situ* measurement of constituent properties

While the mechanical properties of fibers (stiffness, strength) are well known, accurate simulations of the mechanical properties of FRC have to rely on the actual values of the matrix and fiber-matrix interface mechanical properties. Traditional approaches to determine these magnitudes rely on the fabrication of resin coupons to carry out conventional mechanical tests to determine the resin properties [135], and of *ad hoc* specimens, like single fibers embedded in a resin coupon (pull-out test) [136] or a resin drop cured on a single fiber (the microbond test) [137] to determine the fiber-matrix interface strength. However, it is not possible to reproduce the actual composite processing conditions in these *ad-hoc* specimens because the presence of the fiber network introduces local changes in the temperature and pressure distribution during processing that can affect the curing of the resin [3]. Moreover, these *ad hoc* samples cannot be used to reproduce the aging process of the real composite during service, an issue of pressing interest nowadays. Therefore, the best solution relies on using micro and nanomechanical testing techniques that can be applied directly to the real composite material, either as processed or under aging conditions. Hence, the term *in situ* determination of the composite constituent properties used in this context [3].

3.2.1.1 *In situ* micromechanical characterization of matrix properties

The matrix plays a key role in the non-linear response of FRC plies under transverse compression and in-plane shear. The most appropriate method for the *in situ* determination of the elasto-plastic properties of the matrix is the use of nanomechanical testing techniques into the matrix pockets between the fibers on a polished cross-section of the composite. There are a number of studies [138, 139, 140] that have used instrumented nanoindentation to determine the resin properties in FRC. There are usually two main challenges that need to be overcome. First, the nanoindentations must be placed sufficiently far away from the stiff fibers to avoid their constraining effect [3, 138, 140]. This is not difficult in FRC because resin pockets at the interply spaces or at the boundaries of the original fiber tows are easily found (even when the fiber volume fraction is 50% or 60%) due to the inhomogeneous local arrangement of the fibers. Indentations performed in a resin pocket of a FRC are shown in Figure 15(a), while the corresponding load-displacement curves of two of these indentations are plotted in Figure 15(b). One of them shows a sharp stiffening effect induced by a carbon fiber that was very close to the indentation, while the other represents the actual resin response.

The second and most difficult challenge is, however, to extract the mechanical properties of the resin from the nanoindentation curve. Deriving mechanical properties from instrumented indentation is a challenging task and although numerous studies exist in the case of cohesive materials, like metals, there are just a few works that study the case of cohesive frictional material like polymers, where the plastic behavior is strongly influenced by the hydrostatic stress component [141, 142, 143, 144, 145]. Rodriguez et al. [139] established a methodology to determine the constitutive elasto-plastic behaviour of cohesive-frictional materials from instrumented indentation. The methodology was derived by carrying out a parametric study of the indentation process for materials following the Drucker-Prager yield criterion using the finite element method. The Drucker-Prager yield criterion assumes that, for a given stress state $\sigma$, yielding takes place when the Mises equivalent stress $q$ reaches a critical value that depends on the hydrostatic pressure stress $p$ through

$$\Phi(\boldsymbol{\sigma},\beta) = q - p \tan \beta - d = 0 \qquad (1)$$

where $d$ and $\beta$ stand for the cohesion strength and the frictional angle, respectively. The cohesion strength can be related to the uniaxial compressive yield strength $\sigma_{yc}$ according to

$$d = \sigma_{yc} \left(1 - \frac{1}{3} \tan \beta \right) \qquad (2)$$

Thus, four parameters are required to define the elasto-plastic behavior for materials following the Drucker-Prager yield criterion: the elastic modulus $E$ and the Poisson ratio $v$ for the elastic behavior, and the uniaxial compression yield strength $\sigma_{yc}$ and the frictional angle $\beta$ for the plastic behavior. Unfortunately, the output of an instrumented indentation test is the load-displacement curve during the loading and unloading of the indenter, as shown in Fig. 16(a) and this curve alone cannot provide information to determine the four parameters. This is partly due to the uncertainty in the pile-up/sink-in behavior that can take place in the material surrounding the indenter [146], as schematically illustrated in Fig. 16(b).

The hardness $H$ is defined as the indentation load $P$ divided by the contact area of the residual imprint $A_c$

$$H = P / A_c \qquad (3)$$

and, in order to avoid the uncertainty introduced by the pile-up/sink-in behavior, the proposed methodology makes use of two parameters that can be easily determined from the indentation curve: the apparent hardness $H_{ap}$ and the ratio between the elastic and the plastic work of indentation $W_e/W_p$. The apparent hardness $H_{ap}$ is given by

$$H_{ap} = P / A_{max} \qquad (4)$$

where $A_{max}$ is the projected area of the indenter at the maximum displacement $h_{max}$ and $W_e/W_p$ can be easily determined from the areas below the indentation curve, as shown in Fig. 16(a). It is interesting to note that the pile-up behavior can be mathematically described by the parameter $c_p$

$$c_p = \sqrt{\frac{A_c}{A_{max}}} \tag{5}$$

and therefore

$$H = \frac{H_{ap}}{c_p^2} \tag{6}$$

Moreover, the elastic modulus $E$ can be calculated as [147]:

$$E^* = \frac{S\sqrt{\pi}}{2\sqrt{A_c}} \tag{7}$$

where $E^*$ is the reduced modulus that accounts for the elastic deformation of the indenter with an elastic modulus $E_i$ and a Poisson ratio $v_i$ according to:

$$\frac{1}{E^*} = \frac{1-v^2}{E} + \frac{1-v_i^2}{E_i} \tag{8}$$

The methodology was derived for any material following the Drucker-Prager yield criterion, including ceramic and metallic glasses. The apparent hardness $H_{ap}$ is plotted In Figure 17(a) vs. the $W_e/(W_e + W_p) = W_e/W_t$ ratio for materials with varying frictional angles, while the different levels of pile-up expected in each case are depicted in Figure 17(b). Since amorphous polymers typically display $W_e/W_p$ ratios below 0.5, it is evident from this figure that prior knowledge of $E^*$, $\beta$ or $\sigma_{yc}$ is required to solve the inverse problem due to the uncertainty in the level of pile-up. The methodology has been applied to epoxy resins in several studies [139, 178], assuming that the friction angle is of the order of 30º, which is within the range determined by Quinson *et al.* [148] and it is also consistent with the orientation of the fracture planes in FRC plies subjected to transverse compression [177]. From $\beta$, $\sigma_{yc}$ can be univocally determined from $H_{ap}$ and the $W_e/W_t$ ratio using the outcome of the parametrical study gathered in Figure 17(a). Finally, Figure 17(b) provides the pile-up parameter $c_p$, from which $E^*$ can be easily obtained from equation (7). Thus, it is possible to measure the compressive yield stress of the resin at different locations of the composite to determine, for instance, differences in the degree of curing.

If prior knowledge of the frictional angle $\beta$ is not available, a full characterization of the resin requires an additional micromechanical test. One possibility is to use another indenter with a different geometry, such as a cube corner or a spherical indenter [149]. However, this is not always easy because the smaller angle introduces additional uncertainties in the analysis due to frictional forces for cube corner indenters, while spherical indenters generally leave large impressions that are not always easy to fit on the available resin pockets in the cross-section of the composite specimen. An alternative might be the use of other micromechanical tests, like the compression of micropillars milled out of the resin pockets with the aid of a focused ion beam (FIB), as shown in Figure 18 [150].

The interpretation of the micropillar compression tests faces the difficulties associated with the damage induced by the FIB when milling the micropillars. Indeed, the interaction of the Ga+ ions with polymer resins results on a hard skin, of tens of nm in thickness, on the surface of the micropillars (Figure 19b), that strongly affects the deformation of the pillar [151]. The presence of this hard skin results in a strong size effect in the compression strength $\sigma_{yc}$ determined by micropillar compression, as shown in Fig. 19(a). The pillar can be considered as a composite material, where the polymer core is wrapped around by the hard FIB-induced skin. It behaves like a thin-walled pipe constraining the radial expansion of the pillar and artificially increasing the value of $\sigma_y$, due to the hydrostatic stresses. Therefore, it is possible to account for this size effect by, for instance, performing compression of micropillars with varying sizes and extrapolating the results to large pillar sizes (i.e. negligible hard skin thickness), using models that account for this effect [151]. In this way, micropillar compression experiments can be used to obtain the compressive strength of the resin, $\sigma_{yc}$, while the frictional angle $\beta$ and the elastic modulus $E$ are obtained from the nanoindentation experiments. In fact, the combination of nanoindentation and micropillar compression was successfully used to determine the mechanical properties of the epoxy resin ($E$ = 4.5 GPa, $\sigma_{yc}$ = 165 MPa and $\beta$=31°) in a Hexcel AS4/8552 FRC [150].

3.2.1.2 *In situ* micromechanical characterization of interface properties

The most suitable *in situ* technique to measure the interface strength is to use either the push-in [152, 153] or the push-out [154, 155] test on a composite cross-section. In both cases, a nanoindentation system is used to push a single fiber in the cross-section of a specimen until debonding of the interface takes place. The push-out test requires preparing a thin membrane from the composite specimen (of the order of the fiber diameter), whose preparation is time-consuming, but push-in tests can be readily carried out on bulk composites, as illustrated in Figure 20(a). The main advantage of this technique is that it does not require any laborious sample preparation but interpretation is somehow difficult because the length of the debonded interface below the surface is not known. This type of test is very sensitive to the resin elastic properties as well as to the constraint imposed by the surrounding fibers. The initial load-fiber displacement response in the absence of interface debonding is linear, as shown in Figure 20(b), and the departure of linearity marks the onset of interface debonding, as indicated by the images of Figure 20(b). In general, the results are easily analyzed in terms of a simplified shear lag model [153, 156, 157] in which the fiber displacement $u$ is given by:

$$u = \frac{P}{n\pi r E_f} \quad (9)$$

where $P$ is the applied load, $r$ the fiber radius, $E_f$ the elastic modulus of the fiber and $n$ a dimensionless parameter that depends on the elastic properties of the embedding matrix, i.e., the elastic properties of the resin and the configuration of the surrounding fibers. $n$ can be easily determined from the experimental slope of the linear $P$-$u$ response from which the interface shear strength at the onset of debonding $t_c$ can be computed according to

$$t_c = \frac{nP_c}{2\pi r^2} \quad (10)$$

where $P_c$ stands for the critical load at the onset of debonding.

Nevertheless, it was shown [158] that the interface strength obtained with this methodology underestimated the actual properties, particularly in the case of composites with a large fiber volume fraction (> 50%) because it did not take into account accurately the constrain of the surrounding fibers. Moreover, the influence of other factors (elastic anisotropy of carbon fibers, thermal residual stresses or interface friction) was not taken into account in this analysis. Rodriguez *et al.* [158] carried out a full parametric study of the push-in test for carbon fiber and glass fiber reinforced polymer-matrix composites using the finite element method, which took into account the elastic anisotropy of the fibers, the constraint induced by the surrounding fibers and the effect of potential residual stresses arising from processing and/or hygrothermal aging. In order to achieve good reproducibility, they concluded that fiber push-in tests must be performed on the central fibers of highly-packed clusters with hexagonal symmetry, a feature easily found in high-density unidirectional plies (Figure 20c). The main advantage of the finite element simulation is that fiber debonding can be simulated by inserting cohesive elements at the matrix-fiber interface, so that very accurate results can be obtained by using the experimental push-in test results as the basis for the calibration of the constitutive equation of the interface elements. The methodology has been employed successfully in a number of FRC (AS4/8552, AS4/PEEK and S2/MTM44-1) to measure the fiber-matrix shear interface strength [155, 159]. More recently, these micromechanical techniques have been used to assess the strain rate sensitivity of the polymeric matrix [160] or the effect of environment degradation on the strength of the polymeric matrix and of the interface in FRCs. [161].

3.2.2 Computational micromechanics

Unidirectional FRC plies are the building blocks of the multidirectional laminates used in the majority of structural applications. Unidirectional plies are highly anisotropic and can undergo either fiber- or matrix-dominated damage upon loading. At the ply level, computational micromechanics can be used to determine the intrinsic ply properties from the thermo-mechanical properties of the fiber, matrix and interfaces taking into account the fiber shape, volume fraction and spatial distribution within the ply [3, 162, 163]. In addition, computational micromechanics can be used to obtain the ply properties under stress states that are rather difficult to reproduce in the laboratory, such as biaxial or triaxial loading

Damage of unidirectional FRC plies is very complex because five different failure mechanisms can be identified, depending of the loading conditions. The tensile strength of the ply loaded parallel to the fibres, $X_T$, is controlled by the brittle fracture of the fibers while the compressive strength along the same orientation, $X_C$, is dictated by the development of fiber kinking. If the ply is subjected to tensile stresses perpendicular to the fibers, the tensile strength, $Y_T$, is governed by matrix and interface failure, which lead to the formation of a crack perpendicular to the tensile axis. The compressive strength perpendicular to the fibers, $Y_C$, is, however, dominated by the formation of a shear band of localized plastic deformation in the matrix across the ply. Finally, the shear strength parallel to the fibers, $S_L$, is controlled by the matrix and interface shear strength. It is important to notice that these failure mechanisms are linked to each other, and therefore, the maximum load-bearing capability of the ply under any stress state is dictated by a failure locus in the stress space. Computational micromechanics offers a novel approach to determine the failure locus associated with the different failure modes of unidirectional plies by means of computational homogenization of a Representative Volume Element (RVE) of the ply microstructure [164-166].

Thus, computational micromechanics begins by the generation of a RVE of the microstructure of the unidirectional ply. This RVE is obtained from a periodic dispersion of parallel fibers randomly distributed within a parallelepipedic domain. The fiber distribution in the plane perpendicular to the fibers can be generated using either random dispersion algorithms [167] or statistical descriptors of the fiber distribution obtained from cross-sections of the fiber distribution in the FRC [168-170]. In addition to circular fibers, RVEs with different fiber shape (for instance, lobular, polygonal, kidney, etc.) can be generated to ascertain the influence of this factor on the mechanical properties [171]. The planar distribution of fibers is then extruded in the fiber direction to create a three dimensional representation of the microstructure of the unidirectional ply. It was demonstrated [167] that a periodic RVE containing a random dispersion of a few dozens fibers is large enough to accurately reproduce the ply behavior under general loading conditions, which are introduced by using the appropriate boundary conditions in the finite element model.

The RVE geometry has to be discretized with finite elements and the boundary value problem is solved with the finite element method. In addition, the actual failure mechanisms at the microscopic scale have to be introduced in the finite element model. For instance, fiber/matrix debonding can be handled by means of cohesive elements or cohesive surface interactions while the nonlinear deformation and failure of the polymeric matrix behaviour is taken into account by means of continuum models that include plasticity and damage [171]. The numerical simulation of the RVE can provide the average stress-strain curve under different homogeneous stress states (tension, compression and shear) in the different directions (in-plane and out-of-plane), and their combinations. Moreover, the stress and strain microfields within the microstructure are also computed, which reveal the dominant deformation and failure micromechanisms. This information is very useful to design new microstructures with improved properties.

This methodology was pioneered by González and LLorca [165] and it has been successfully applied to predict the stiffness and strength of unidirectional plies of FRC as a function of material constituents. For instance, the in-plane stress-strain response of $[0/90]_{3s}$ laminates reinforced with high modulus (M40J) carbon fibers embedded in a MTM57 epoxy resin (Advanced Composite Group, ACG) was predicted using the computational micromechanics approach previously described from the properties of the respective constituents [172]. The contour plots of the accumulated plastic strain in the RVE of the M40J/MTM57 composite subjected to in-plane shear along the fiber direction and perpendicular to the fibers are presented in figures 21a) and 21b), respectively. In the case of shear along the fiber direction, $\tau_{12\parallel}$, damage develops in the form of shear bands in the matrix, parallel to the fibers that links the interfacial voids formed by fiber/matrix debonding (Fig. 21a). However, plastic deformation and interface debonding are homogeneously distributed within the RVE when the shear is applied perpendicular to the fibers, $\tau_{12\perp}$, and fiber rotation and scissoring is the main mechanism responsible of the strain hardening observed upon large strains (Figure 21c). The shear stress - shear strain response of the unidirectional ply can be obtained by averaging the shear behavior in the directions parallel ($\tau_{12\parallel}$) and perpendicular ($\tau_{12\perp}$) to the fibers and it was in very good agreement with the experimental shear stress-strain curve of the unidirectional ply, Figure 21c).

This technique has been successfully used to predict the main mechanical properties a the ply level in unidirectional FRC as a function of material constituents as fibers, matrix and fiber/matrix interfaces, including the environmental conditions as the moisture absorbed [171], as well as the study of the role played by defects produced during manufacturing, as the porosity on the mechanical performance of the material [173]. One of the main advantages of computational homogenization using RVEs is the capability to simulate the mechanical behavior under loading conditions that are difficult to reproduce in the laboratory. This includes, for instance, the response of the material submitted to biaxial or triaxial loading conditions including normal and shear stress combinations, and opens the way to determine the failure locus of the unidirectional ply. For instance, Naya et al. [171] determined the failure locus of unidirectional plies of the AS4/8552 FRC in the $\sigma_2$ - $\tau_{12}$ plane by means of computational micromechanics, Figure 22. The failure mechanisms of the undirectional ply in tension and compression perpendicular to the fibers are shown in Figures 22(a) and 22(b), respectively. Failure in transverse tension was triggered by interface decohesion and was followed by the fracture of the matrix ligaments between interfacial voids. Failure in compression perpendicular to the fibers developed by a combination of fiber/matrix debonding and the formation of a shear band in the matrix across the RVE. The failure locus can be obtained by applying different combinations of normal ($\sigma_2$) and shear ($\tau_{12}$) stresses until failure, and it is shown in Figure 22c). The failure locus predicted by computational micromechanics from the properties of the constituents and the fiber volume fraction in the RVE was in very good agreement with that obtained with phenomenological failure criteria (such as, for instance, Puck or LarC). Similar results were reported for combinations of transverse compression and out-of-plane shear loading [174, 175] in carbon/epoxy FRC and transverse compression and in-plane shear in carbon/PEEK FRC [176].

Thus, computational micromechanics has become a powerful and versatile simulation tool that can be used to study the effect of the different microstructural parameters on the material behaviour without the need of extensive and costly experimental campaigns. This virtual material design approach can be used to optimize the microstructure or, for instance, to explore the influence of fiber shape on the mechanical response of the material [120]. The contour plots of the accumulated plastic strain in the matrix and the development of interfacial damage are shown in Figures 23(a) to 23(d) for RVEs containing fibers with different shape (circular, polygonal, bi-lobal oriented and bi-lobal randomly distributed). The fiber volume fraction was 50% in all cases. Failure was caused in all cases by a crack perpendicular to the loading axis that was triggered by interface decohesion, but the transversal strength of the plies, $Y_T$, depended on the fiber shape because of the differences in the stress concentrations around the fibers [171]. Similarly, computational micromechanics was also used to study size effects effects associated with matrix cracking initiation in cross-ply composites. In this case, models containing individual fiber distributions corresponding to 90º plies were embedded between two supporting layers corresponding to the plies oriented in the 0º plies. The laminate was subjected to uniaxial stress and the progression of matrix cracking in the laminate as well as the crack density or distance between cracks was obtained from the simulations. This approach was able to account for size effects associated with ply thickness in composite laminates [173].

In addition to the strength and the failure locus of unidirectional plies, computational micromechanics can be used to determine the intraply toughness by means of embedded models [177, 178]. To this end, the propagation of a crack within the unidirectional ply is simulated by means of the finite element method. Full details of the composite microstructure (including the matrix, reinforcements and interfaces) are resolved in the fracture region, while simple constitutive equations (based on any suitable homogenization approximation) and coarser discretizations are used in the rest of the model to save computer time. An example of this approach to compute the fracture toughness of a unidirectional ply of a FRC made up of glass fibers in an epoxy matrix is shown in Fig. 24(a). The model simulates the crack propagation from a sharp notch during a three point bend test. The details of the fiber distribution were fully resolved in front of the notch, and the actual damage mechanisms (fiber/matrix debonding as well as plastic deformation and fracture of the matrix) were included in this region while the rest of the beam was modeled as a homogeneous, transversally isotropic elastic solid. Crack propagation was successfully simulated using this approach (Fig. 24b) and the results of the numerical simulations in terms of the crack path and the fracture toughness were in very good agreement with the experimental results [178].

3.2.3 Computational mesomechanics

The second step of the multiscale modelling approach (Figure 9) involves the simulation of the mechanical behavior of a laminate built by stacking plies with different fiber orientation by means of computational mesomechanics. The geometrical model explicitly includes each of the plies (which are considered as homogeneous solids) as well as the interfaces between them (Figure 25a). The laminate is discretized using solid elements for the plies while cohesive interface elements, or cohesive surface interactions, are used to model the interply interfaces. In this way, intraply and interply damage can be introduced separately, but the complex interaction between them is taken into account.

The simulation of intraply damage is more complicated. The plies are modeled as homogenized solids. The onset of intraply damage for any complex loading scenario is predicted by the failure locus obtained by means of computational micromechanics, as described above. In addition, experimentally-validated phenomenological models have also been used to predict the failure locus [179-183]. As a result, the combination of stresses which leads to the initiation of damage, and the particular damage mode activated (matrix cracking, fibre kinking, etc.), is known. Nevertheless, laminate failure is not always associated with the initiation of damage and the evolution of damage (including the complex interaction among the different intraply and interply failure mechanisms) has to be included in the simulation. Different approaches have been successfully used to solve this complex problem. In the case of discrete ply models, Figure 25(b), cohesive elements were used to include fiber splitting, matrix fracture as well as interplay delamination [184], while other authors modeled discrete cracks within the framework of the extended finite element method (Figure 25(c) ([185-189]). However, the most extended approach to model intraply damage in composite laminates is by smearing damage within the elements by means of continuum damage mechanics, Figure 25(d) [14, 190-194].

The development of computational mesomechanics strategies to simulate the mechanical behavior of laminates has found several difficulties. For instance, it is difficult to account correctly for the interaction between the matrix cracks and delaminations in angle-ply laminates. In addition, continuum damage mechanics models were not able to correctly predict the intraply damage patterns and very often crack patterns were not consistent with the constraints imposed by the fiber orientations [187]. To overcome these problems, several solutions were proposed. For instance, aligned meshes in combination with continuum damage mechanics were used to model intraply damage while adjacent layers were coupled by means of cohesive surfaces. This strategy was successfully applied to predict the low velocity impact response of angle-ply laminates [193] and angle-ply coupons subjected to plain and open hole tension or compression [195]. Discrete ply models [184, 196] offer similar solutions but inserting cohesive elements between adjacent bricks to model fiber fracture and matrix cracking rather than smearing the damage in the element by means of continuum damage mechanics. These techniques solve the complex problem of matrix crack interaction with delamination but the complexity of the meshing techniques for angle-ply laminates in the presence of holes and coupon details prevent their expansion to more complex geometries. Future developments should make use of kinematically-enhanced finite elements formulations that can deal with individual matrix cracks without the need of complex meshes, Figure 25(c) [185-189].

Computational mesomechanics has been used to predict the maximum load-bearing capability of laminates under complex loading conditions. For instance, the failure mechanisms of an AS4/8552 laminate under under low velocity impact are shown in Figures 26(a) and 26(b) for impact energies of 19.7 and 50.8 J, respectively [193]. The finite element model reproduced accurately the geometry and boundary conditions of the impact test, which follow the recommendations of the corresponding ASTM standard for the drop-weight impact test. Only internal delamination and matrix cracking occurs at the maximum impactor penetration when the impact energy is 19.7 J (Figure 26a), while fiber fracture is also predicted at 50.8 J (Figure 26b), leading to the perforation of the specimen, in agreement with the experimental results. In addition, the permanent specimen indentation after the impact was captured by the model that took into account the influence of the non-linear shear behaviour of the composite, element erosion after failure and intraply frictional effects.

Computational mesomechanics simulations, in which the physical deformation and failure mechanisms at the intraply and interplay level are taken into account explicitly in the analysis, provide accurate predictions of the failure strength of composite laminates. In most cases, model predictions do not differ from the experimental results by more than 10% for typical configurations used in structural applications of FRCs (plain and open hole tension and compression tests) and layups with soft and stiff response depending on the content of 0º plies. As a result, multiscale physically based virtual testing tools open new opportunities to optimize and push materials towards their maximum performance through a better understanding of the competition between failure mechanisms [4, 195].

3.2.4 Computational mechanics

Although computational micromechanics and mesomechanics are very powerful tools to simulate the mechanical behaviour of FRC, they are usually restricted to plies, coupons or small panels due to the computational cost. Therefore, the multiscale simulation scheme has to be extended to the structural level incorporating the physical deformation mechanisms observed at lower length scales.

Within the computational mechanics framework, the laminate is modelled by means of shell elements or continuum shell elements that contain as many integration points through the thickness as the number of plies in the laminate (Figure 27(a)). The numerical analysis is thus limited to bidimensional in-plane stress states and no internal delaminations are explicitly included in the model. These models are very efficient from the numerical viewpoint and can capture the structural failure modes of large structures including instabilities by buckling and large displacements [197, 198]. Moreover, more sophisticated models were developed to account directly for internal delamination in solid-like-shell elements [199] and standard shells [200]. In this case, debonding cracks are modelled as jumps of the displacement fields introduced on-the-fly during the simulation allowing to incorporate delaminations in a very efficient form.

The laminate is treated as a homogeneous material whose mechanical behaviour until failure is dictated by a continuum damage mechanics model simplified for in-plane stress states. Accurate models for the onset and evolution of damage at the laminate level [181, 190, 191, 201-204] are included in order to ensure the fidelity of the numerical simulations. Such methodology has been applied successfully to study the response of open hole composite coupons subjected to tension and compression [205], and the impact response of laminates and sandwich plates [22, 206, 207]. In these latter cases, ply delamination becomes an important energy dissipation mechanism, especially for low energy impacts dominated by out-of-plane stresses. Johnson *et al.* [208] proposed an engineering solution based on the stacking of a set of shell elements governed by a continuum damage model at the ply level coupled together by cohesive interactions, which was implemented in an explicit finite element formulation in PAM-CRASH [209] and validated for low velocity impact tests performed in carbon/epoxy laminates. Similar results were obtained in the case of high velocity impact tests on sandwich panels manufactured with carbon/epoxy skins and PEI and Nomex cores. The core material is treated as an additional physical layer using a homogenized constitutive equation, known as crushable foam, that allows taking into account the densification of the foam upon crushing, a very important factor from the viewpoint of energy dissipation.

An example of a large structural simulation using the computational mechanics approach is the case of a bird impact against a composite component part as the leading edge of an aircraft presented in Figure 27(b) [210]. Similar results of bird impact in composite leading edges were presented in [4]. The impact is simulated using a coupled Lagrangian Eulerian approach. The composite structure is modeled using standard Lagrangian continuum shell elements that include damage by fiber fracture, fiber kinking, matrix compression and matrix splitting following the Hashin criterion. Morever, an additional criterion is used to delete elements and allow perforation once the shell element is fully damaged. The Eulerian mesh is used to account for the large deformations of the bird during impact. The corresponding contour plots of the damage variables for fiber fracture and matrix cracking are shown in Figures 27(c) and (d), respectively. The bird impact produced a strong damage localization. This information was very relevant to assess the damage localization and extension in order to optimize the stacking sequence of the laminate to optimize the structural behaviour of the composite component. Similar results were presented in [4].

**4. Structural composites with enhanced functionalities**

*4.1 Electrical conductivity*

The interest in improving the electrical properties of structural composites and increasing their electrical conductivity is motivated by different fabrication and application requirements. They include the prevention of unwanted electrostatic discharges, electromagnetic interference shielding or damage suppression during lightning strike. The corresponding level of electrical conductivity spans from $10^{-4}$ S/m to around $10^{2}$ S/m, which contrasts with the insulating properties of the polymers used in structural composites (about $10^{-10}$ S/m).

There is a wide range of conducting polymers that have been synthesized and studied over the last decades, but they are intrinsically unsuitable as matrices for FRC because they are not thermoplastic and are difficult to dissolve. Although they can reach high conductivities ($10^{5}$ S/m) at high doping levels, they are sensitive to moisture, and their semi-conducting transport mechanism implies an exponential drop in conductivity with decreasing temperature.

A popular approach to increase the electrical conductivity of FRC has been to add conductive fillers to otherwise insulating polymer matrices. Most studies used graphitic nanofillers (graphite platelets, carbon black, carbon nanotubes and graphene), as they are in general easier to manufacture than metallic fillers. The vast majority of these studies are on systems containing the filler and polymer matrix, but in the absence of macroscopic reinforcing fibres. Electrical conduction in such systems is described by percolation theory based on the concept of excluded volume. Initially, the addition of fillers has no effect on the composite conductivity, until the percolation threshold ($V_{fc}$) is reached at a critical volume fraction, leading to a continuous network of conductive particles and to a large increase in conductivity according to

$$\sigma = \sigma_0 \left(V_f - V_{fc}\right)^t \qquad (11)$$

where $\sigma$ is the composite conductivity, $\sigma_0$ a proportionally coefficient, and $t$ an exponent that depends on the dimensionality of the system and usually takes values between 1 and 5. At high volume fractions of fillers, the conductivity levels off to a limiting value of $\sigma_0 V_f^t$.

The percolation threshold depends strongly on the aspect ratio ($s$) of the filler, roughly as $1/s$. Under typical laboratory conditions $V_{fc}$ < 0.01 wt% for carbon nanotubes (CNTs) with $s \approx$ 100, whereas $V_{fc}$ is > 1 wt. % for spherical particles. This effect, and the high conductivity of highly crystalline CNTs, have made them a popular nanofiller choice.

Figure 28 summarizes literature data for conductivity of polymer matrices with different volume fraction of nanocarbons. It shows that modest volume fractions of CNT can bring composite conductivity to the levels required to avoid electrostatic discharges, and that conductivity saturates at around $10^3$ S/m. Similar results are in general observed with graphene-like fillers and CNTs, albeit with higher $V_{fc}$ for graphene on account of its planar geometry, and a lower $\sigma_0$ because of the degradation of properties after the oxidation/reduction of graphene to render it processable.

The current challenge lies in producing highly nanofiller networks in a composite containing also reinforcing fibres, that is, to produce hierarchical composites of nanofiller/macroscopic fibre/polymer matrix. A first obstacle is in the fabrication alone, mainly due to the large increase in polymer viscosity upon addition of nanofillers. Viscosity increases rapidly with increasing nanofiller content and aspect ratio (proportional to $s^2/\ln s$), which implies that conventional thermosets reach viscosities far greater that the acceptable limits for an infusion process (100-500 cPs), even when the nanofiller volume fraction is small. In addition, the presence of a fiber fabric can act as a filter that blocks the flow of conductive fillers, thus leading to a heterogeneous structure with insulated areas. In fact, it is not clear that the low percolation threshold of high aspect ratio fillers is beneficial in the case of hierarchical composites (i.e. containing macroscopic fibres) processed by infusion. The difficulties in adapting high aspect ratio nanofillers to composite fabrication routes has often meant that low $s$ particles are preferred to increase matrix conductivity in spite of their lower percolation threshold. Suppliers of composite raw materials such as resins and pregpregs offer products with enhanced conductivity through the addition of conductive nanoparticles.

While fillers can increase matrix conductivity, it would seem that the main issue in FRC is the percolation between fibers. Assuming a similar anisotropy ratio as in graphite (1/100), the transversal conductivity of a typical carbon fiber would be $\approx 10^2$-$10^3$ S/m, which is already in the range required for most applications. However, a FRC has a much lower conductivity through the thickness, often by orders of magnitude. The issue is then one of connectivity, starting probably between tows and then between plies. In this respect, spherical particles or low aspect ratio CNTs that can penetrate deeper into a fabric structure and interconnect carbon fibers would seem to be a better alternative.

The difficulties discussed above can be partially overcome by reducing particle flow distances or placing nanoparticles in selected areas of dry fibers or dry fabrics and then infiltrating the structure with the polymer. The objectives are to preserve the low viscosity of the resin and thus ensure adequate infusion, and to locate the fillers at critical interfaces to produce conductive paths transversal to the fiber axis and the ply plane.

Recently, Herceg *et al.* [213] have introduce large $V_f$ of CNTs in carbon fiber fabrics to produce hierarchical composites. The method consists in preforming a thermoset/CNT powder which is then integrated into a fiber fabric and consolidated by compression moulding (Figure 29). This technique reduces flow distances, thus overcoming issues associated with an increased viscosity, while also avoiding filtering problems. Through-thickness conductivities as high as 53 S/m were reported in these hierarchical laminate composites. They also showed improvements in interlaminar properties due to the presence of CNTs and the development of a heterogeneous microstructure.

The two other popular methods include electrophoresis [214, 215] and *in situ* growth of the CNT on the surface of the fibers [216-219]. Both have succeeded in improving interlaminar properties and increasing through-thickness conductivity for different fibers (carbon, glass and alumina). One of the envisaged challenges in the case of electrophoresis is to achieve high CNT deposition rates and to simplify the method to make it similar to a sizing process. For *in situ* growth, there has been significant progress in avoiding the degradation of carbon fibers as a consequence of catalyst pitting during CNT growth [220].

An interesting possibility is to use interleaves containing CNTs that can then be integrated into the lay up and consolidated as an additional interleaf/pre-preg layer using standard consolidation methods. One of the routes explored so far consists in producing prepregs of polymer resin that contain a small amount of CNTs above percolation. The prepreg can then be introduced between carbon fiber plies. Lonjon *et al.* [221] for example produced prepregs with 0.4 wt.% CNTs that led to through-thickness conductivities of 0.21 S/m compared with 0.07 S/m for the control sample.

A more promising route is to use sheets of CNTs aligned in the plane as interleaves, a strategy that has already led to significant increases in interlaminar properties. Those produced directly from the gas-phase during CNT growth by chemical vapour deposition are particularly attractive since this process is already at a semi-industrial scale. Wang et al. [222] have recently used commercial sheets of CNTs synthetized by Nanocomp to produce hierarchical composites (Figure 30). The addition of the CNT veils led to increases in the through thickness conductivity from 4 to 21 S/m, albeit with a small drop in tensile properties [222]. Xu et al. [223] deposited CNT fiber veils directly on UD T700S carbon fabric. In addition to improvements in interlaminar shear strength, an increase in through-thickness conductivity from 7 to 20 S/m was reported.

The conductivity data for the examples discussed in this section are depicted in Table 1, together with a representative overview of the different methods used to produce hierarchical composites by addition of CNTs, and the corresponding through-thickness conductivity values obtained.

In the context of electrical protection against lightning strike, Gonzalez et al. [224] have integrated commercial CNT fiber veils from Tortech as a protective conductive layer on $[0]_8$ woven carbon fabric and compared them against samples with the standard copper mesh. Panels were manufactured by vacuum infusion using the CNT veil as an additional layer integrated in the preform. Preliminary results showed that the samples containing CNT sheets of 80 g/m$^2$ provided similar level of protection as the copper mesh when subjected to 100 kA simulated lightning impacts according to the EUROCAE ED84 standard (Figure 31). X-ray computed tomography confirmed that the CNT sheets were ablated but arrested the progression of internal delamination in the 2-3 plies in the area of the lightning attachment. These results are encouraging, as the CNT veils were not optimized for this particular application and had relatively low in-plane conductivity (below $10^5$ S/m). The success in limiting delamination after low-energy lightning strike highlights the fact that this is a complex phenomenon requiring not only electrical conductivity, but also maximum current density and thermal conductivity under extreme conditions.

There have been great advances in the improvement of electrical properties of structural composites, mostly directed at increasing their conductivity through the thickness. The addition of nanocarbons, especially CNTs, has been a successful route. Whereas initial work in the field focused on their simple dispersion in the matrix, it became clear that more efficient strategies were required in order to reduce CNT flow distances and to place the nanofillers at relevant interfaces. There are now various routes to produce hierarchical composites with high $V_f$ of CNTs from dispersion or as interleaves and they are *a priori* compatible with standard FRC processing routes. There is unlikely a one-fits-all solution; the reference carbon fiber composites themselves have conductivities that span two order of magnitude (see Table 1). Instead, advanced solutions will be developed *ad hoc* for different composite fabrication processes (RTM, VARI, etc.) taking into account the fiber, matrix, infusion method and final application.

*4.2 Structural energy harvesting and storage.*

There is growing interest in the composite community in integrating energy management functions, such as harvesting, storage and distribution in structural composites. This trend is particularly relevant in the transport sector, where materials properties, weight of components and fuel consumption go hand in hand, and thus, where greener energy sources and more efficient energy management are sought. In this context, there has been extensive work on developing multifunctional structures and composites, that is, systems that can perform at least two functions: one mechanical and one associated with energy management. Broadly speaking, the *ethos* behind *multifunctional structures* consist in the integration of energy management devices into the load-bearing composite structure, but without the device bearing any substantial load. This was initially achieved by embedding the energy-active phase between composite lamina in a way that produces the minimum disruption to the laminate properties and results in a net mass/volume saving. In contrast, *multifunctional composites* consist of structures where the composite structure performs both mechanical and energy management functions, offering potentially higher weight reductions. This section gives an overview of recent work on multifunctional composites and their associated challenges. Reviews on multifunctional structures can be found elsewhere [225].

4.2.1 Basic properties of materials for multifunctional composites.

Structural composites for transport applications predominantly use carbon fibers. The optimisation of the structure and fabrication of carbon fibers has been very successful, leading to fibers with very high stiffness and strength. Multifunctional composites exploit additionally the electrical conductivity of carbon fibers which enables using them as electrodes. Thus, the current produced at the harvesting material through a variety of mechanisms (photovoltaic, piezoelectric, triboelectric, thermoelectric, etc.) is transferred to the carbon fiber. The scheme in Figure 32a shows the transfer of an electron to a carbon fiber anode after irradiation of a semiconductor. In energy storage configuration, the charge is accumulated as ions via two mechanisms. Ions in the electrolyte are electrostatically bound to the carbon fiber surface in an electric double-layer supercapacitor (Figure 32b). Using carbon fibers as a battery anode, ions are produced via an electrochemical reaction and intercalated between graphitic planes in the carbon fiber microstructure (Figure 32c). In a pseudocapacitor, a thin layer of metal oxide or a polymer deposited on the current collector (in this case the carbon fiber) transfers charge to it as it undergoes a redox reaction.

The process of charge transfer (from harvester or redox reaction) and electrostatic (non-Faradaic) charge storage are interfacial processes, and it is obvious from the simple schemes in Figure 32 that energy densities (stored and harvested) in multifunctional composites increase with the accessible surface of the electrode. Carbon fibers have an intrinsically small specific surface area (SSA) of around 0.2 $m^2/g$, compared for example with activated carbon (>1000 $m^2/g$) which is typically used in commercial energy storage devices. The SSA of carbon fibers can be increased upon activation, but these treatments are difficult to scale up and can reduce dramatically the mechanical and electrical properties of the fiber.

On the contrary, nanomaterials have intrinsically large surface areas. The SSA of an individual CNT is > 1000 $m^2/g$, and even a conventional array of CNTs in the form of a sheet has a SSA above 100 $m^2/g$, 500 larger than carbon fibers. In the context of structural energy-managing devices, continuous macroscopic fibers of CNTs have emerged as a very promising electrode material. These fibers have altogether a different microstructure than that of carbon fibers. They consist of a network of long CNT associated in bundles and predominantly aligned parallel to the fiber axis. This network structure is similar to that of yarns [232] and enables the coexistence of a large porosity and extended crystalline domains. Figure 33 shows an example of the hierarchical yarn-like structure of a CNT fiber and a carbon fiber.

The combination of mechanical properties (specific strength > 1GPa/SG, specific modulus > 30 GPa/SG) and large SSA (> 100$m^2/g$) of CNT fibers make them attractive for applications requiring both load-bearing and charge storage/transfer. CNT fibers have been extensively studied for over a decade. Their structure-properties are still very much under development, but the high level of control in laboratory synthesis [23,24] and the emergence of semi-industrial samples [235, 236] are enabling studies on their use for composites [226-230] and large-area electrodes [237]. Figure 34 shows literature data for the tensile properties of polymer composites based on CNT fibers and compared with common carbon fiber composites. While there is clearly large room for improvement, it is apparent that the composites based on CNT fibers are approaching the structural level. CNT fiber electrodes have been extensively studied as a component of energy harvesting/storage devices with augmented mechanical properties (flexible, tough, stretchable) [238-240] because of their combined large surface area and mechanical and electrical properties.

Properties of carbon fibers, CNT fibers and their composites are summarized in Table 2. Overall, the SSA of carbon fibers is orders of magnitude lower than that of nanocarbons, and remains a factor of > 10 lower after mild chemical activation whereas nanocomposites have tensile strength/modulus that cannot yet rival with those of carbon-fiber reinforced composites. It is thus evident the properties of carbon fibers and nanocarbons are complementary in the context of multifunctional composites. Indeed, there are a few examples of hierarchical architectures that combine nanocarbons (and activated carbon) with carbon fibers. Overall, there is a whole range of architectures, from those based purely on CNTs to those based on carbon fibers that can lead to a wide spectrum of mechanical-energy properties in multifunctional devices. Combination of nanocarbons and carbon fibers for multifunctional energy-related devices are likely to receive significant attention in the coming years as the properties and size/format of the two types of materials can be combined for stress transfer and charge transfer/storage applications.

4.2.2 Structural energy storage

Energy storage in multifunctional composites was initially developed for its military applications, but the imminent electrification of transport is providing enormous momentum to the research in this area. Energy storing devices are increasingly being used in automobiles and small aircrafts to power devices, in regeneration systems and to power the vehicle itself. According to The More Electric Aircraft (MEA) initiative, for example, electrical power going into systems alone (i.e. non-propellant) are currently at around 1400 kW and 600 kW for the Boeing B787 and Airbus A380, respectively [241].

Electric aircraft and automobiles will most likely use a combination of supercapacitors and batteries. This is not only because they inherently require both high power (thrust) and high energy (long-range), but also for regenerative systems and, in general, to extend battery lifetimes. Although these devices are often coupled, there are various different challenges in making structural supercapacitors or batteries.

4.2.2.1 Structural capacitors and pseudocapacitors

A supercapacitor consists of two parallel plates separated by a thin membrane that allows diffusion of ions but not flow of electrons. Energy is stored by the accumulation of ion at the electrodes upon charging. The diffusion of ions is relatively fast, compared for example to rates of redox reactions, and hence supercapacitors are characterized by having a high power density.

Structural laminate composites have a similar architecture as supercapacitors. Work on exploiting this similarity to develop structural composites has been pioneered by Imperial College and collaborators in the STORAGE EU project. Figure 35 shows a scheme of a laminate consisting of two outer carbon fiber layers sandwiching a glass fiber layer. When this layer arrangement is embedded in an ion-conducting matrix, the carbon fiber layers act as electrodes and the glass fiber fabric as separator, and the final composite material can store energy as a capacitor.

In general, there are two main challenges on a materials level to make efficient structural supercapacitors. The first one is the requirement on the reinforcing fiber to present the high SSA of a typical electrode without degradation of the mechanical properties. The second one is to find a matrix that enables ion diffusion and maintains a high elastic modulus, which are conflicting properties. Both challenges are analyzed below

The first challenge in the development of supercapacitors has been the low capacitance of carbon fibers as a consequence of their low SSA. Synder *et al.* [243] carried out an electrochemical study on structural commercial carbon fibers and other graphitic materials. They determined capacitance by cyclic voltammetry (CV) using a two-electrode coin cell with 1.0 M $LiPF_6$ in 30% ethylene carbonate/70% ethyl methyl carbonate at 20 mV/s. Capacitances ranged between 0.4 and 3.5 F/g. For reference, activated carbon can have a capacitance of 100F/g.

Work by the STORAGE consortium has explored various methods to increase SSA with minimum degradation of mechanical properties. These include chemical activation with potassium hydroxide (KOH) (SSA = 21 m$^2$/g) [242], addition of CNTs as a sizing (SSA = 35 m$^2$/g) and the direct growth of CNTs on the carbon fiber surface (SSA = 35m$^2$/g) [244]. More recently, they developed a method to embed carbon fiber arrays into monolithic carbon aerogels. The process consists in impregnating woven carbon fabric with a resorcinol−formaldehyde polymer, then cure at 50-90°C for 24hrs and finally carbonize at 800°C for 30min in a nitrogen atmosphere. The resulting carbon fiber-monolithic carbon aerogel structure has a SSA of 160 m$^2$/g and a number of improvements in composite mechanical properties [245]. Examples of these high SSA carbon microstructures are shown in Figure 36. It is worth noting that carbon fibers with SSA as high as 1100 m$^2$/g are commercially available, but their poorly graphitized structure renders them unsuitable for the structural function.

The carbon fiber-monolithic carbon aerogel samples have been further subjected to electrochemical studies. Capacitance values in three-electrode cell was determined by cyclic voltammetry in 3M KCl at 5 mV/s. The introduction of monolithic carbon aerogels into carbon fabric provided a notable increase in capacitance from 0.06 F/g to 5.9-14.3 F/g, normalized by electrode material (i.e. carbon fiber + monolithic carbon aerogel) [245].

CNT fibers with semi-structural properties have also been subjected to electrochemical studies. The capacitance of as-produced CNT fibers is in the range of 23 to 79.8 F/g. It can be increased after chemical functionalization [246], but the effects on their tensile properties are unclear. Compressive properties of fibres are also critical for structural composites. CNT fibres have modest properties in compression because of the high aspect ratio of their constituents CNTs and the low shear strength between them. Preliminary values for compressive strength have been obtained from tensile recoil (175 – 416 MPa) [247, 248] estimated from composite compression testing (928 MPa) [16] and by Raman spectroscopy measurements (≈3.5 GPa). Here, we take a conservative intermediate value of 0.6 GPa/SG for CNT fibres produced by the direct spinning process [249].

The relevant properties of the fiber-based electrode materials discussed above are compared in Table 3. It includes: AS4 carbon fiber, which is widely used in the composite industry; Toho Tenax HTA, used in [245]; as well as CNT fibers produced at the authors' laboratory and by other research groups. Electrochemical conditions are included for reference. The data in Table 3 provide a good overview of the library of macroscopic materials currently available as reinforcement/electrode in structural supercapacitors (and pseudocapacitors) in spite of the assumptions and different measurement conditions. It shows that carbon fibers and CNT fibers have complementary properties and confirms the enormous potential in combining these two types of materials.

In the simple schematic shown in Figure 32, it is assumed that the polymer matrix can simultaneously be structural and conduct ions. These are in fact conflicting requirements. Stiffness in polymers increases with increasing degree of cross-linking, crystallinity or other arrangement of the polymer chains that reduce their mobility. In contrast, ionic transport in polymeric electrolytes decreases with decreasing polymer chain mobility, hence why non-structural electrolytes are usually liquids whereas non-conducting matrices are rigid thermoplastics or thermosets. The synthesis and electrochemical properties of solid polymer electrolytes are very active areas of research. However, most work in the field is devoted to solid polyelectrolytes that have sufficient mechanical integrity to avoid leakage and withstand large bending or axial strains, typically rubbery materials with modulus in the range of MPa. The solid polymer electrolyte (SPE) for structural composites needs to have stiffness around 2-3 orders of magnitude higher.

Various polymer chemistry tools have been used to realize structural SPE. A popular strategy consists in producing a 3D network that combines structural cross-linked elements and branched chains that enable transport of ions, typically from lithium salts or ionic liquids [244]. The range of properties of a multifunctional polymer electrolyte is depicted in Figure 37 depending on the ratio of structural/conductive monomers. The structural monomer In Figure 37(a) is tetraethylene glycol dimethacrylate (SR209), used as a crosslinker. The ethylene oxide chains in the methoxy polyethylene glycol (550) monomethacrylate monomer (CD552) provide ionic conductivity, and particularly lithium ion conductivity which is of interest for batteries [253]. The plot clearly illustrates the trade-off between mechanical and transport properties of the polymer matrix electrolyte.

Another strategy consists in preparing blends of ionic liquids and structural epoxies that segregate and form two interpenetrating networks [254]. An example of the resulting bicontinuous structure can be inferred from the electron micrograph in Figure 37c). The material shown consists of the remaining structural epoxy after extraction of the electrolyte. This bicontinuous structure can enable simultaneous stress transfer and ion transport. Its morphology and ratio of mechanical/ion transport properties can be adjusted by varying the Li salt concentration (Figure 37b).

The two strategies discussed above show encouraging results, yet they are not exempt of challenges. The addition of ionic liquids to epoxy leads for example to acceleration of epoxy curing [244]. Perhaps more challenging is the fact that the SPE properties measured on SPE membranes or other formats have to be realized in a device in the vicinity of the electrode, which can therefore lead to different properties. The complex morphologies obtained by phase-segregation might be disrupted once they are produced inside a carbon fiber fabric and more so in a porous scaffold such as a carbon aerogel or a CNT network. It is also noteworthy that improvements of multifunctional matrices can be achieved by addition of fillers to the matrix. It is, however, unlikely that this approach is suitable when combined with electrodes with pores in the range of the filler size.

Another strategy exploited for the production of SPE is electropolymerisation. Leijonmarck et al. recently demonstrated the electrodeposition of thin (<500 nm) uniform coating of lithium salt-containing methacrylate-based SPE on the surface of carbon fiber. This material was later successfully used as a battery anode [255]. One of the main advantages of this approach is in reducing the distance for ion diffusion by producing thin SPE layers on the electrode surface.

Electrodeposition of polymers has the added benefit of producing thin conformal coatings on porous materials, therefore solving some of the challenges of producing the desired SPE microstructure already in the device architecture in the presence of the electrode. This method has been used, for example, to coat highly porous CNT fiber-based electrodes with polyaniline, a semiconducting polymer that undergoes electrochemical redox reactions and is therefore used for pseudocapacitors (some examples are discussed below).

4.2.2.2 Examples of devices towards structural supercapacitors

At present, the most advanced structural supercapacitor consists in woven carbon fiber fabrics embedded in carbon aerogels (Figures 38a and 38b) [245]. The balance of energy-mechanical properties depends on the composition of the electrolyte, particularly the ratio of ionic liquid to epoxy, as well as of the choice of separator, and ranges from purely mechanical to purely electrochemical [245]. Early work produced a good combination of properties using a glass fibre separator and an electrolyte of 10 wt. % ionic liquid (1-ethyl-3-methylimidazolium bis(trifluoromethylsulfonyl) imide, (EMITFSI) in a mixture of polyethylene glycol diglycidyl ether (PEGDGE) resin and a triethylenetetramine (TETA) hardener.

The gravimetric capacitance, normalized by electrode mass, was 0.07F/g from chronoamperometry measurements at 0.1V voltage for 60s (Figure 38c). Assuming the same properties at 2.7V, they translate into electrode-normalized power and energy densities of 17 mWh/kg and 672 mW/kg. These structural capacitors have energy and power densities of around 9 mWh/kg and 340 mW/kg considering a fiber plus carbon aerogel volume fraction around 50 vol. %. These results show the difficulty in translating the relatively high capacitance of the carbon fiber/carbonaerogel system measured in three-electrode-cell configuration (up to 14F/g) to structural device properties ($\approx$ 0.03F/g).

Although with modest electrochemical properties, this structural supercapacitor has impressive in-plane shear properties (Figure 38d), which are matrix dominated and critical for structural elements that need to operate in compression. The shear modulus was 895 MPa and the shear strength 8.71 MPa, which are about 17% and 33% of the properties of the purely mechanical control samples. This is a good achievement considering that the elastic modulus of this matrix is only around 6 MPa as a consequence of the addition of ionic liquids. Interestingly, the presence of the carbon aerogel provided stiffening of the conductive matrix in addition to the expected large increase in surface area.

More recently, Greenhalgh *et al*. [256] reported improvements in the electrochemical properties of these systems using a polyester separator and a commercial epoxy resin combined with EMITFSI and LiTFSI. Scaled up laminates (630 mm$^2$) reached a specific capacitance of 0.27 F/g, a power density of 18.7 W/kg and an energy density of 0.2 Wh/kg, combined with high in-plane mechanical properties (shear modulus of 536MPa and shear strength of 2.8 MPa).

Snyder *et al.* [257] produced comparatively simpler structural supercapacitors based on plain and activated carbon fibre fabrics, a polymer separator and a SPE based on conducting CD552, high modulus ethoxylated pentaerythritol tetraacrylate (SR494) and 1M lithium

bis(trifluoromethanesulfonyl)imide (LiIm). Their structural supercapacitor device had an estimated capacitance of 93m F/g, with a composite specific in-plane modulus of 12 GPa/SG and a specific shear modulus of 0.31 GPa/SG.

Reports on semi-structural supercapacitors based on CNT fibers are scarce, but separate mechanical and electrochemical studies on these materials indicate that they can make a large contribution to the field. Benson *et al.* [252], for example, electrodeposited PANI on commercially-produced CNT fiber sheets, which resulted in a capacitance increase from 25 to 250 F/g normalized by electrode. Power and energy density, measured in a two-electrode coin cell using liquid electrolyte, gave values as high as 22 kW/kg and 21 Wh/kg normalized by electrode mass. But the relevance of this study is that the PANI-coated CNT fiber sheets exhibited a specific tensile strength as high as 0.38 GPa/SG, a specific tensile modulus of 15 GPa/SG and a fracture energy of around 77J/g. These are, however, properties of single electrode which are yet to be demonstrated in all-solid devices.

Senokos et al. [251] have recently carried out a study to assess the potential of CNT fiber materials for all-solid supercapacitors. They developed a simple pressing method to combine sheets of CNT fibers with polyelectrolytes based on thermoplastics and TFSI, and produce all-solid supercapacitors as large as 100 cm$^2$ (Figure 39a). The work includes testing of commercial CNT fiber materials as well as providing design parameters to optimize device properties. Their semi-structural design has currently low fibre volume fraction (5 - 10%) and therefore modest mechanical properties (elastic modulus of 790 MPa and tensile strength of 53MPa) (Figure 39b), but notable power and energy densities of 46 kW/kg and 11.4 Wh/kg, respectively. As an indication of device properties, these values are normalised by mass of active material and membrane to give 3.7 kW/Kg and 0.9 Wh/Kg. A simple reduction of separator thickness, not yet optimized, should increase tensile properties by a factor of ≈ 5, which combined with improvements in the assembly of multiple filaments into a unidirectional array could lead to a specific elastic modulus of 18GPa/SG and a specific tensile strength of 0.37 MPa/SG. In a demonstration of the robustness of the CNT fiber/polyelectrolyte interface, the devices could be bent and folded without degradation of their electrochemical properties (Figure 39c).

A summary of mechanical and electrochemical properties for the materials and devices discussed above is provided in Table 4. It shows that carbon fiber-based solutions have reached high mechanical properties, particularly in compression and shear. They are determined by the matrix and it is thus more challenging to make them compatible with energy storage. At the other end of the multifunctional spectrum are supercapacitor devices based on fibers of CNTs. They typically result in high power and energy densities but modest mechanical properties, more suitable for energy storage and electronics that are tough [238], flexible/stretchable or wearable [239]. The opportunity would seem in combining both types of structures in architectures where both the carbon and CNT fibers take part in load bearing and energy storage, but where these functions are appropriately distributed in the two materials according to their inherent properties.

O'Brien has proposed a quantitative analysis of weight savings by multifunctional structures [259] used now also by other research groups [244]. It is based on efficiency factors defined as the ratio of a property of the multifunctional solution to that of the baseline. Weight savings are possible when the multifunctional efficiency factor, $\eta_{mf}$, equal to the sum of mechanical ($\eta_m$) and energy-storage ($\eta_e$) efficiency factors is greater than 1. To illustrate the potential of combining different architectures available, note that taking the elastic modulus of carbon fiber/carbon aerogel system in Table 3, $\eta_m = 0.94$, while the power density of the CNT fiber solution, normalized by the power density of a conventional supercapacitor (Maxwell BCAP0010) gives $\eta_e = 0.8$. The two systems are unlikely to be fully additive, but they still seem to be better positioned combined than separate.

4.2.2.3 Structural batteries

The basic architecture of structural batteries is similar to that of structural supercapacitors: two separated electrodes acting as reinforcement, a polymer matrix that is both stiff and has high ionic conductivity, and current collectors (and often separators) such as glass fibers. Structural batteries have so far been mainly of Li-ion type, most commonly with a transition metal oxide used as a cathode and graphitic carbon as anode. Li oxide is reduced during charging and Li+ ions diffuse through the polymeric electrolyte to the anode, where they intercalate between graphitic layers. The active materials are often mixed with conductive and agglomerating additives and/or supported on a porous material in order to improve the electronic conductivity and the mechanical integrity of the electrodes.

An example of the earlies structural battery architectures developed is shown in Figure 40. It had a laminate arrangement similar to that in a structural composite, with carbon fiber fabric as anode, $LiCoO_2$ or $LiFePO_4$-coated metal mesh cathode and a glass fiber separator. Half-cell measurements on coin-cell samples gave capacities of 120 mAh/h and 100 mAh/g for anode and cathode, respectively. Yet, a fully functional battery based on these materials was not demonstrated [260].

Liu *et al.* [261] prepared a composite battery structure based on the dispersion of electrolyte and cathode materials in a matrix of high-molecular weight PVDF and PEO. The cathode consisted of carbon whiskers and $LiCoO_2$. It reached a specific capacity of 90 mAh/g for $LiCoO_2$, elastic modulus of 650 MPa and strength of 12 MPa. A full device was assembled using an anode of activated carbon (coke) and carbon whiskers, with polymer matrix as separator and Al and Cu grids as current collectors. It achieved 35 Wh/kg at a C/20 rate. The elastic modulus of the device, obtained from three-point-bending, reached 3.1 GPa, although it was presumably dominated by the mechanical properties of the Cu and Al current collectors.

The electropolymerisation method discussed above to prepare carbon fibers with a SPE sizing was developed using polymers with high Li-ion diffusion and therefore suitable for batteries. Leijonmarck et al. assembled a battery using coated fibers as anode, lithium metal as cathode and glass fibers as separators. The capacity of this material at C/5 was 102 mAh/g [255]

Based on the electropolymeration method, a new concept of structural battery has been recently introduced by Asp and coworkers (Figure 41) [262]. In the proposed structure, the carbon fiber constitutes the anode's active material, which is coated with a thin layer of electrically insulating polyelectrolyte. The anode and electrolyte ensemble are in ionic contact with a PVDF-based matrix containing $LiFePO_4$ and activated carbon, which act as a cathode. The breakthrough in this design is the reduction of ion diffusion lengths by making a thin SPE that can potentially simultaneously act as a separator, therefore reducing ohmic losses. The challenge is the controlled production of a good dispersion of cathode material that is percolated by a conductor that thus connects it to the current collector.

In spite of the extensive research using nanocarbons for battery electrodes [263], most of work on multifunctional batteries has targeted flexibility in bending [238, 264] or stretchability [265]. There are however, examples of CNT fiber-based cathodes of interest for structural batteries. Evanoff et al. [266] developed a route to coat CNT fibers with silicon (Si). Si has the highest lithium storage capacity, but undergoes a large (>300%) volumetric expansion during insertion. Whereas such expansion is unsuitable for brittle monolithic materials, the network structure of CNT fibers makes them more suitable to absorb the expansion without damage. Si-coated CNT fiber electrodes have reached 500 mAh/g during 150 cycles using a liquid electrolyte, combined with a specific strength of 100 MPa/SG.

The realization of structural batteries that outperform monofunctional components will require combined improvements on electrolyte, electrodes and device architecture. Work on solid polymer electrolytes has explored the use of block-copolymers, grafting, phase-segregated bicontinuous structures, and electrodeposited thin coatings. The latter strategy is particularly attractive as it enables selective deposition of a thin layer of SPE on the electrode surface, thus potentially reducing ion diffusion lengths and the overall volume fraction of SPE.

Carbon fiber is likely to remain the preferred choice of reinforcement/anode material. Their lithiation capacity depends on the fiber type. It can approach the theoretical value for graphite (372 mAh/g.), but it drops rapidly after cycling [243]. This reduction is more pronounced in tows than in individual filaments [267]. It is very encouraging that the tensile properties of carbon fibers are not dramatically affected by lithium intercalation. Tensile modulus remains unaltered and tensile strength drops by less than 40% after lithiation [268]. Altogether, a better understanding of the structure-properties in lithiated carbon fibers is desired. This is likely to come from a combination of simulation and *in situ* spectroscopy and diffraction techniques.

Translating the properties of electrodes into a full device is very challenging. The sole fabrication of structural battery devices is complex and imposes design constraints. The possibility of undesired electric short circuits, for example, leads to the use of separators, which lead to larger ion diffusion lengths (large ohmic drop) and reduced electrochemical properties. In this sense, for the electrodeposition of SPE or similar thin-layer techniques to be viable they need to ensure very uniform coating of the electrode.

A further opportunity for development is the combination of materials. Cathode materials with higher capacity than most common insertion materials but which undergo large volumetric expansions, for example sulphur or $Li_2S$, probably require to be supported on a nanocarbon network scaffold. A similar nanocarbon material could help solve the problem of interconnecting the cathode material and linking it to the current collector. Though still under development, inorganic solid-state electrolytes [269] might prove to be an attractive solution for combined ionic conductivity and stiffness, particularly in compression.

Other topics of interest include: the development of Li-air batteries, which have the benefit of not requiring a cathode and have already been tested with CNT fibers and molten polyelectrolytes [270]; exploiting the benefits of using a graphitic current collector to avoid Cu dissolution in the electrolyte; understanding the mechanisms by which SPE mitigate formation of dendrites and therefore extend battery safety and lifetime; and taking advantage of the complex architecture in structural batteries for short circuit detection [271] and eventually mitigation.

4.2.3 Energy harvesting

The field of structural composites that harvest energy is much less developed than energy storage. One of the main reasons is their more complex device architecture, fabrication and testing. Whereas carbon fibers play an active role in energy storage (Figure 32), it is used only as a current collector in most energy harvesting devices. The integration of the active phase, typically a semiconductor, often requires to grow it directly onto the carbonaceous electrode via sol-gel or gas-phase deposition. A counter electrode has then to be brought in contact to close the circuit.

Current developments are mainly on the level of a reinforcement material showing energy harvesting capabilities. The examples discussed in this section have been selected to show improvements in structural energy harvesting from photovoltaics, piezoelectrics and others. Triboelectric and thermoelectric are not discussed here.

4.2.3.1 Main sources of energy and associated challenges

Piezoelectric harvesting

Piezoelectricity is based on the polarization of non-centrosymmetric crystals under an applied stress. An archetypical material is ZnO, with wurtzite hexagonal structure consisting of alternating planes of $Zn^{2+}$ or $O^{2-}$, which produces a permanent dipole upon deformation (Figure 42). The constructive addition of crystal dipoles results in a macroscopic potential difference. Higher aspect ratio of the piezoelectric leads to larger deformations and higher piezopotentials and thus nanowires are the preferred architecture.

The process to extract current from the deformation of a piezoelectric nanowire is illustrated in Figure 43, which serves to highlight important features of the process. The nanowire is attached to a conductive substrate in this case and it is deflected with a Pt atomic force microscope (AFM) tip that also closes the electric circuit and enables measurements of current (*I*) and/or voltage (*V*) in the strained state. Figure 43a shows how the deflection of the nanowire produces a potential across its width. The low work function of Pt forms a Schottky barrier at the tip/ZnO interface that serves as a diode. As a result, current flows through the system only in forward mode, that is, when the tip is on the compressive side of the nanowire. Using an Al tip, which forms an ohmic contact, results in negligible piezovoltages (Figure 43b). Thus, diode rectification (i.e. Schottky barrier) at one of the contacts is seen as a prerequisite for energy harvesting. It prevents electron flow from the metal to the semiconductor and thus leads to their accumulation at the interface, from where they can be discharged through the external circuit [273]. An insulating layer, such as a polymer, can play a similar role as the barrier, although with a higher resistance [273].

The basic operation principle discussed above has been exploited to make a vast range of energy harvesting materials to power numerous devices [274]. Most of these have been flexible [275] or wearable [275], as these applications combine large deformations and modest mechanical properties. Yet, there are emerging examples of structural composite materials with energy harvesting functions.

Malakooti *et al.* [276] have produced a laminate composite based on woven fabrics of high-performance fibres and ZnO nano/micro pillars in an epoxy matrix, and carried out a thorough analysis of its piezoelectric properties. The details of the microstructure are shown in Figures 44(a) to (b). It includes a layer of aramid fibers with radially-protruding ZnO pillars between two carbon fiber layers, one of which has a thin Au coating in order to produce a rectifying interface. The composite structure was shown to produce piezoelectric voltage/current signals under vibration in a cantilever geometry, with a maximum output power of around 3.7 nW for 5 x 50 mm$^2$ beam samples (Figure 44d). Interestingly, the addition of ZnO led to substantial increases in tensile strength and stiffness (Figure 44c), attributed to enhancements in the properties of aramid fibre after ZnO growth as well as in inter-ply stress transfer.

These results show that the modification of structural composites to enable piezoelectric energy harvesting can in fact lead to improvements in some mechanical properties. However, further work is required to increase specific output power, which is still in the nW/g range even under optimum load resistance and resonant vibrational frequency conditions. This might require further electronic studies to inspect the resulting composite cross sections (Figure 44c) as electronic junctions.

Photovoltaic harvesting

The introduction of photovoltaic functionalities in structural composites has an obvious constraint in that the thickness of a typical photovoltaic cell and particularly its active area (i.e. without current collector or encapsulation) is in the thin film range ($< \approx 1$ μm). Thus, only the outer most surface can contribute to energy harvesting for an opaque structural composite, and the active material inside would correspond again to a thin layer even in the case of semi-transparent structural composites. This is in contrast with energy storage or piezoelectric energy harvesting which benefit from a large interface between the active component and the electrode/reinforcement. Yet, the high efficiency and energy/power output of photovoltaic processes makes them very attractive. A simple estimate gives specific power of $> 50$ W/kg for commercial solar cells, orders of magnitude higher than piezoelectrics.

There are various reports where solar cell modules are integrated in structural composites. Although these are multifunctional systems, they provide an interesting benchmark against multifunctional composite materials. An archetypical example of this approach consist in laying a Si photovoltaic thin film on a conventional fibre lay-up and curing it [277]. The fabrication is reasonably straightforward, as the photovoltaic layer can be treated as another layer of a pre-preg structure and the whole lay-up cured for example in a vacuum bag. The advent of flexible thin solar cells supported on polymer films could further simplify the process.

Photovoltaic structural composite systems have been subjected to mechanical tests. Contrary to structural energy storing devices, the load-bearing structure of the composite remains largely unaffected by the introduction of the energy-harvesting layer. Instead, a clear challenge is in the degradation of the properties of the photovoltaic materials under modest strains. Tensile deformation >1% for Si or even > 0.3% for metal oxides reduces the photovoltaic efficiency significantly due to a drop in carrier mobility and eventually to the formation of microcracks (see [277] for a discussion).

There has been comparatively much less work on developing structural photovoltaic materials that can carry both structural and energy harvesting functions. Most reports relevant to the topic consist of studies in which semiconductors are grown on the surface of structural carbon fibers and subjected to basic photovoltaic measurements. Xu et al. [278] used a gas-phase deposition method to grow polycrystalline Si on individual carbon filaments. They could produce a junction of p- and n-doped poly-Si connected to the carbon fiber by subsequent doping of the Si and formation of appropriate contacts. Photogenerated carriers produced upon light absorption in the Si are separated by the built-in field in the depletion region of the p-n junction, after which electrons drift to the n-doped side and recombine at the carbon fiber through the external circuit. This report showed one of the first examples of the use of carbon fibers in a photovoltaic device, although the efficiency was very low (0.04%) partly on account of the high series resistance of the device. Recent theoretical work suggests that large improvements in efficiency can be achieved by optimisation of the thickness of the different poly-Si regions [279]. Further studies have included the growth of CdS nanowires on carbon fibers to produce a range of solar cell types [280]. However, they have reached low power conversion efficiencies (< 0.04 %).

When using a carbon fiber as current collector in a structural photovoltaic, the semiconductor is directly grown on the carbon surface. The growth methods used, however, are likely to produce a semiconductor with low crystallinity and a large number of defects compared to photovoltaic-grade Si, which thus reduce efficiency due to recombination of charges before there are separated and transferred through the carbon fiber and the external circuit. The devices have an overall large series resistance (>10 $\Omega$ cm$^2$), which further reduces efficiency. This can be due to a high resistance at the carbon fiber-SC interface. The ideality factor, $\eta$, parametrises how closely charge transmission in a junction follows the ideal Schottky diode. $\eta$ can be as high as 3.2-9.7 [280] for carbon fiber-CdS heterojunctions whereas $\eta \approx 1$ for conventional solar cells. This comparison indicates that there is large room for improvement of the carbon fiber-SC electronic interface.

CNT fibers have been extensively used as current collector for solar conversion devices [28]. One of the geometries explored so far is as front electrode deposited onto a monolithic semiconductor structure [281, 282]. It is not clear that such architecture lends itself to structural materials, but the photovoltaic properties are high, with ideality factors of 1.15 [282], saturation current of 35.6 mA/cm$^{-2}$, open circuit voltage of 0.6V and reaching efficiencies of 13.1% [283]. Similarly, CNT fibers have been coated with perovskites and carrier blocking layers to produce flexible solid-sate photovoltaic devices [284]. This includes an arrangement of two twisted strands in contact that can operate as a device and retain their efficiency (3%) after multiple bending cycles [285]. Recent improvements put efficiency at 7.1% while preserving some flexibility [286].

An interesting possibility for structural solar conversion devices is to shift to a dye-sensitized solar cells (DSSC). DSSC are photoelectrochemical cells in which photocarriers are first generated by the oxidation of a dye upon light absorption. Electrons are transferred to a SC, typically a metal oxide and into an external circuit and finally reintroduced by a counter electrode and transferred through the electrolyte to regenerate (reduce) the dye. In the context of multifunctional composites, DSSC have the advantage over traditional photovoltaic of being compatible with roll-to-roll manufacturing techniques. Additionally, the use of a nanostructured porous metal oxide lends itself to produce a network architecture with superior ductility compared to a monolithic semiconducting phase. However, a great disadvantage of DSSC from a structural point of view is the use of a liquid electrolyte.

Carbon fibers were first used as anodes in ZnO DSSC, but with low performance (350 mV, 11.2 mA/cm$^2$, and 0.28) [287]. A seed-growth solvothermal process was later used to grow TiO$_2$ nanowires directly on carbon fibers (Figure 45). These filaments could be used as flexible photoanodes in a DSSC, with 2.48% efficiency under visible light irradiation [288]. A schematic and a photograph of a device, as well as data for current density and output power under different illumination conditions are shown in Figure 46. Used as counterelectrodes, carbon fibers coated with platinum and used in a DSSC with a cobalt electrolyte reached power conversion efficiencies as high as 8.97% [289].

There are also numerous studies on the of CNT fibers as electrodes in DSSC, mainly in the context of wearables [290]. Some of the first studies consisted of intertwining two CNT fibers: a photoanode containing $TiO_2$ and dye molecules and a counter electrode to close the circuit. The schematic of the device can be found in Figure 47, while the actual microstructure of the intertwined fibers and of the woven array is shown in Figure 48. Immersing these intertwined CNT fibers in an electrolyte solution, led to efficiencies close to 3% [291]. The use of a Ti anode and an organic thiolate/disulphide redox couple led to further improvements in efficiency to 7.3% [292].

In spite of their high efficiency, a clear drawback of DSSC is the use of a liquid electrolyte. In an effort to move towards solid electrolytes Li et al. produced DSSC based on CNT fibers deposited on rubber and wound around a $TiO_2$/Ti photoanode. They used a gel electrolyte consisting of a thermoplastic Poly(vinylidene fluoride-co-hexafluoropropene) (PVDF-HFP) and an ionic liquid 1-butyl-3-methylimidazolium bis(trifluoromethanesulfonyl)imide (BMITFSI). The efficiency was 5.47%, only 9% lower than the reference cell with ionic liquids [293].

The development of composites with combined semi-structural mechanical properties and light harvesting capabilities is clearly challenging. Traditional semiconductors such as Si have high stiffness and have high power conversion efficiencies on account of their high crystallinity and purity, but their optoelectronic properties are rapidly degraded at small strains. DSSC are comparatively more flexible and can operate with polyelectrolytes, although with the trade-off between stiffness and ionic conductivity discussed above. Reports of high efficiency and flexibility using perovskites are encouraging, as these are solid-state devices. This might be a consequence of the intertwined configuration used so far, which could have design elements that enable stress transfer between the different elements without degradation of the photoactive phase.

From a mechanical point of view, the graphitic phase containing the semiconductor is expected to act as reinforcement, scaffold and current collector to provide a net weight-save. It would have to have low light absorption if used as a front current collector, as a low density network of CNTs. However, this would reduce the fiber volume fraction and its contribution to the mechanical properties. Alternatively, the graphitic phase can take the roles of the panel backing and back contact and thus increase its contribution to load-bearing. From the literature discussed above, it is reasonable to assume a thickness of 10 μm for the graphitic phase, 10 μm for the photoactive phase and a negligible contribution from the front electrode.

This arrangement would seem attractive mechanically as it would put the volume fraction of fibre above 20% and possibly benefit from the stiffness of the photoactive inorganic phase (the specific stiffness of $TiO_2$ is around 35 GPa/SG [294], whilst leading to a device with a thickness comparable to that of a conventional thin ply of a FRC.

**5. Concluding remarks**

This review has summarized the state-of-the-art and the current challenges in three areas of FRC: manufacturing techniques, multiscale strategies for optimum design and the incorporation of novel functionalities (mainly electrical conductivity and energy management) into structural composites. These three topics will play a major role in the future development of FRC in so far they will boost the application of composites in more applications from which they have been hindered as a result of high manufacturing costs, non-optimum properties or limited functionalities.

From the viewpoint of novel manufacturing strategies, it should be pointed out that AFP technology is only applied nowadays in the aerospace sector and it is sill in its infancy. Further growth to reach maturity will require the strong interaction among AFP machine manufacturers, composite material suppliers, researchers working of composite design optimization as well as final users. Further developments in robotics and numerical control will add more versatility and increase the possibilities of AFP. In addition, some of the possibilities offered by AFP technology involve the modification of the tape/tow material characteristics at several levels: tow/tape characteristics (geometry, tackiness, etc.), material type (thermosets, thermoplastics, dry fabrics, etc.) microstructure (fibre type, fibre volume fraction, fibre hybridization, etc.), functionalization (thermal & electrical conductivities, sensorization, etc.), etc. These issues will have to be addressed by the suppliers of matrix, fibers and prepregs.

In addition, the application of FRC based on thermoplastic has been growing continuously in automotive and aerospace sectors because of their higher mechanical properties and damage tolerance, chemical resistance, infinite shop life, weldability of components, reuse of material and shorter manufacturing cycles compared to their competitor thermosets. Thermoplastic materials offer a high potential not only regarding material properties, but also in order to reduce processing, logistic, operation and a life cycle costs. The opportunity for the automation of the production of thermoplastic composites is to couple AFP with techniques such as a gas torch, infrared or laser as heating source, a process that took the designation of *in situ* consolidation of thermoplastics [295].

The current design, validation and certification approach for structural composites relies on tools and methods build on traditional, constant stiffness, laminates. The building block pyramid of structural certification is based on the concept of design allowables, i.e. statistically processed outcomes of simple tests on simple coupons for a fixed and relatively small number of laminates. This approach is hard to implement for design concepts based on a large number of possible configurations, e.g. ply angles, in which the material is changing from point to point, creating in essence an infinite amount of materials, or were a certain amount of manufacturing effects/defects is tolerated. Creating design allowables for all these possibilities is not a feasible option so a new and practical approach has to be adopted if the application of dispersed stacking sequence and steered-fiber laminates proliferates. A promising strategy to approach this challenge is the use of virtual testing based on a multiscale analysis approach using micro mechanics, mesomechanics and macromechanics. Moreover, the bottom-up approaches available for multiscale modeling of FRC offer tremendous potential to tailor the composite properties by means of designing the structure at the ply and laminate level. However, these capabilities only become real if these techniques for composite optimization are integrated with the manufacturing process strategies (either AFP or liquid moulding technologies). Thus, the combination of both virtual processing and virtual testing is critical to carry out virtual design and optimization of FRC.

From the viewpoint of electrical properties of FRC, incorporation of nanocarbons is the dominant strategy to increase the interconnection between macroscopic fibers. After a decade of work on these hierarchical composites, there are now various nanocarbon integration routes compatible with composite manufacturing, either through the impregnation of tows with thermoset/CNT mixtures or by using sheets of CNTs are interleaves. Both strategies have shown increases in composite conductivity through the thickness. The latter method has also been demonstrated to be effective for low energy lightning strike protection. Such milestones have been possible by realising that the potential of nanomaterials in structural composites requires placing them at relevant interfaces, in the right volume fraction and incorporated using suitable (often new) methods specific to each composite fabrication route.

In reviewing trends in power management in structural composites, this section has highlighted results obtained using either carbon or CNT fiber as current collector/reinforcement. The multifunctional devices based on these materials are currently at opposite ends of the spectrum. Carbon fiber-based devices have semi-structural properties, most notably in the fiber array plane, but low efficiencies for energy storage and harvesting. CNT fiber-based devices, on the other hand, have comparatively much higher energy/power storage and higher energy conversion efficiencies, but the mechanical properties of these devices is largely unknown. Future work should be directed at combining carbon with CNT fibers and in establishing accurately the mechanical properties of large-scale arrays of CNT fiber.

A common challenge for energy storage and DSSC is in developing electrolytes with high ionic conductivity and high stiffness, either through phase-segregation, polymer grafting or by reducing diffusion lengths by electrodeposition. One of the difficulties lies in making these polyelectrolytes compatible with conventional structural epoxy resins without affecting the curing kinetics or the properties of the polyelectrolyte.

An alternative for structural energy management in structural composites is to use a solid-state ionic conductor (storage) or photoactive phase (e.g. perovskite). These materials, however, are comparatively less developed and present their own challenges in terms of chemical stability.

A critical aspect for the implementation of structural energy-managing composites is compatibility of fabrication with existing composite production technologies. In this sense, the integration of new phases (e.g. semiconductor) onto the reinforcing fibers will require the use of gas or liquid deposition processes that can ideally be carried out on-line similar to a sizing process. Similarly, assembly of different layers of a device should be compatible with roll-to-roll technologies. The starting point is encouraging. $TiO_2$, for example, is widely available as nanoparticles in dispersion in large quantities at very low costs. Several polyelectrolytes are based on thermoplastics (e.g. PEO) and therefore compatible with composite fabrication methods. Furthermore, optimized synthesis of ionic liquids in bulk quantities gives cost estimates below 6 USD/kg for selected ionic liquids [296].

Finally, it is worth highlighting that the whole potential of multifunctional structural composites to produce net benefits requires a broad perspective. At the material level, for example, an emerging trend consists in integrating energy-related functions in a single material, which can reduce parasitic weight and minimize charge transfer/diffusion lengths [297]. As an example, an integrated DSSC and supercapacitor can be produced in the context of wearables by supporting energy harvesting and storing materials on intertwined filaments of CNTs and Ti [298]. Similarly, there are emerging reports of energy harvesting from piezo-electrochemical effects during lithiation/delithiation of carbon fibers which could operate a low frequencies and already in the battery structure [299]. Output power is at present of the order of 1mW/g [300] but likely to increase as the physics of the process is better understood. At the systems level, multifunctional structures can provide additional weight saving by reducing the weight associated with wiring, connections and other optimization. Large commercial aircrafts (e.g. Boeing 747) carry around 2000 kg of thin copper wiring [301]. Substantial indirect reductions in weight can be expected ($\approx$ 0.07 kg/m per wire) by introducing self-powered systems or reducing the distance between power source-output as a result of using multifunctional structures. Furthermore, if multifunctional structures/materials can be adapted to composite fabrication methods, there are likely to be associated benefits from reduction in production times, cost and installation which might temporarily outweigh a materials multifunctionality figure of merit below unity.

As a final note, we emphasize that structural composites with multifunctional properties naturally call for the use of nanomaterials in combination with macroscopic reinforcing fibers. The expected result is a composite with a hierarchical structure and the capability to carry out various functions simultaneously. The complexity of this challenge calls for controlled assembly of the different phases and careful engineering of the different interfaces. This will inevitably lead to modification of existing composite fabrication processes.

*Acknowledgements*

The authors acknowledge the support of the Comunidad de Madrid (program DIMMAT, P2013/MIT-2775), the European Union Seventh Framework Programme (MAAXIMUS, grant agreement 213371; CRASHING, Clean Sky Joint Undertaking, grant agreement 632438; CARINHYPH, grant agreement 310184; MUFIN, grant agreement 322129), the European Research Council (Starting Grant STEM, grant agreement 678565 supported by H2020 program), and the Spanish Ministry of Economy and Competitiveness (MAT2012-

37552-C03-02, MAT2015-62584-ERC, MAT2015-69491-C03-02). JJV and CSL also acknowledge the Ramon y Cajal fellowships from the Spanish Ministry of Economy and Competitiveness (RyC-2014-15115 and RyC-2013- 14271). In addition, the financial support and the continuous collaboration with Airbus, Airbus Defense & Space, BE/Aerospace, Hexcel and Fokker is gratefully acknowledged. Last -but not least-, the authors are indebted to the graduate students and post-doctoral research associates of their research groups for their work along these years.

**FIGURE CAPTIONS**

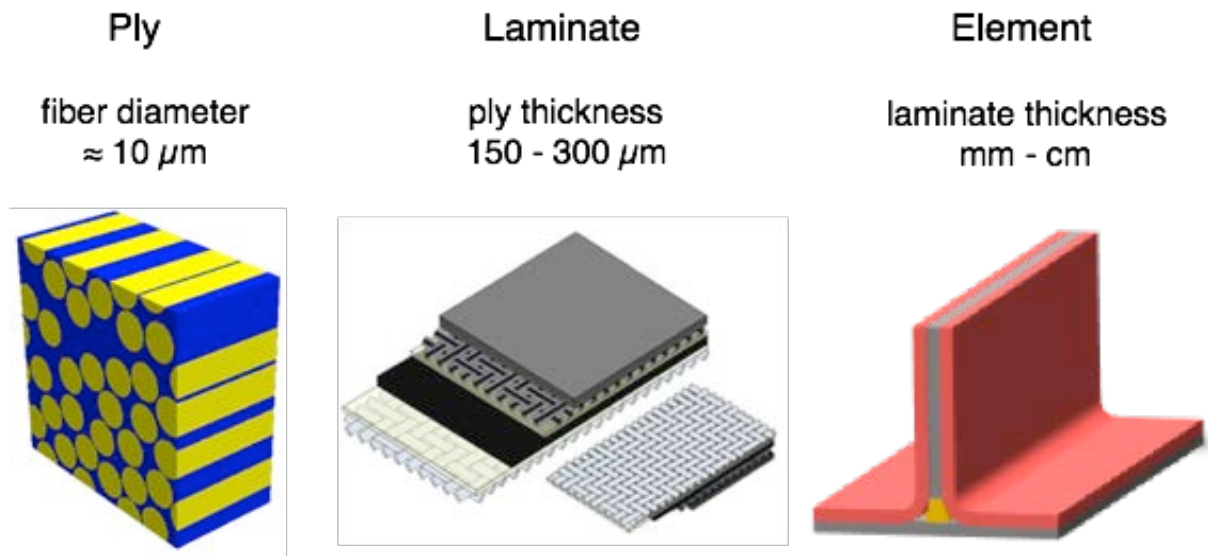

**Figure 1.** Hierarchical structure of FRC and associated lengths scales for each level.

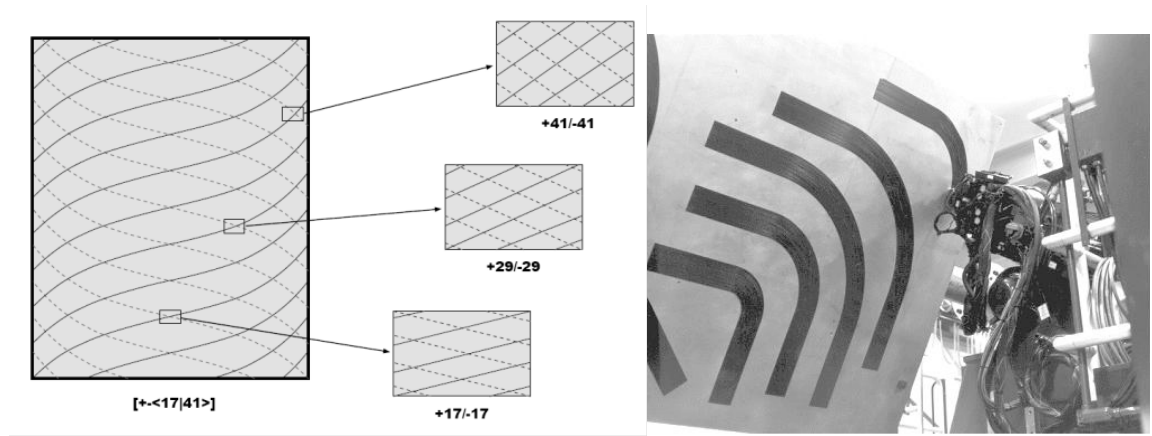

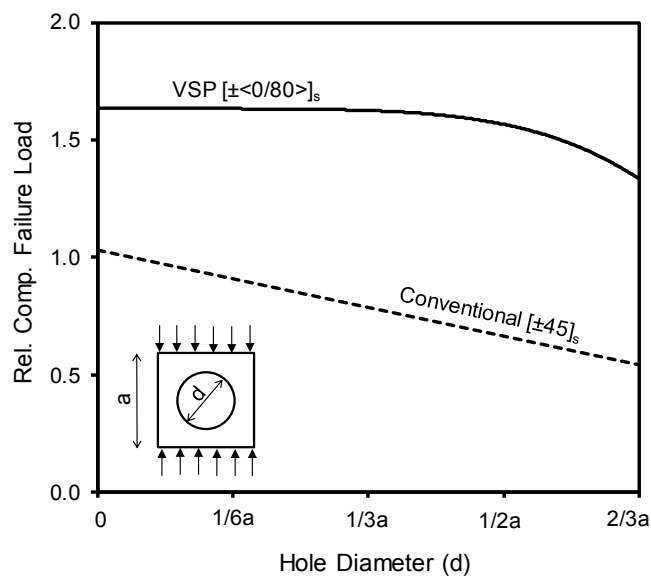

**Figure 2.** (a) An automated fiber placement machine demonstrating the placement of steered tow courses [20]. (b) Representation of a 2-ply VSP non-conventional composite panel showing different layups at different points in the laminate [20]. (c) Compression failure performances of traditional ([±45]s) and VSP [±<0/80>]s panels.

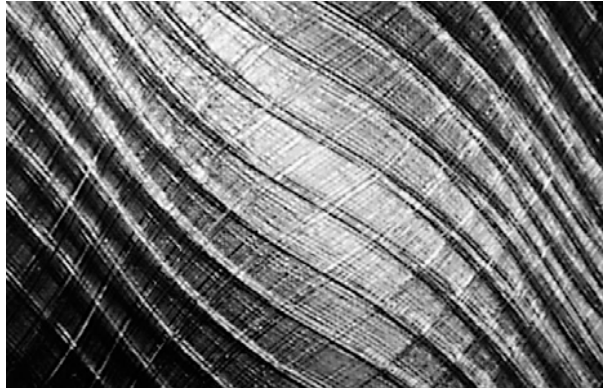
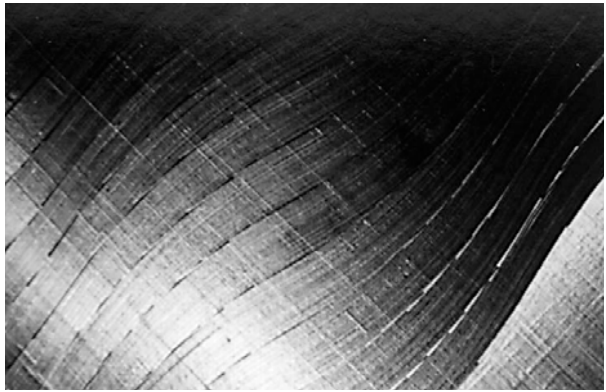

**Figure 3.** Fiber-steered panels with two types of manufacturing defects: (a) overlaps. (b) tow-drops. Adapted from [20].

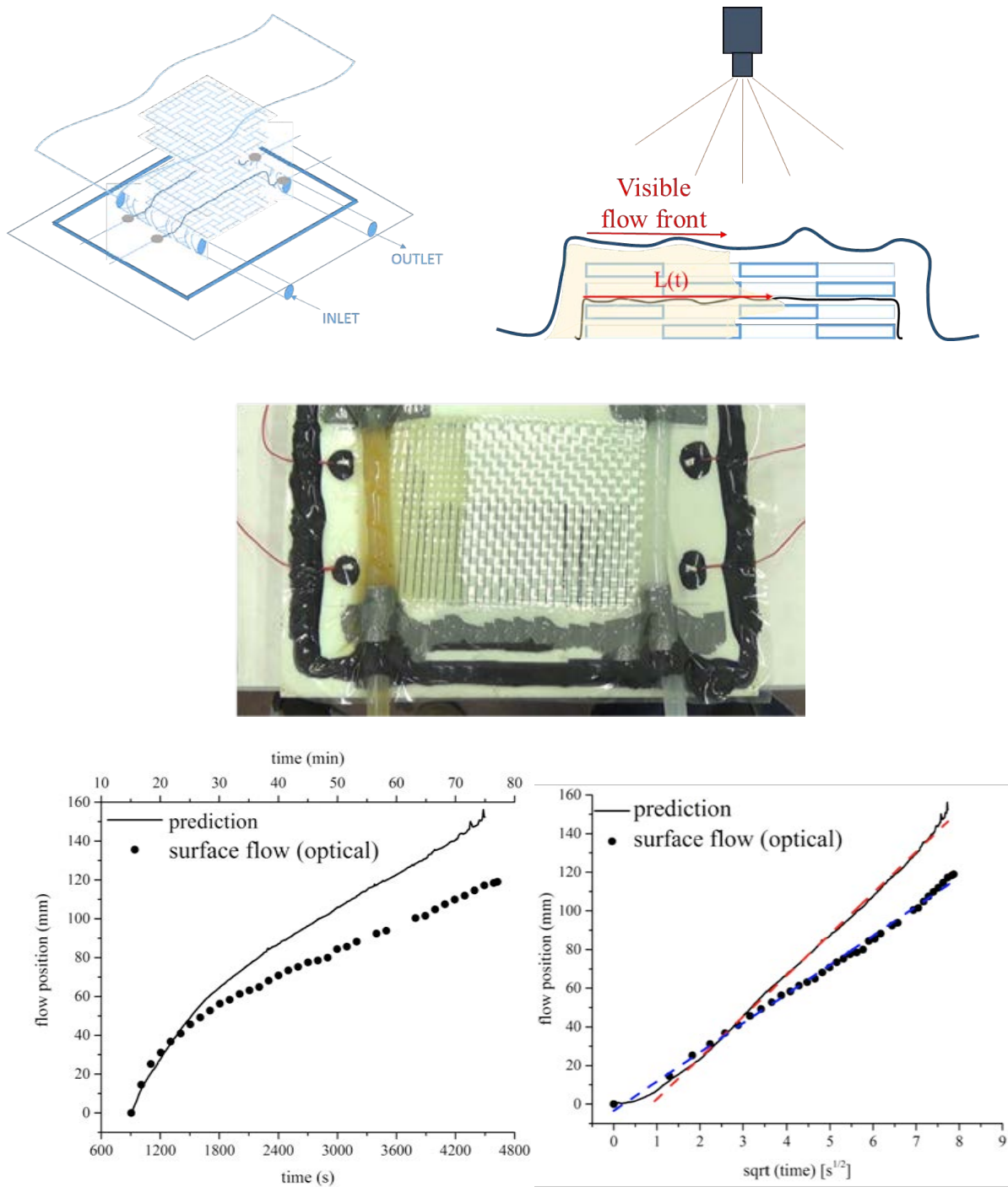

**Figure 4.** Monitoring flow progress during VARI of an E-glass woven fabric laminate infused with a vinyl-ester resin using a CNT fibers yarn sensor. a) Schematic illustration of the experimental set-up. The CNT fiber yarn is placed between two glass fiber fabrics parallel to the expected resin flow direction. The (surface) flow is independently determined by a visual correlation system. c) Photograph of the vacuum infusion set-up, showing electrical contacts outside the fiber preform. d) Comparison of predicted flow based on CNT fiber sensor and surface flow from optical images. e) Evolution of the flow front position a sa function of the square root of time, l(t)~$\sqrt{t}$. The linear relationship indicates that the macroscopic flow is dictated by Darcy's law [33].

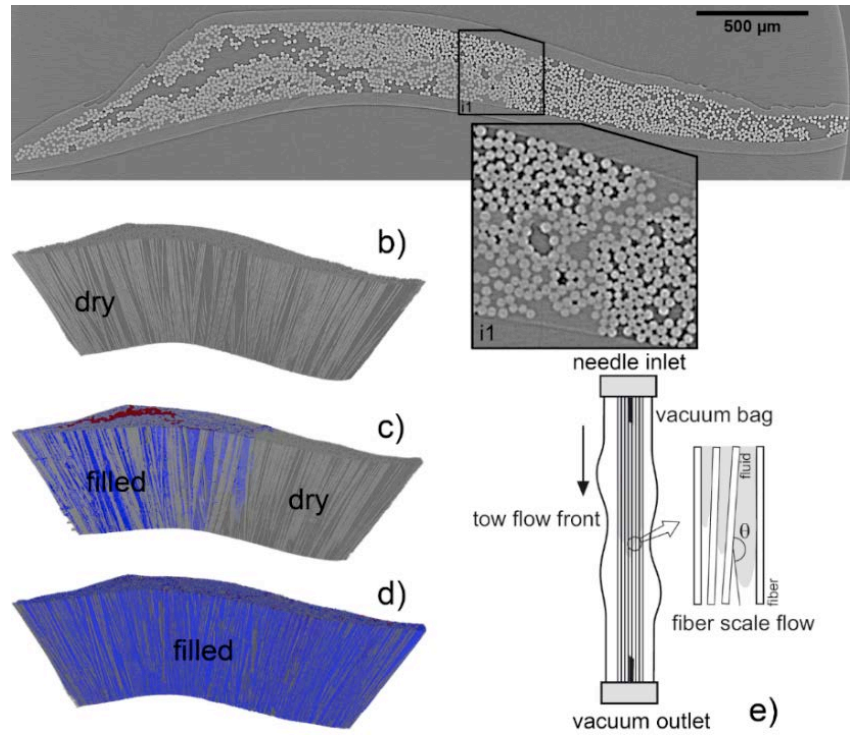

**Figure 5.** a) Cross section of the fiber bundle within the vacuum bag. Fibers and voids are clearly visible. b) 3D of the fiber tow prior to the infusion. c) 3D of the partially impregnated fiber tow. d) Fully impregnated fiber tow. e) Sketch of the micro-flow at the fiber level during the impregnation. [58]

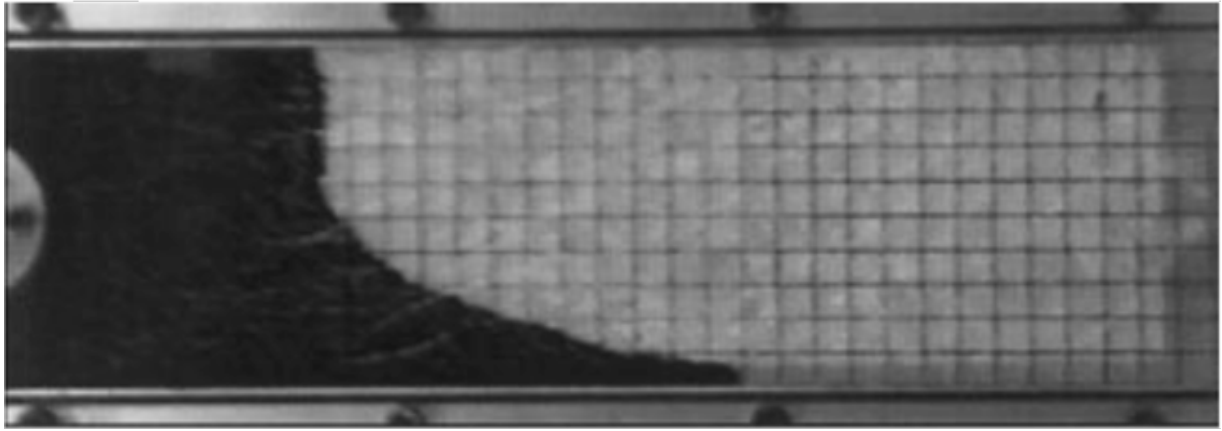

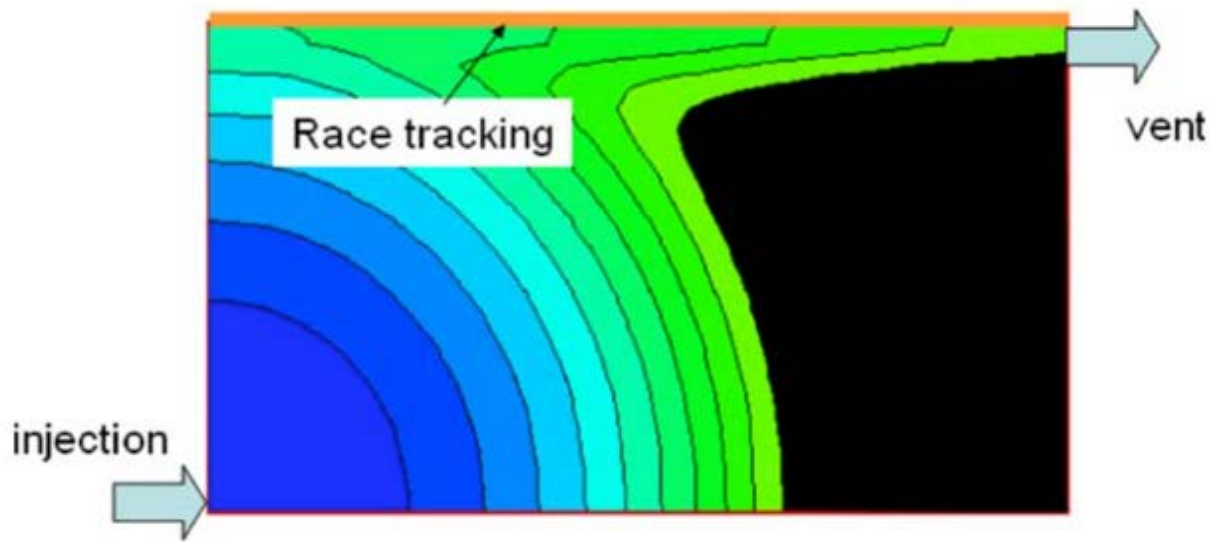

**Figure 6.** a) Experimental RTM set-up with acrylic mould cover visualizing race-tracking effects in the lower edge of the mould [40], b) Race tracking simulation along the top edge of the mould modifying the flow progression of the resin through the venting port producing dry spot formation [72].

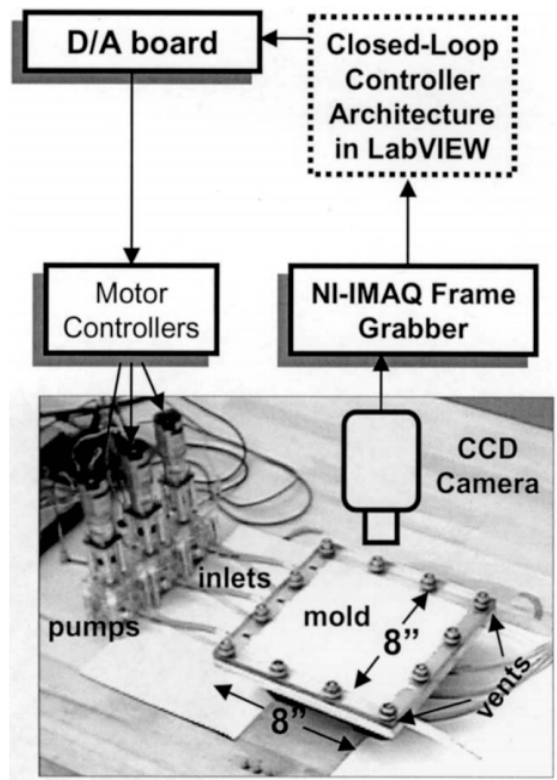

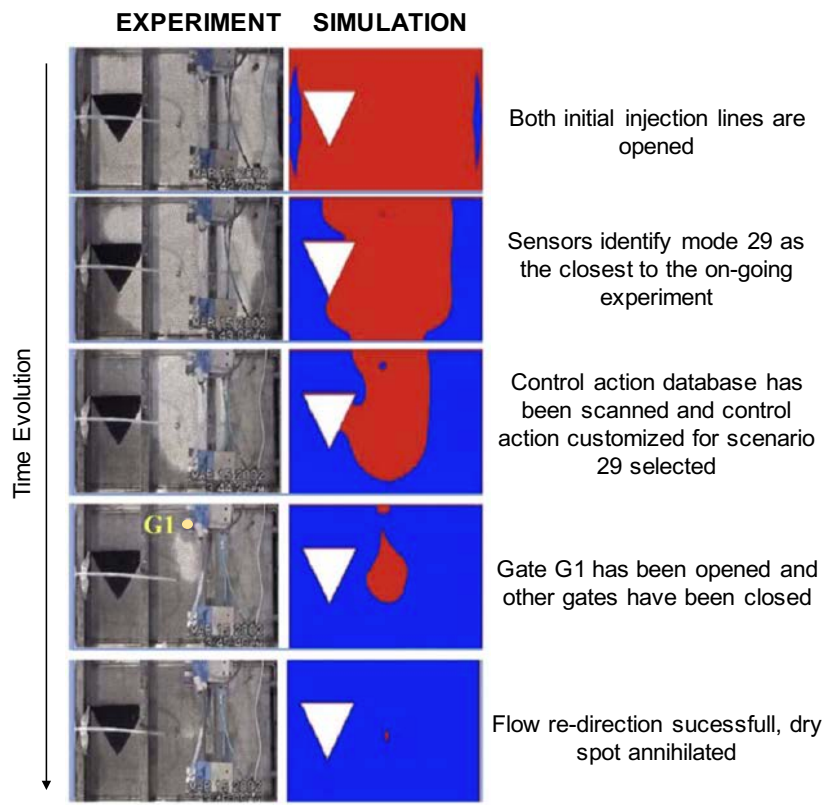

**Figure 7.** a) Schematic of a closed-loop experimental set up for RTM [69], b) Experimental and simulated flow fronts obtained using active control [74].

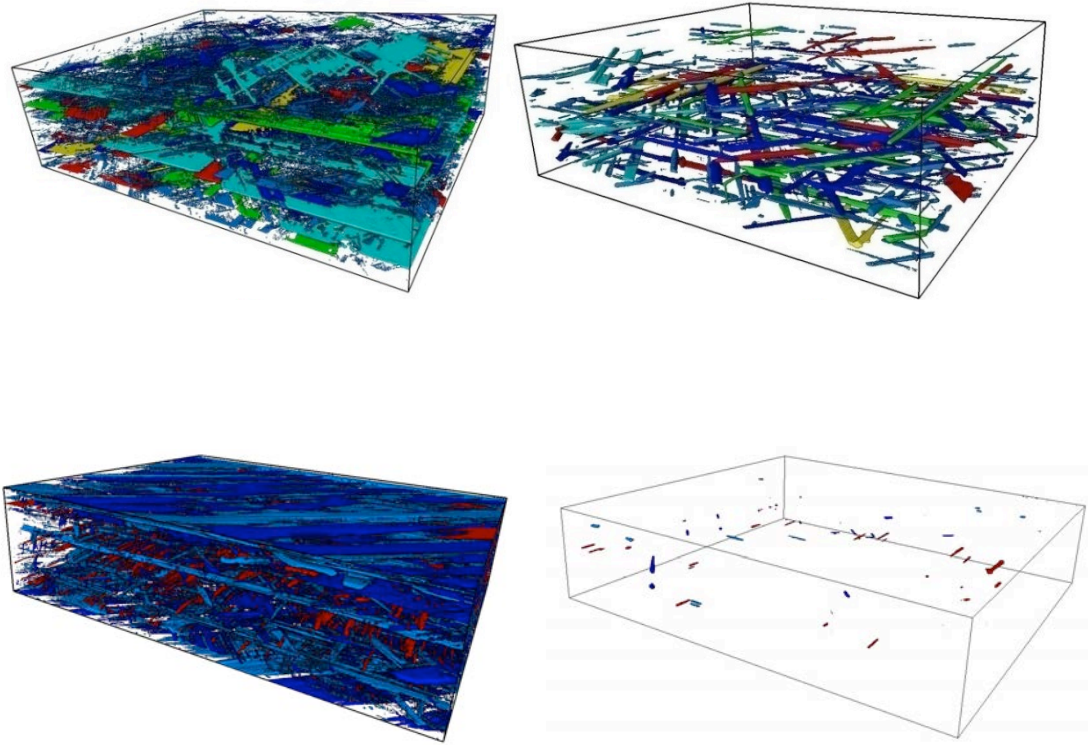

**Figure 8.** XCT images of the void distribution in 20 x 20 mm$^2$ samples extracted from a 400 x 400 mm$^2$ HexPly M56 [+45/0/-45/90]3s laminates. a) hand lay-up, fresh. b) Hand lay-up, cured. c) AFP, fresh, d) AFP, cured. Each color represents one pore. Regions with the same color stand for an interconnected pore [83].

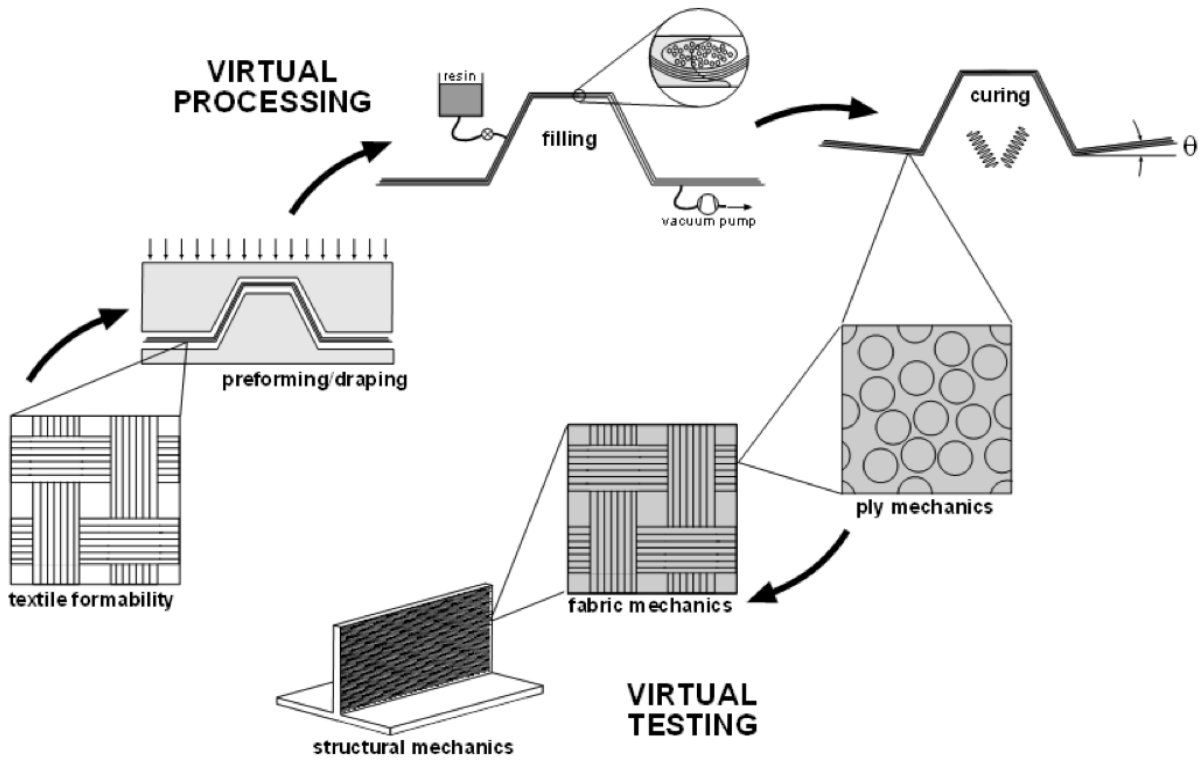

**Figure 9.** Schematic of the multiscale modelling strategy for virtual testing and virtual processing of FRC.

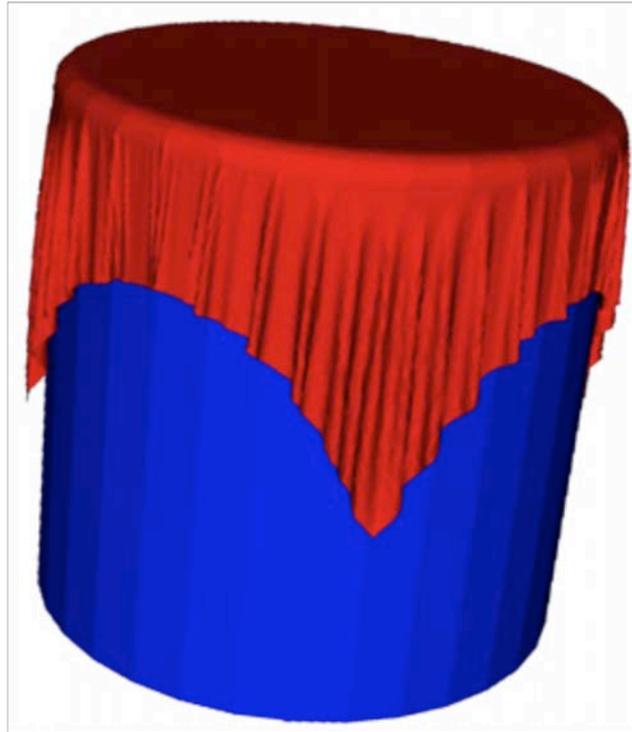

**Figure 10.** a) Discrete model for a woven composite using spring elements (shear, torsional, stretching and flexional) [99], b) Draping a woven fabric onto the surface of a dome using semi-discrete finite element modelling [94].

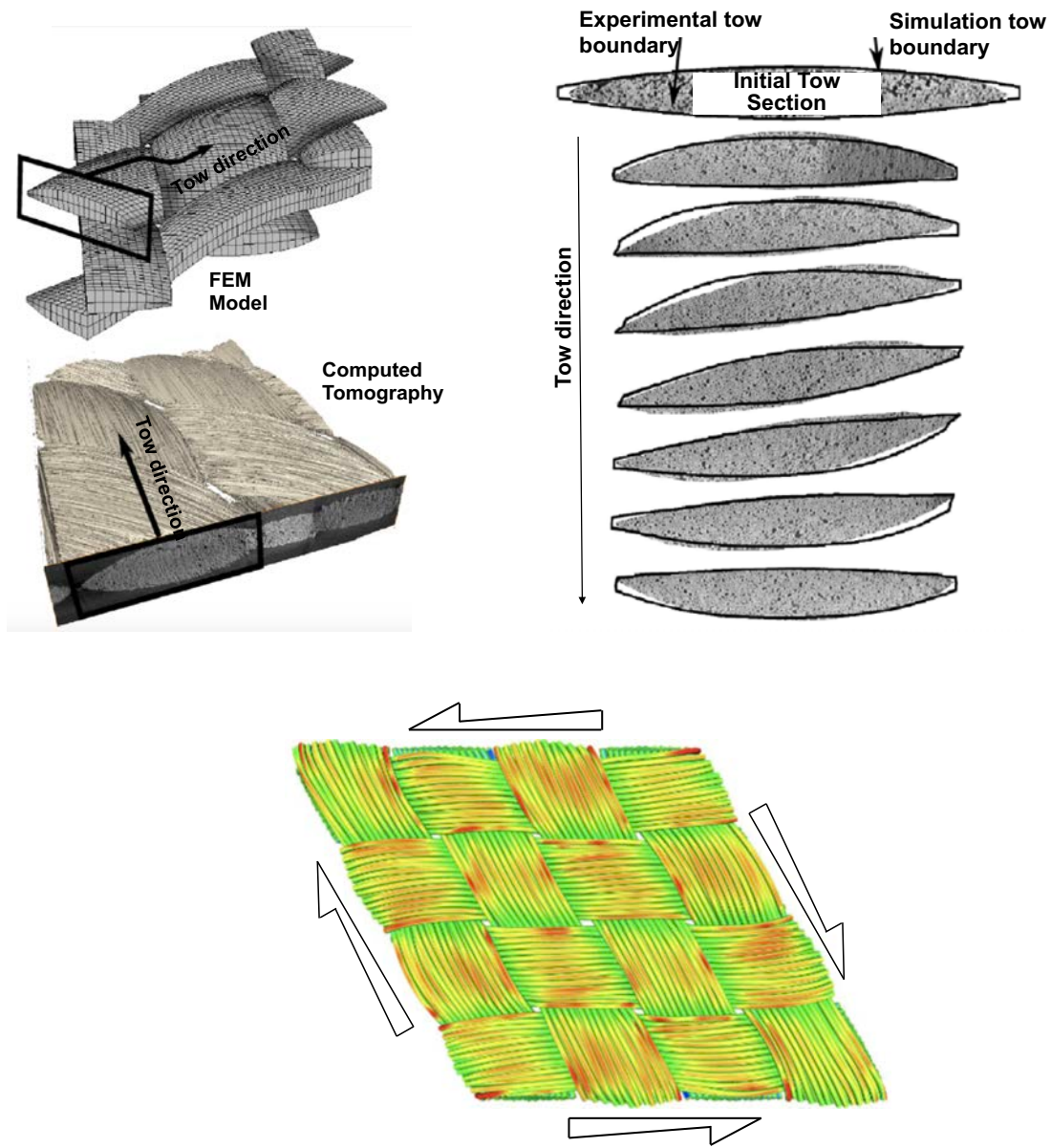

**Figure 11.** a) Comparison of internal deformation obtained with mesomechanical modelling of a woven fabric and XCT reconstructions. b) Yarn cross section deformations along the unit cell [98], c) Bending deformation of a plain weave using interactions between individual fibers [100].

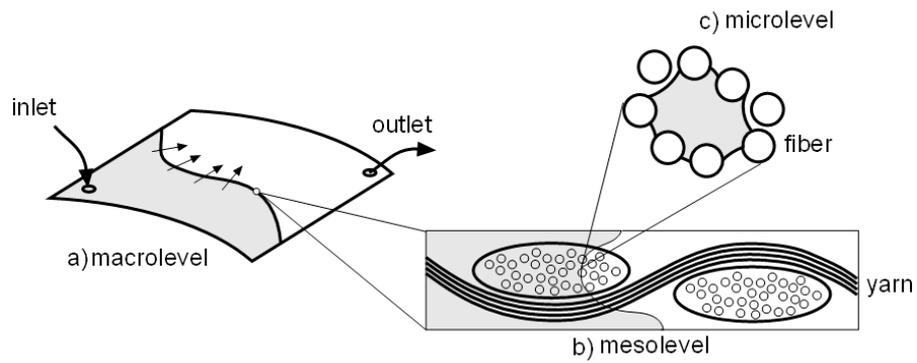

**Figure 12.** Multiscale simulation strategy of fabric infiltration. a) Macrolevel with homogenized permeability, b) Mesolevel with yarn-to-yarn structure, c) Microlevel at which individual fibers are resolved.

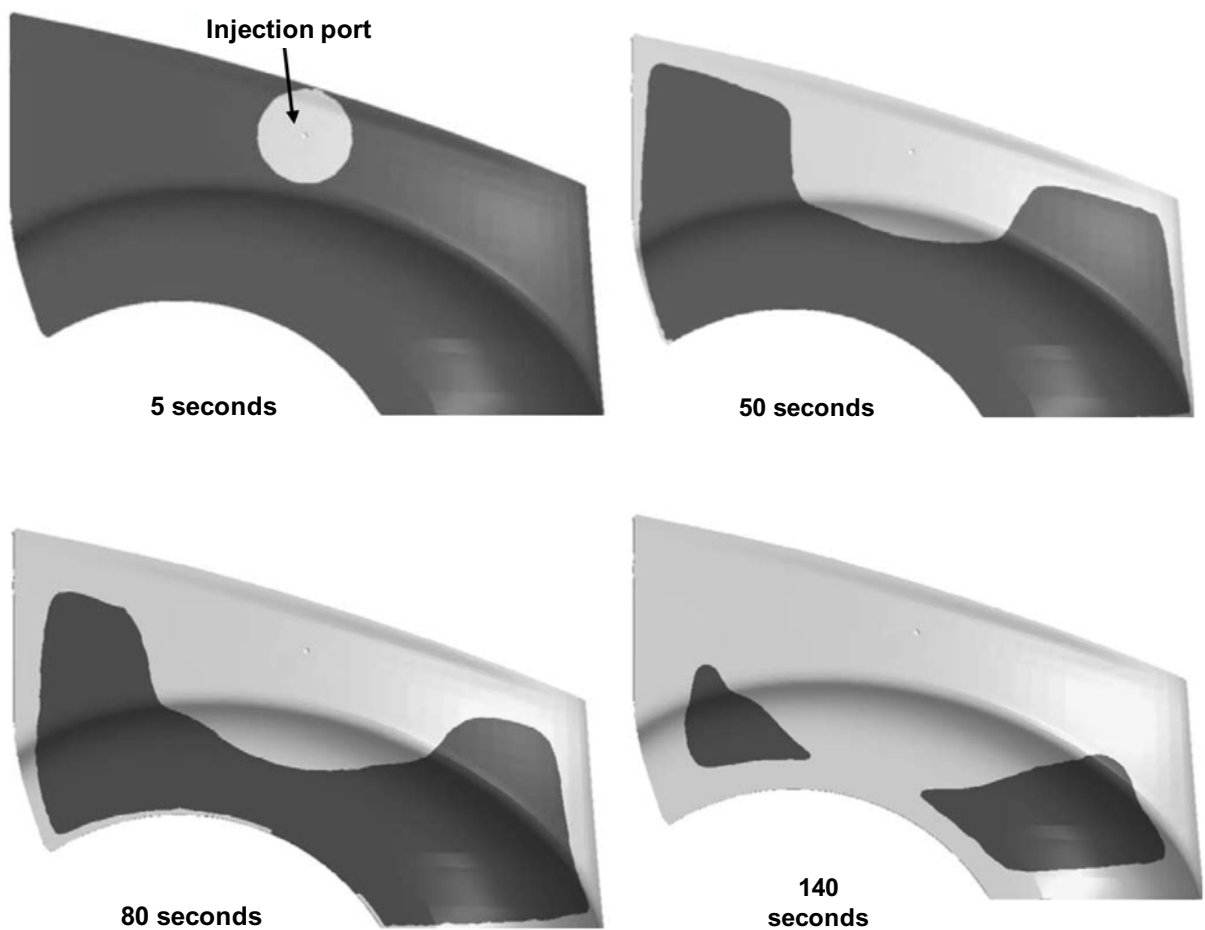

**Figure 13.** Front flow positions after 5, 50, 80 and 140 s during RTM filling of an automotive fender with runner. Channel effects are remarkable after the resin reaches the boundary. Finally, two dry spot areas are located after mould filling [102].

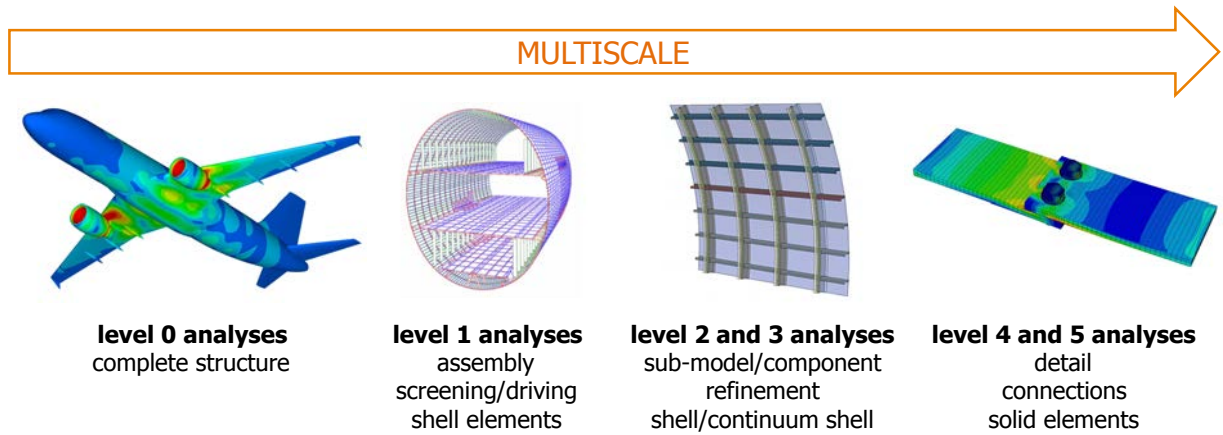

**Figure 14.** Traditional top-down multiscale analysis strategy for design of composite structures adopted by the aerospace industry [133].

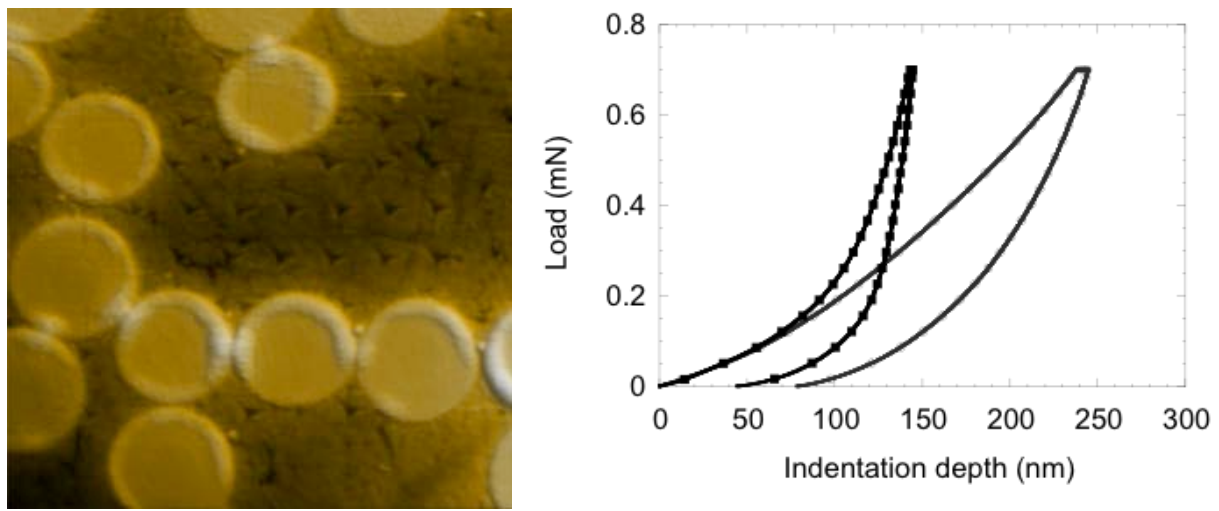

**Figure 15.** (a) Atomic force microscopy image of matrix of indentations placed in resin pockets found in the composite cross section. (b) Nanoindentation load-displacement curves of two of these indents showing the constraint effect of the surrounding fibers when the indentation is very close to one of the fibers. Adapted from [3].

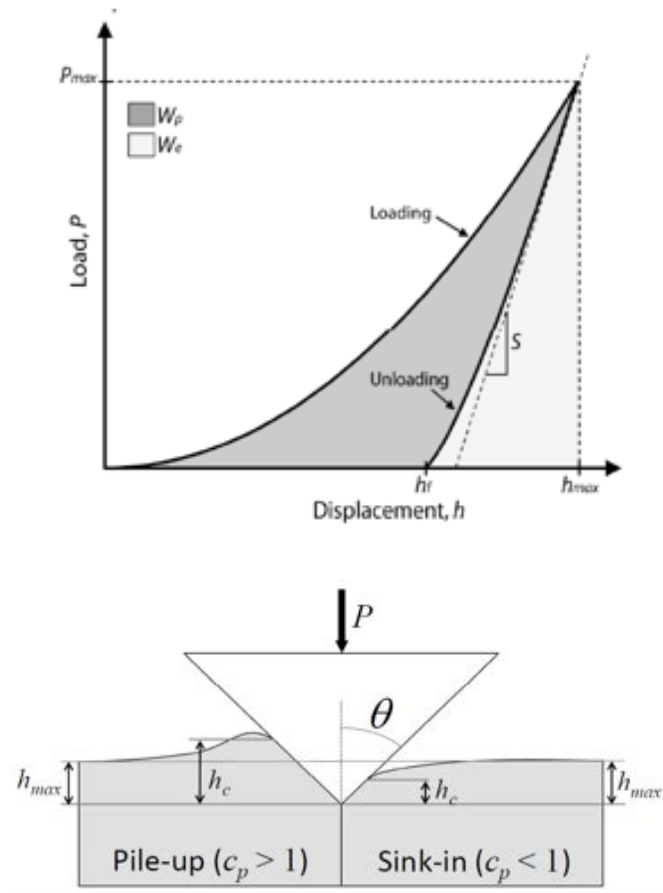

**Figure 16.** (a) Representative indentation load-displacement curve. (b) Schematic illustration showing pile-up/sink-in behavior affecting the contact area of indentation.

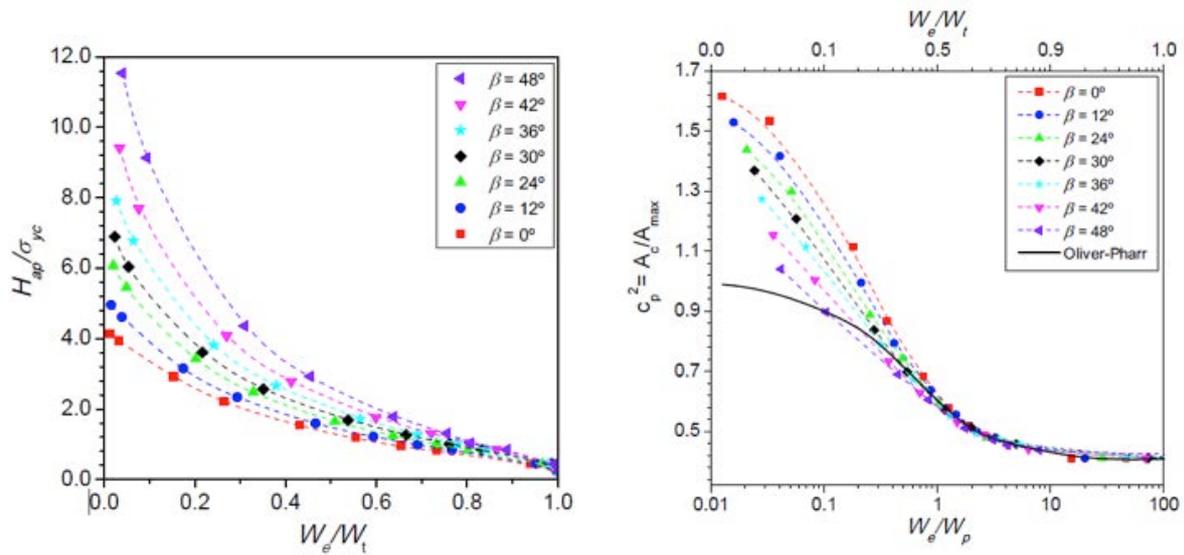

**Figure 17.** (a) Apparent hardness *vs.* $W_e/(W_e + W_p) = W_e/W_t$. (b) Pile-up effect for materials with different pressure sensitivity. The continuous line represents the pile-up estimated by the Oliver and Pharr method (from [139]).

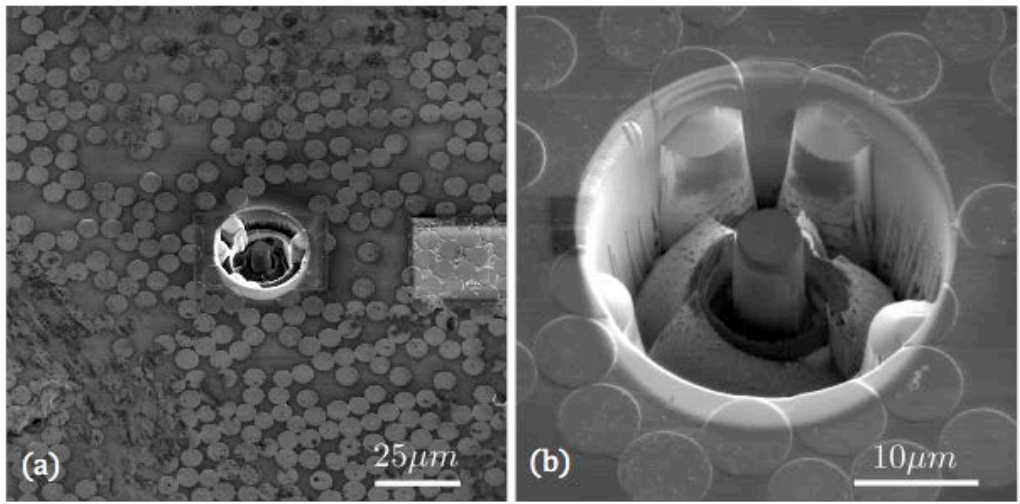

**Figure 18.** (a) Scanning electron micrograph image of the transverse cross-section of a FRC showing a micropillar milled on a resin pocket. (b) Close up image of the milled pillar (from [150]).

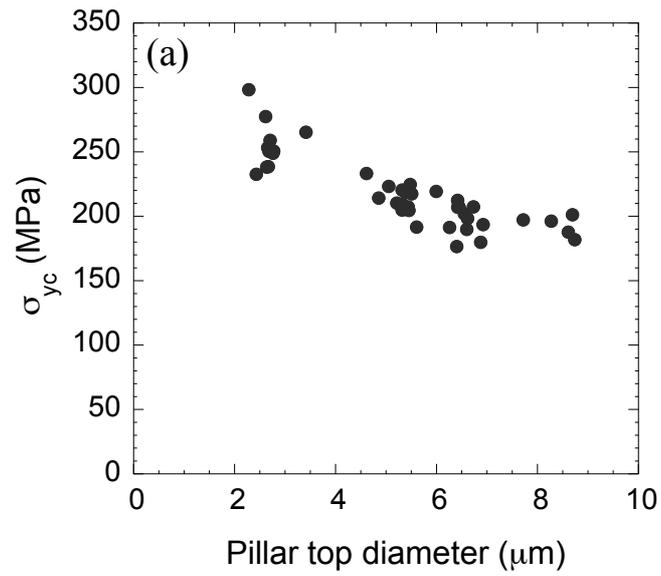

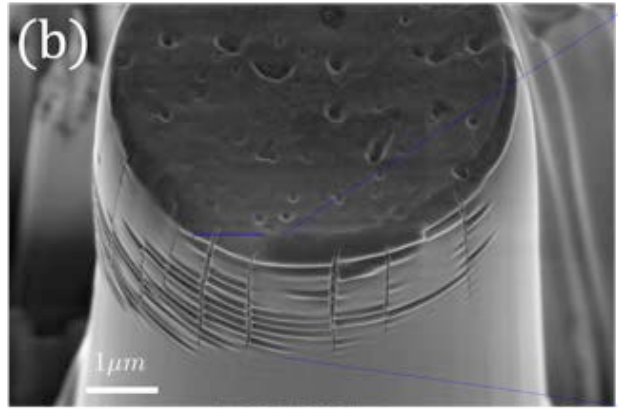

**Figure 19.** (a) Compressive yield strength, $\sigma_y$, of epoxy micropillars as a function of the micropillar diameter. (b) Scanning electron micrograph of the deformed pillar showing the hard skin during (from [150]).

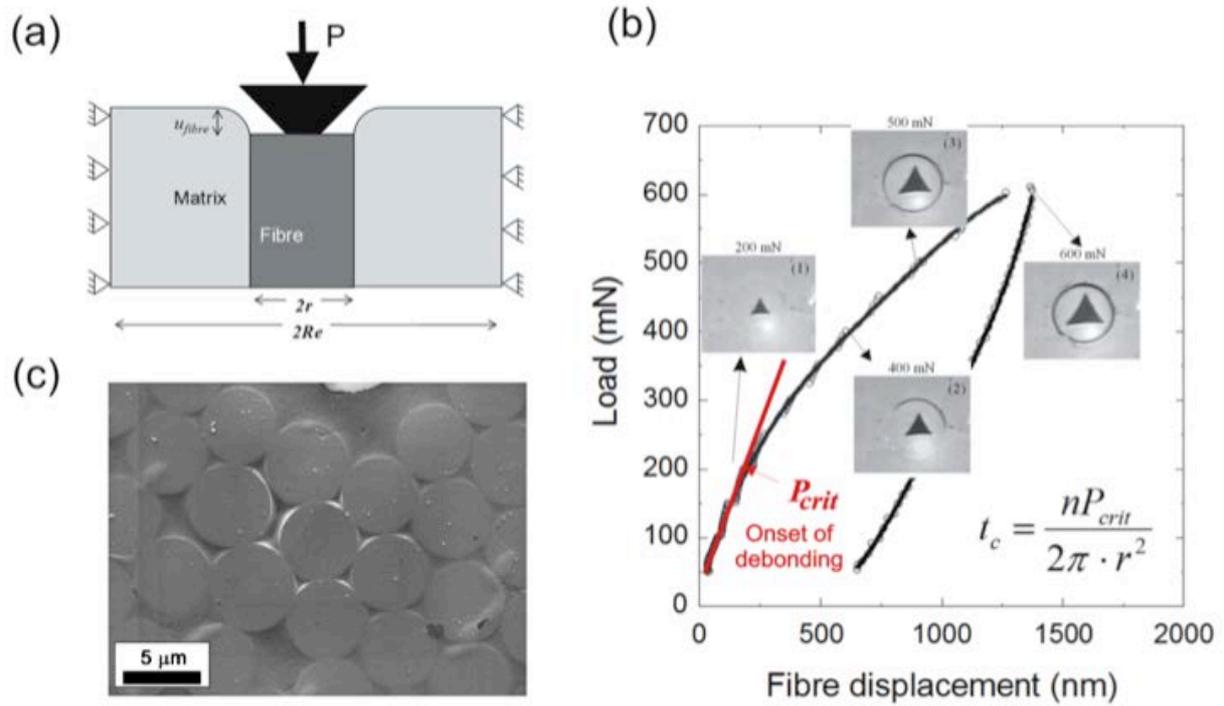

**Figure 20.** Figure 6. (a) Diagram illustrating the push-in test. (b) Load-displacement curve resulting from a push-in test. (c) SEM image of a fiber displaying hexagonal packing of close neighbours (Adapted from [157] and [158]).

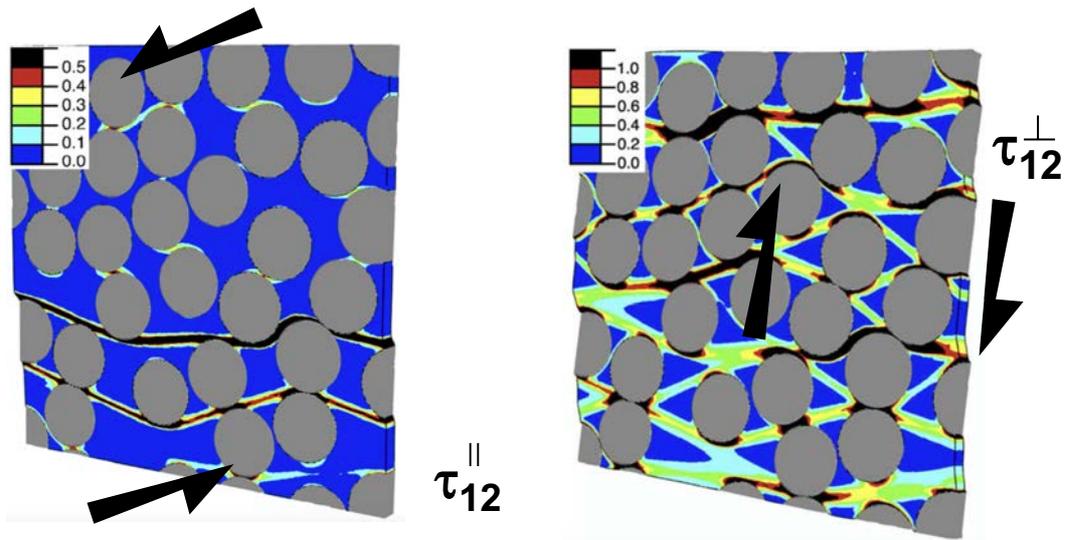

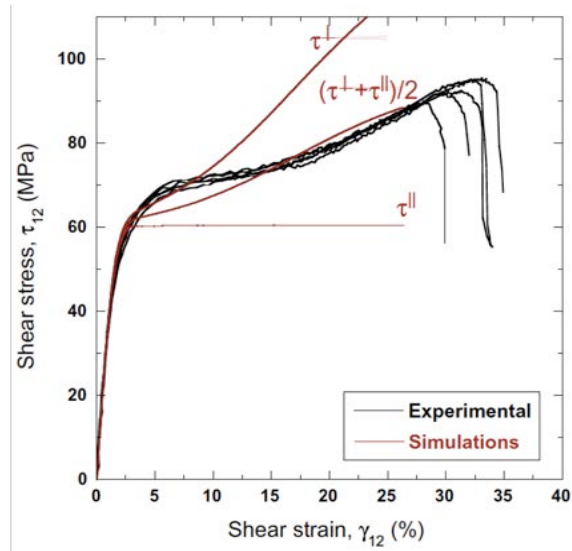

**Figure 21.** Contour plot of the accumulated plastic strain the matrix of an RVE of a unidirectional FRC subjected to in-plane shear. (a) Shear parallel to the fiberss ($\gamma_{12}\approx6\%$). (b) Shear perpendicular to the fibers direction ($\gamma_{12}\approx20\%$). (c) Numerical predictions and experimental stress-strain curves of the unidirectional M40J/MTM57 FRC subjected to in-plane shear. Adapted from [172].

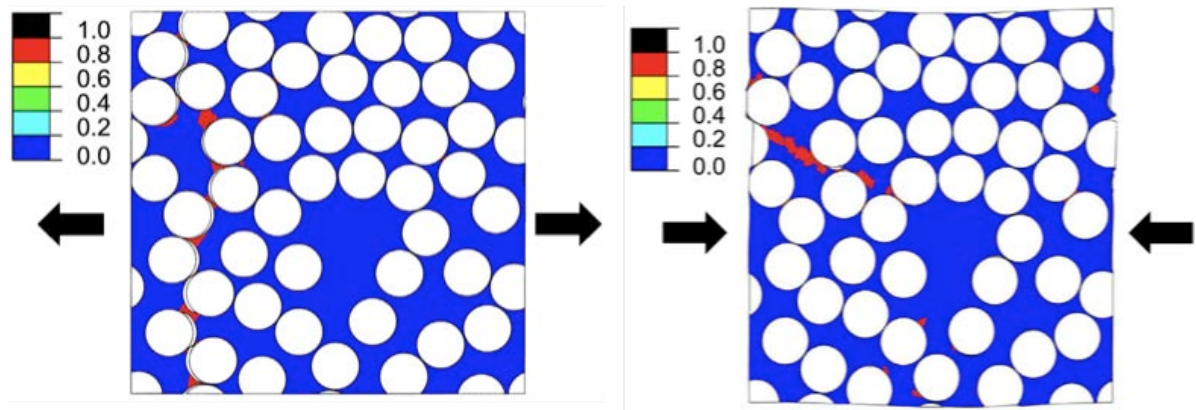

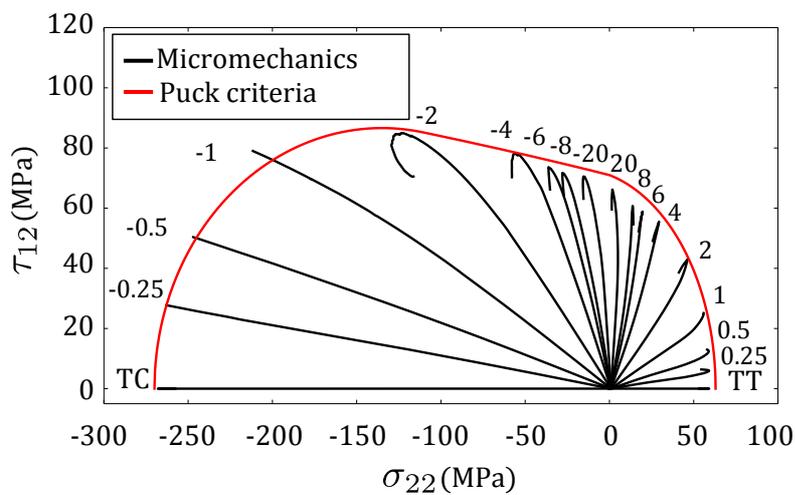

**Figure 22.** (a) Contour plot of the damage variable in the matrix in an RVE of a unidirectional ply of AS4/8552 FRC subjected to transverse tension in the horizontal direction. (b) *Idem* in the case of transverse compression in the horizontal direction. (c) Computational micromechanics predictions of failure locus of AS4/8552 FRC in the $\sigma_2$-$\tau_{12}$ plane and comparison with the the predictions of the phenomenological Puck criterion. Adapted from [171].

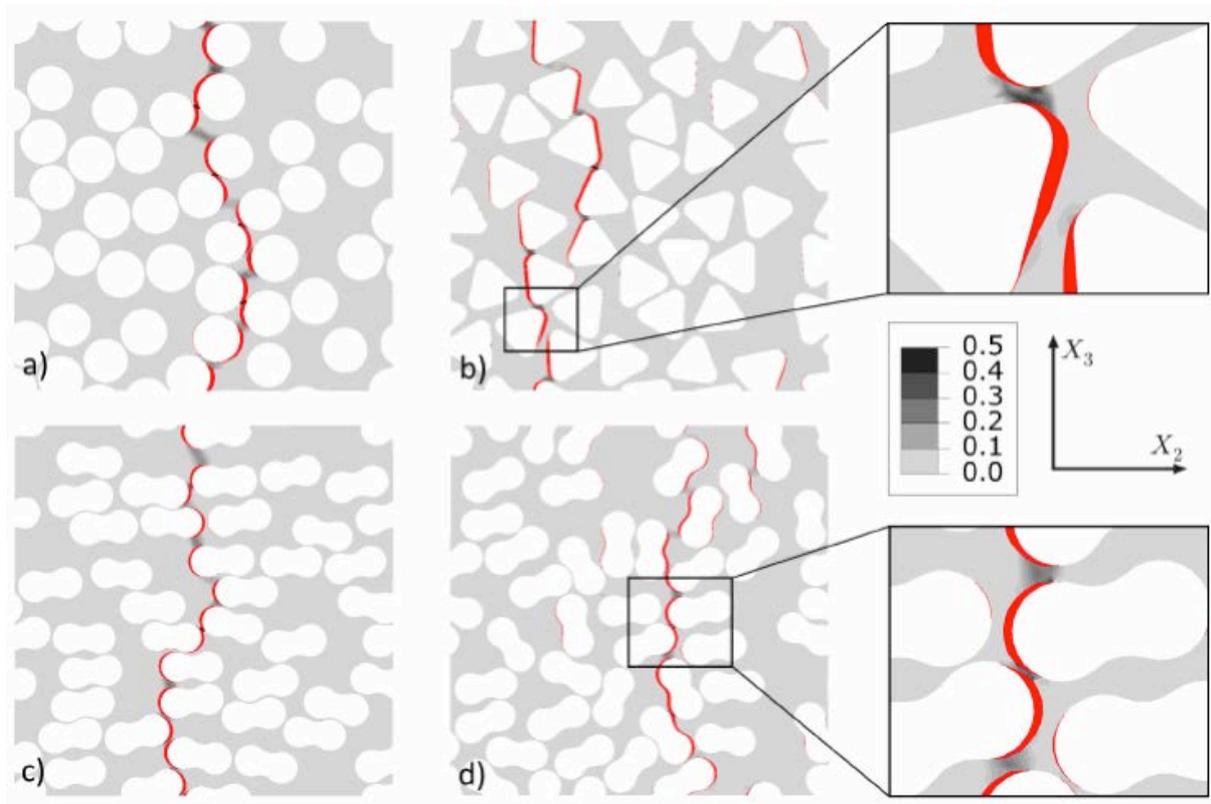

**Figure 23.** Contour plot of the accumulated equivalent plastic strain in the matrix (shown in grey scale) and interfacial damage (shown red) under transverse horizontal tension in RVEs containing fibers with different shape. (a) Circular fibers. (b) 3-polygonal fibers. (c) 2-lobed aligned fibers. (d) 2-lobed fibers. The fiber volume fraction was 50% in all cases. Adapted from [120].

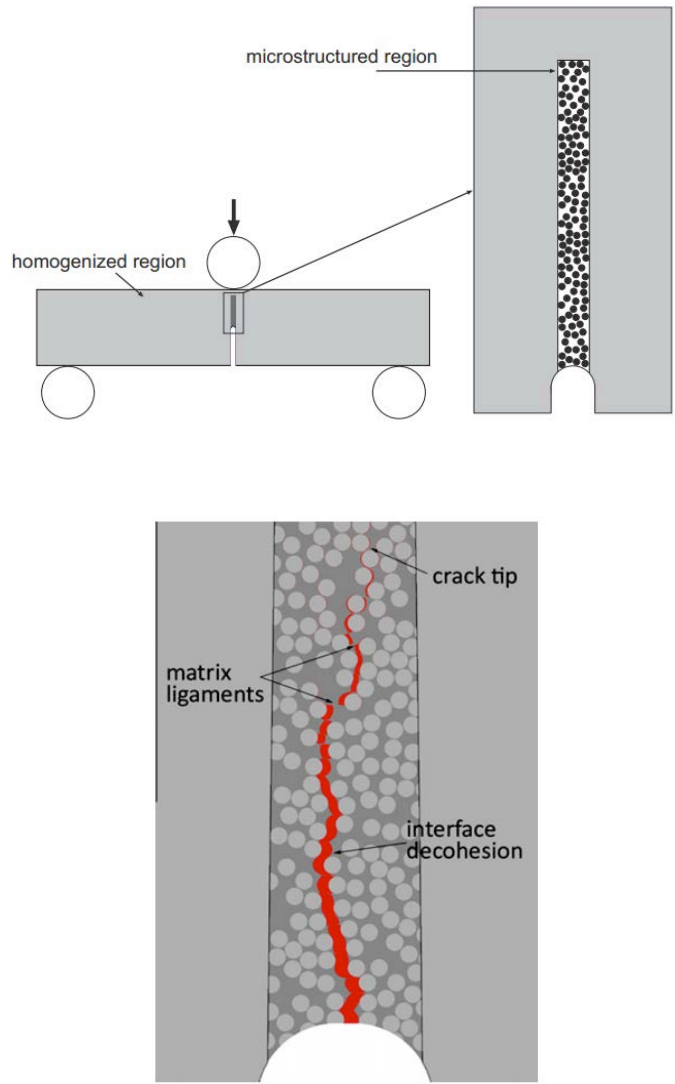

**Figure 24**. (a) Schematic of the embedded cell model to simulate the propagation of crack from a sharp notch in a three-point bend test of a unidirectional ply (fibers are perpendicular to beam plane). (b) Detail of the damage micromechanisms leading to crack propagation according to the numerical model. The crack path is shown in red. Adapted from [178].

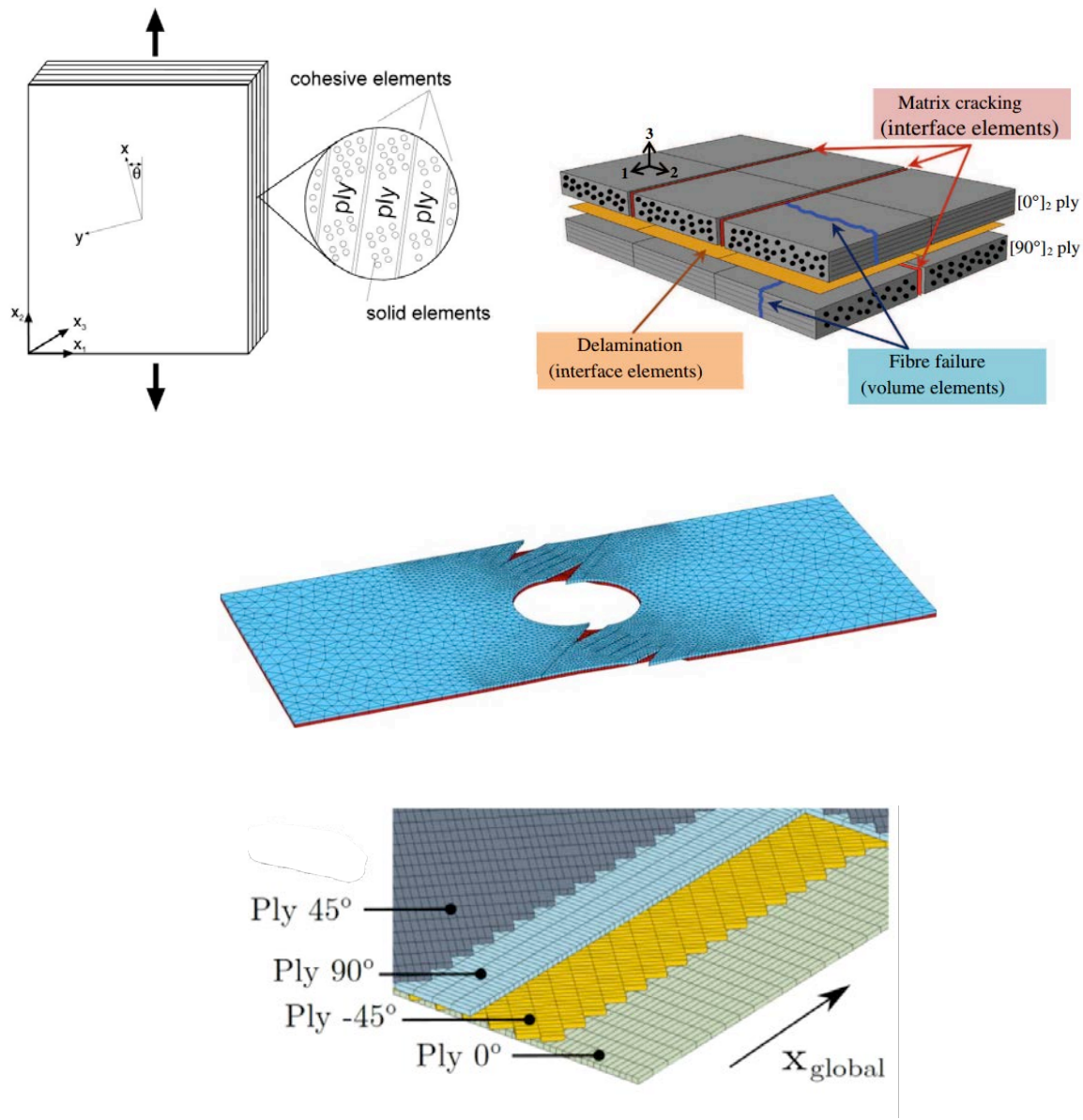

**Figure 25.** (a) Schematic of computational mesomechanics approach to simulate the behaviour of composite laminates. (b) Discrete ply models [184] for the analysis of a two layer composite. Fiber fracture, matrix splitting and ply delamination were introduced by inserting cohesive elements at the respective interfaces. (c) Simulation of the mechanical behavior in tension of an open hole [±45]s laminate, showing the interaction between matrix cracks and ply delamination [187]. Matrix cracks were introduced in the model through the phantom node approach. (d) Aligned meshes method for the analysis of [45/90/45/0] laminate including intraply damage at the Gauss point level by means of continuum damage mechanics and interply delamination using cohesive surface interactions.

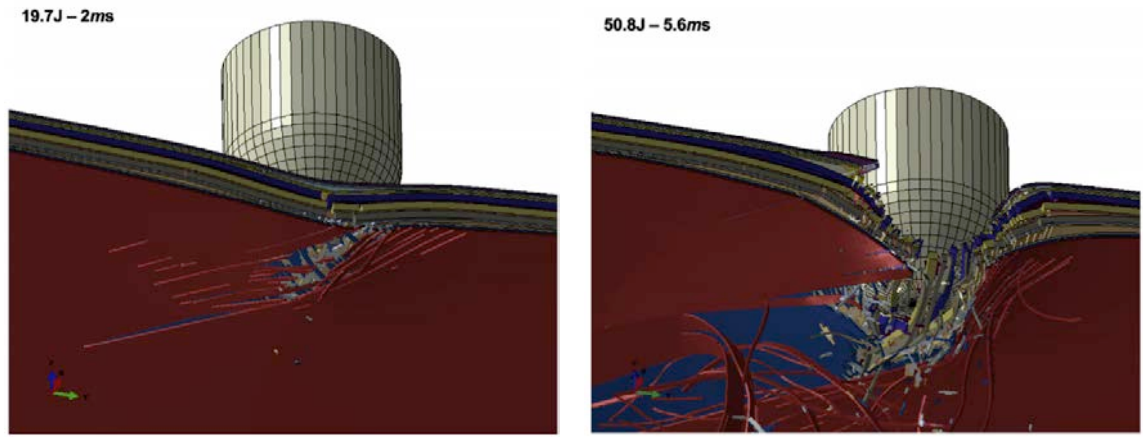

**Figure 26.** (a) Numerical simulation of a low velocity impact on a carbon/epoxy [±45/90/0/45/0$_4$/-45/0$_2$]$_s$ laminate. Impact energy 19.7 J (b) *Idem* with impact energy of 50.8 J. (a) and (b) show the competition between intraply and interply damage at the maximum impactor penetration [193].

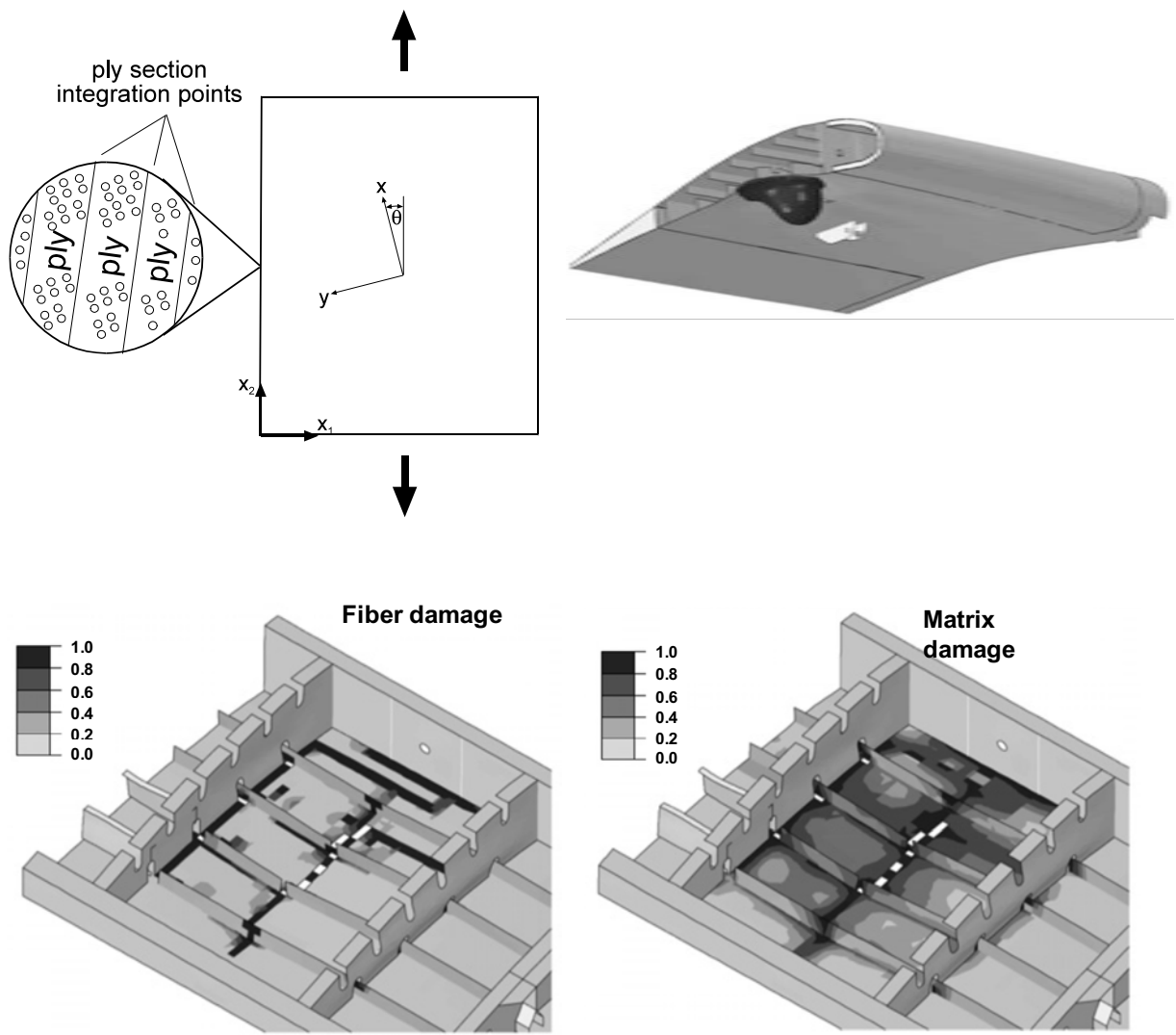

**Figure 27.** (a) Schematic of the computational mechanics approach to simulate the behaviour of FRC structures. (b) Simulation of bird impact on a composite structure using Eulerian-Lagrangian approach. (c) Contour plot of the fiber fracture damage variable. (d) Contour plot of the matrix damage variable [210].

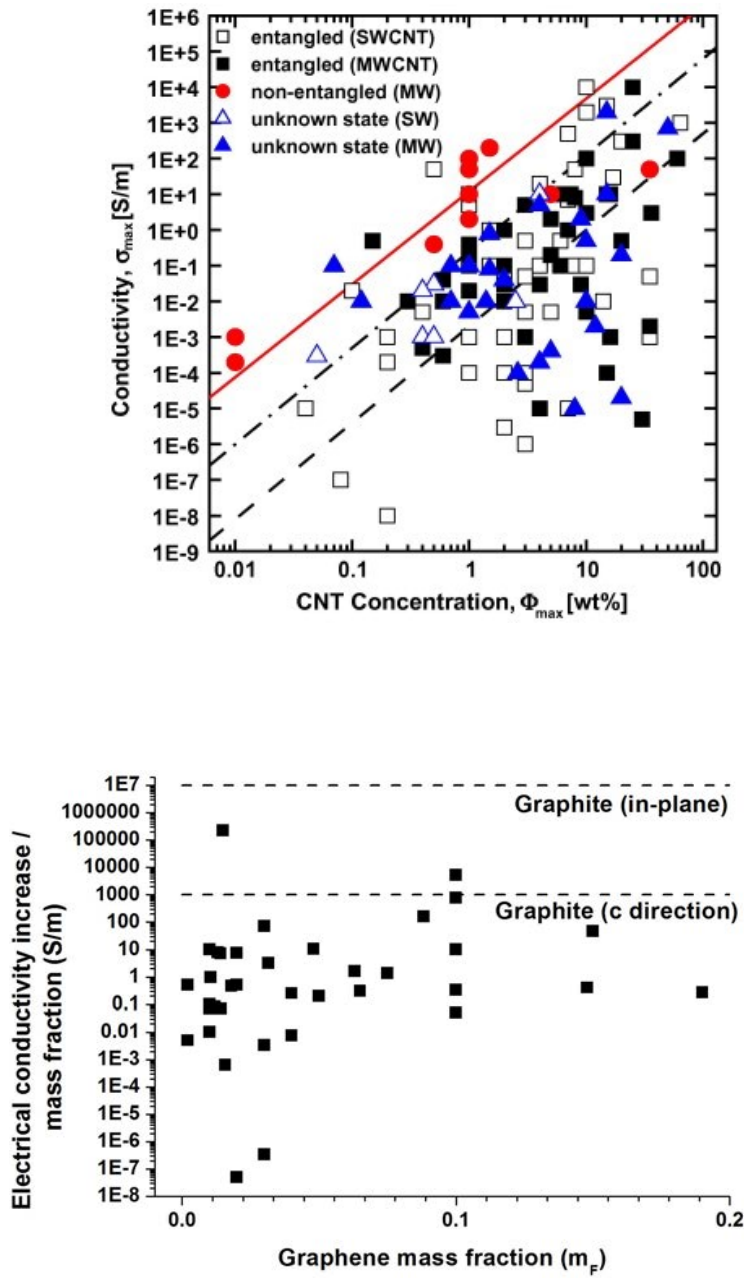

**Figure 28.** Literature data for electrical conductivity of nanocomposites with CNTs or graphene as a function of volume fraction. Selected data as of 2009 and 2012, for CNTs [211] and grapheme [212], respectively. Nanocarbons lead to low percolation thresholds typically below 0.1 wt. % and maximum values of conductivity around $10^3$ S/m. Reprinted with permission from Elsevier and Wiley.

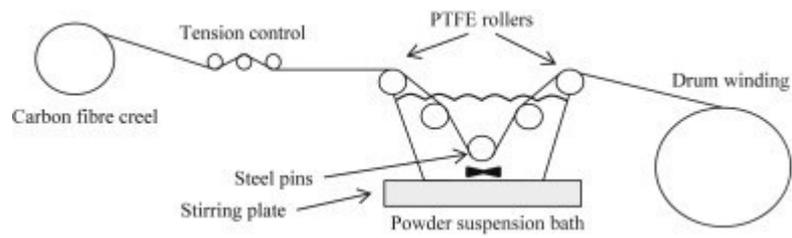

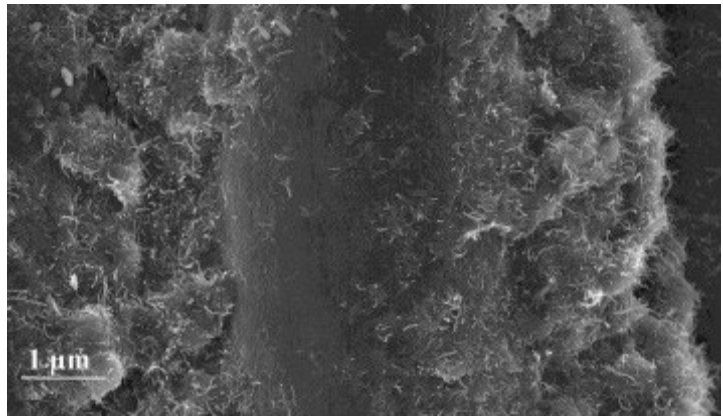

**Figure 29.** Hierarchical composites with a high $V_f$ produced using a method that minimises flow distances. a) Schematic of the process to impregnate carbon fiber tows with a thermoset matrix containing a high $V_f$ of short CNTs. The structure can consolidated into a hierarchical structure by hot compression. b) Scanning electron micrograph showing the distribution of CNT near the carbon fiber. Adapted from [213] with permission from Elsevier.

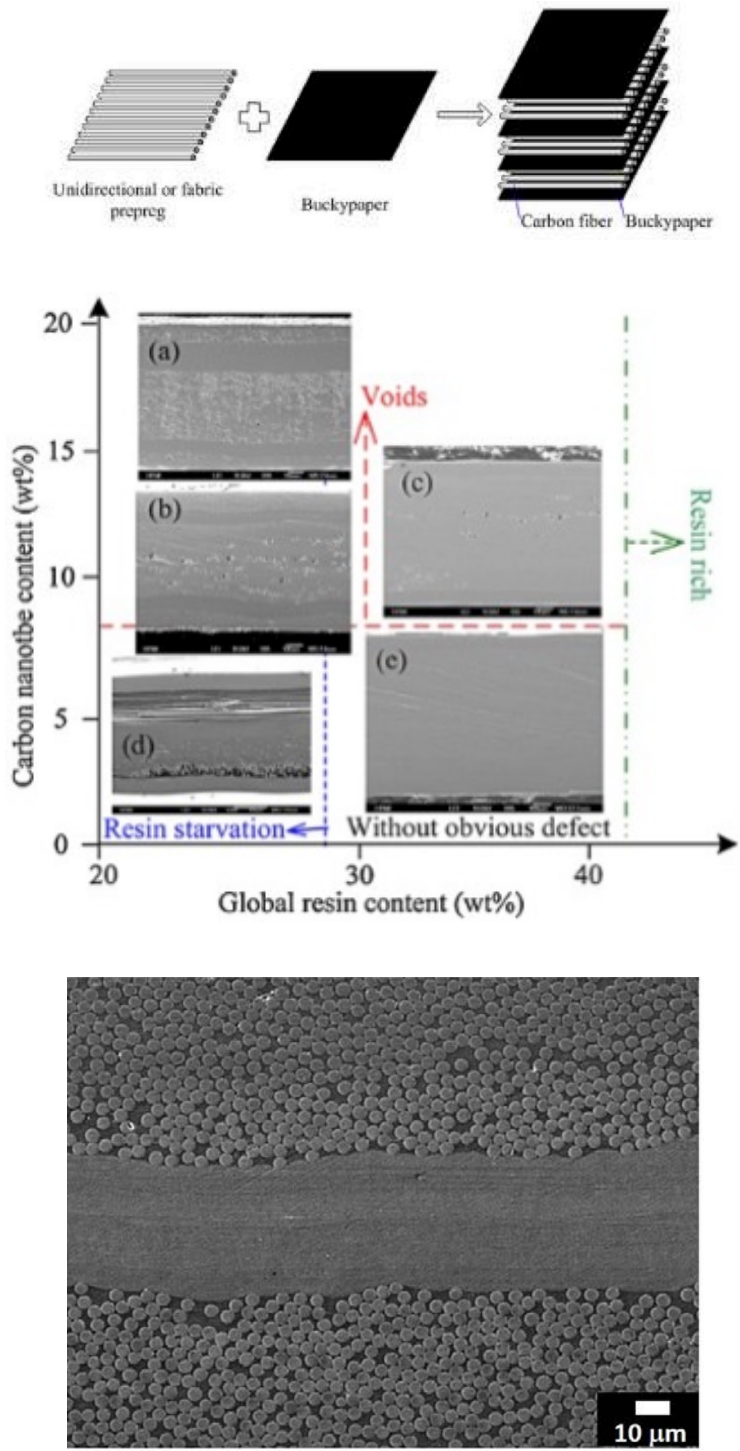

**Figure 30.** Hierarchical composites with CNT fiber veils (bucky paper). a) Schematic of the lay-up with alternating layers of unidirectional carbon fiber prepreg and CNT veils. b) Map of volume fractions to avoid the formation of porosity. c) Cross-section of a composite with 7.99 wt.% CNT veils (middle region). Adapted from [222] with permission from Wiley.

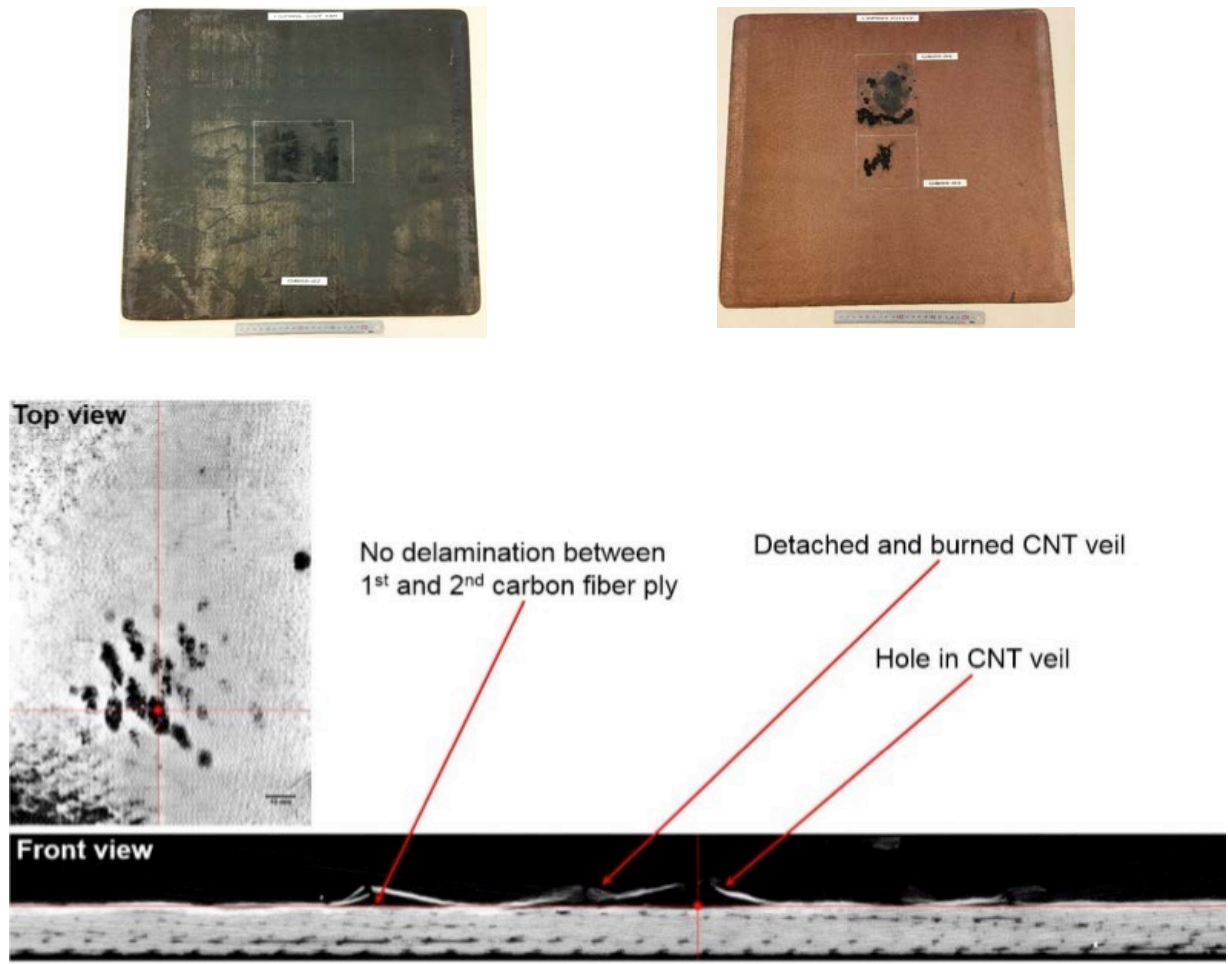

**Figure 31.** Photographs of composite panels subjected to low-energy lightning strike tests. The panels contain a protective layer of 80 g/cm2 consisting of CNT a fiber sheet (a) or a commercial copper mesh (b). CNT sheets were locally ablated but prevented the progression of internal delamination (c) [234].

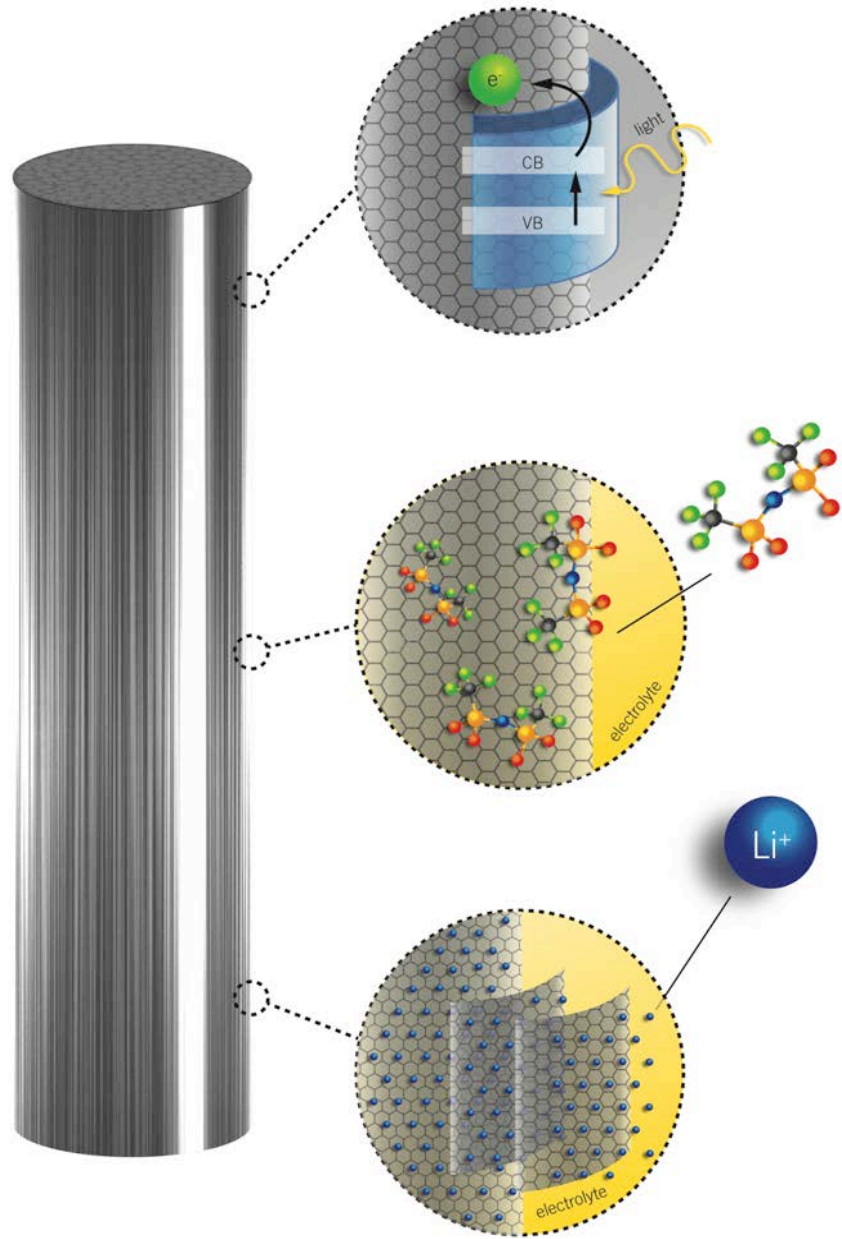

**Figure 32.** Schematic of the use of a reinforcing fiber as electrode/current collector. (a) Charge transfer to the fibre after photoexciting an electron from a semiconductor valence (VB) band to the conduction band (CB). (b) Energy storage at the fiber surface by electrostatic adsorption of ions. (c) Energy storage by intercalation of reduced ions inside the fiber.

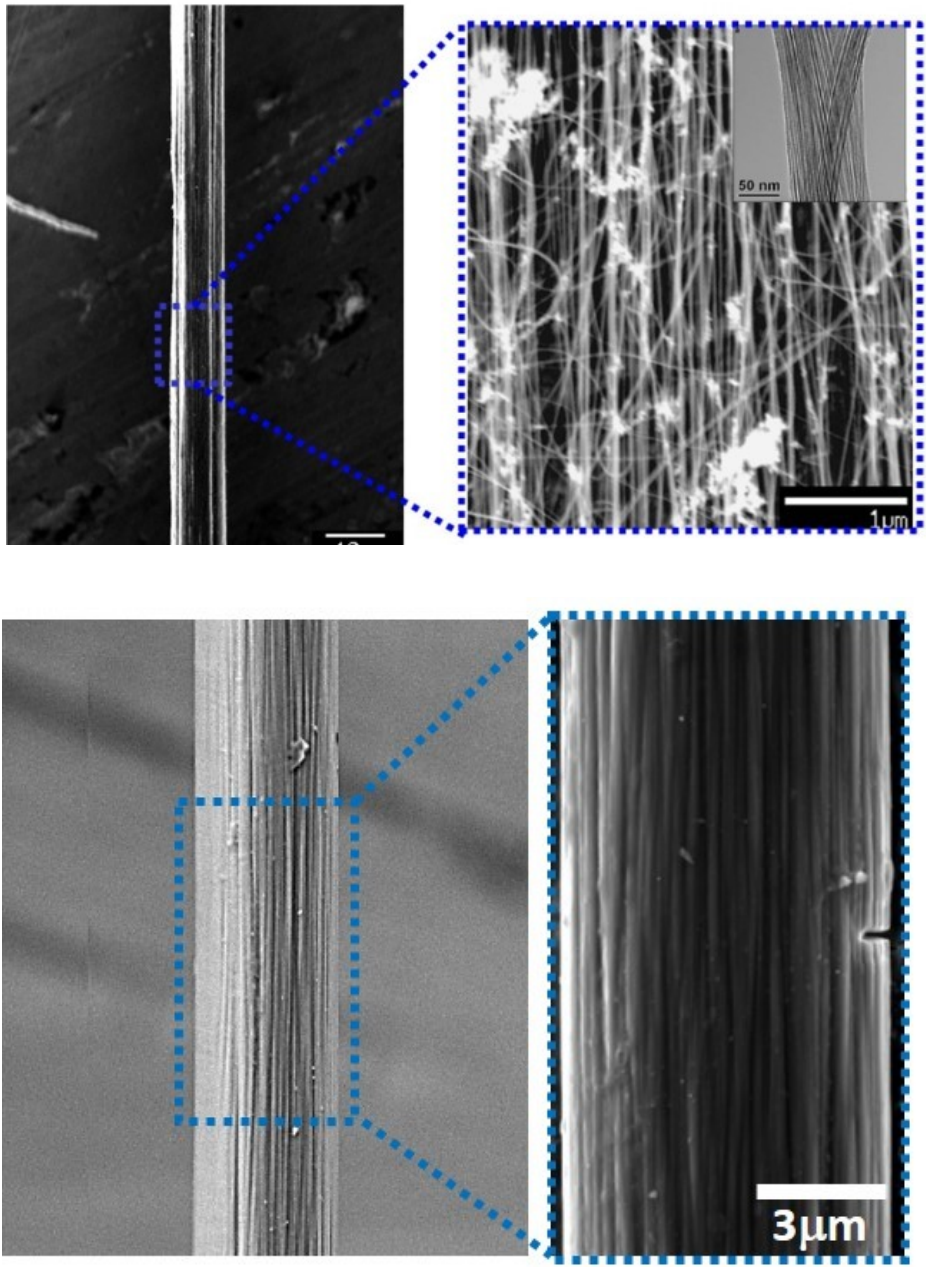

**Figure 33.** Comparison of a CNT fiber (a) and carbon fibers (b). CNT fibers have a porous yarn-like structure and a high SSA above 200 m$^2$/g, whereas carbon fibers are monolithic and have a SSA of around 0.2 m$^2$/g.

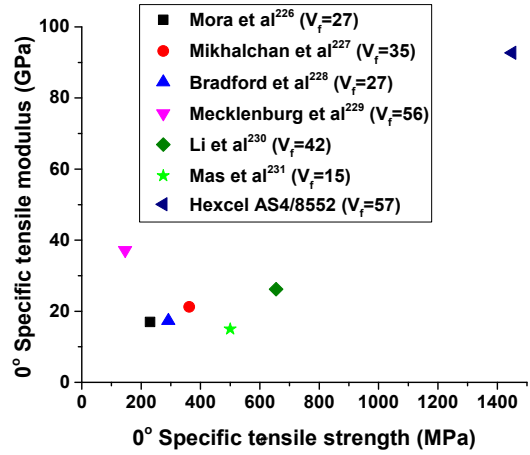

**Figure 34.** Comparison of tensile properties of unidirectional composites based on standard carbon fibers and macroscopic fibers made up of CNTs. Data from [226-231].

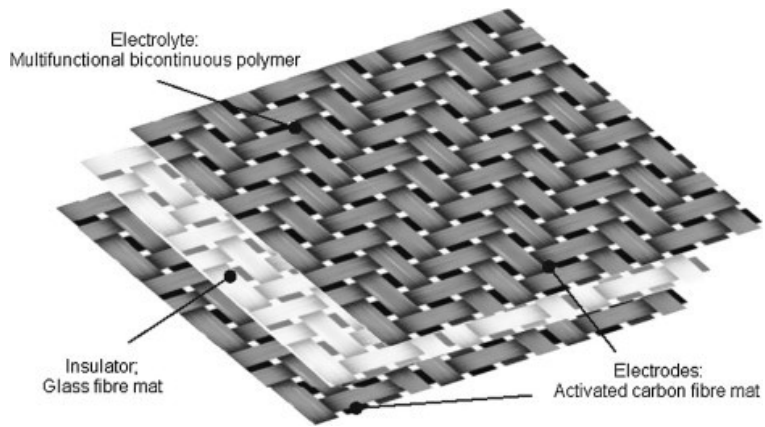

**Figure 35.** Schematic of a fibre laminate composite that can operate as a capacitor [242]. Reprinted with permission from Elsevier.
111

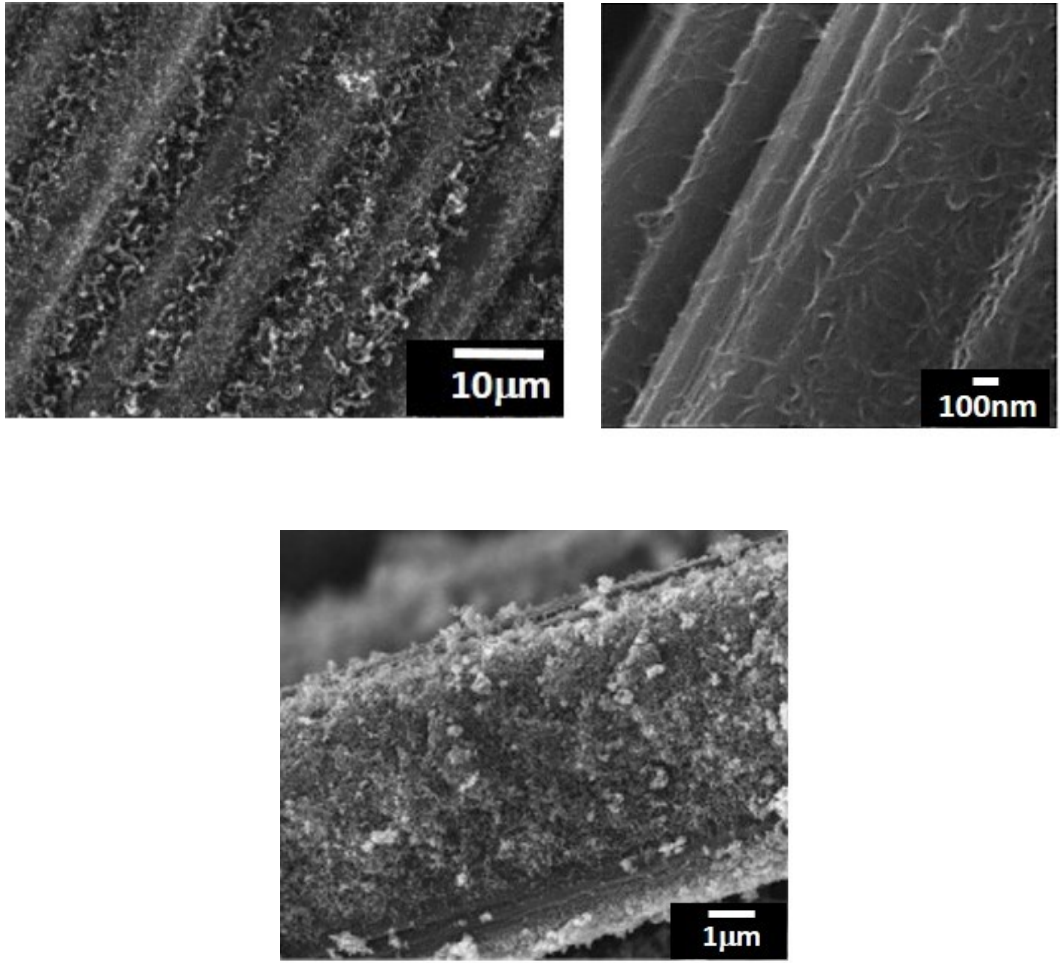

**Figure 36.** Microstructure of different carbon fiber-based electrodes used for structural supercapacitors. a) CNTs directly grown on carbon fiber. b) carbon fiber with a CNT-containing sizing. c) carbon fiber with a carbon aerogel grown around it [244]. Adapted with permission from Elsevier.

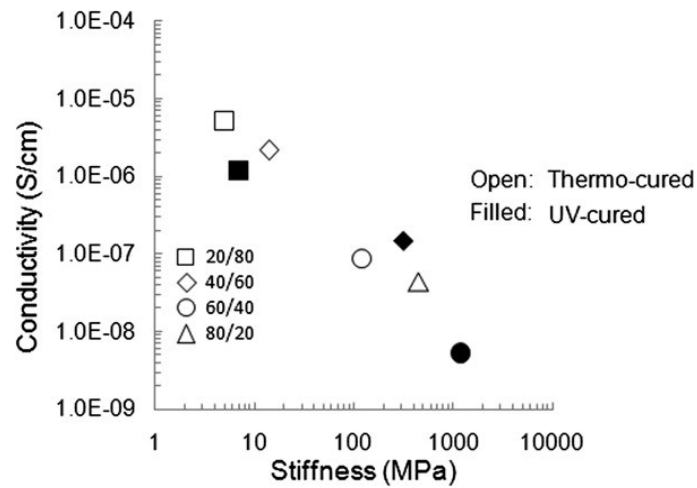

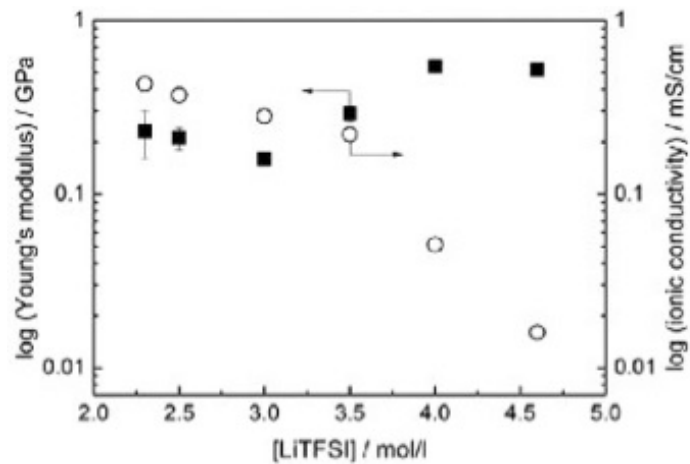

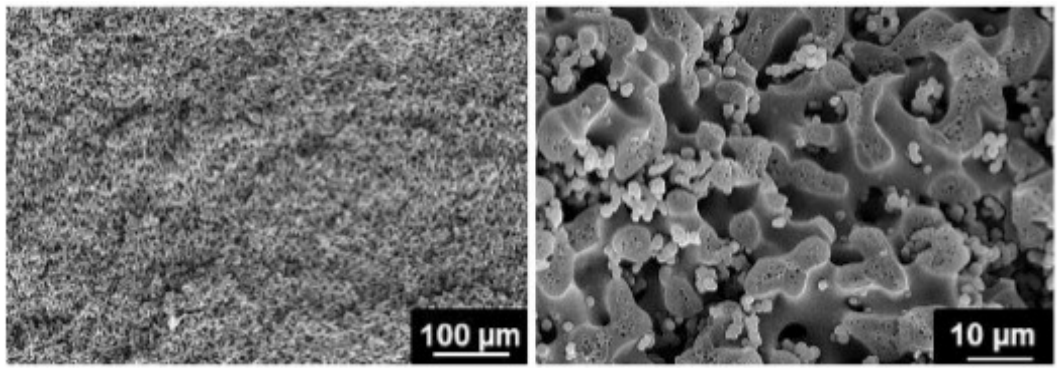

**Figure 37.** Ionic conductivity and stiffness of different polymer electrolytes. a) Different ratios of structural/conductive monomer (SR209/CD552) and SR209/SR55043. N.B. SR550 is similar to CD552 but with a lower number of ethylene oxide groups per monomer. [253] Reprinted with permission from Taylor & Francis. b) Properties of a mixture of structural epoxy Cytec MTM57 and 1-Ethyl-3-methylimidazolium bis(trifluoromethylsulfonyl)imide (EMIMTFSI EMIM-TFSI) at a 50:50 ratio as a function of LiTFSI salt content. c) Scanning electron micrographs of the SPE in b) produced with 4.6 mol/L LiTFSI. Image after extraction of the electrolyte [254]. Adapted with permission from [254]. Copyright (2013) American Chemical Society.

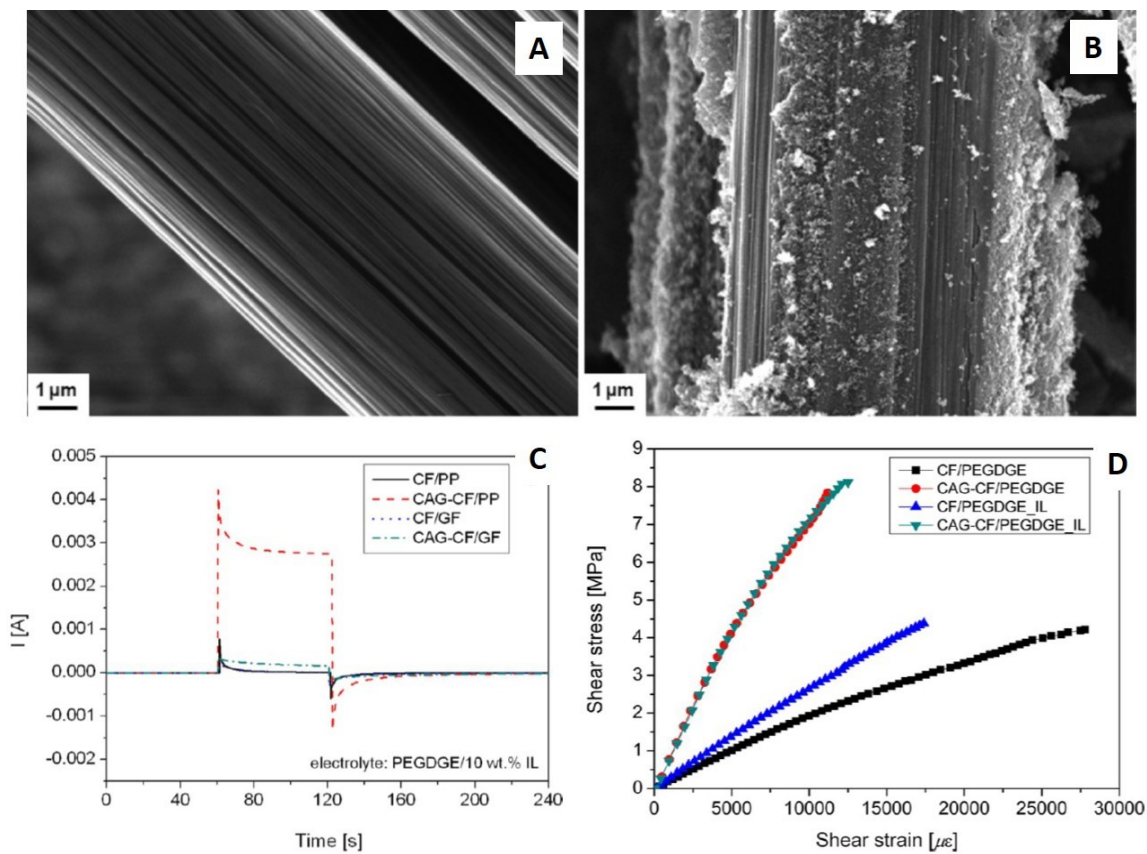

**Figure 38.** Structural supercapacitors based on woven carbon fabrics embedded in carbon aerogels and a SPE based on PEGDE epoxy and EMITFSI ionic liquid. (a) Electron micrograph of the bare carbon. (b) Fracture surface showing the carbon aerogel around the carbon fiber. (c) Choronoamperometry results showing increase capacitance (higher current) for the carbon aerogel-modified carbon fabric. (d) Comparison of in-plane shear properties of structural and multifunctional architectures. Adapted with permission from [245]. Copyright (2013) American Chemical Society.

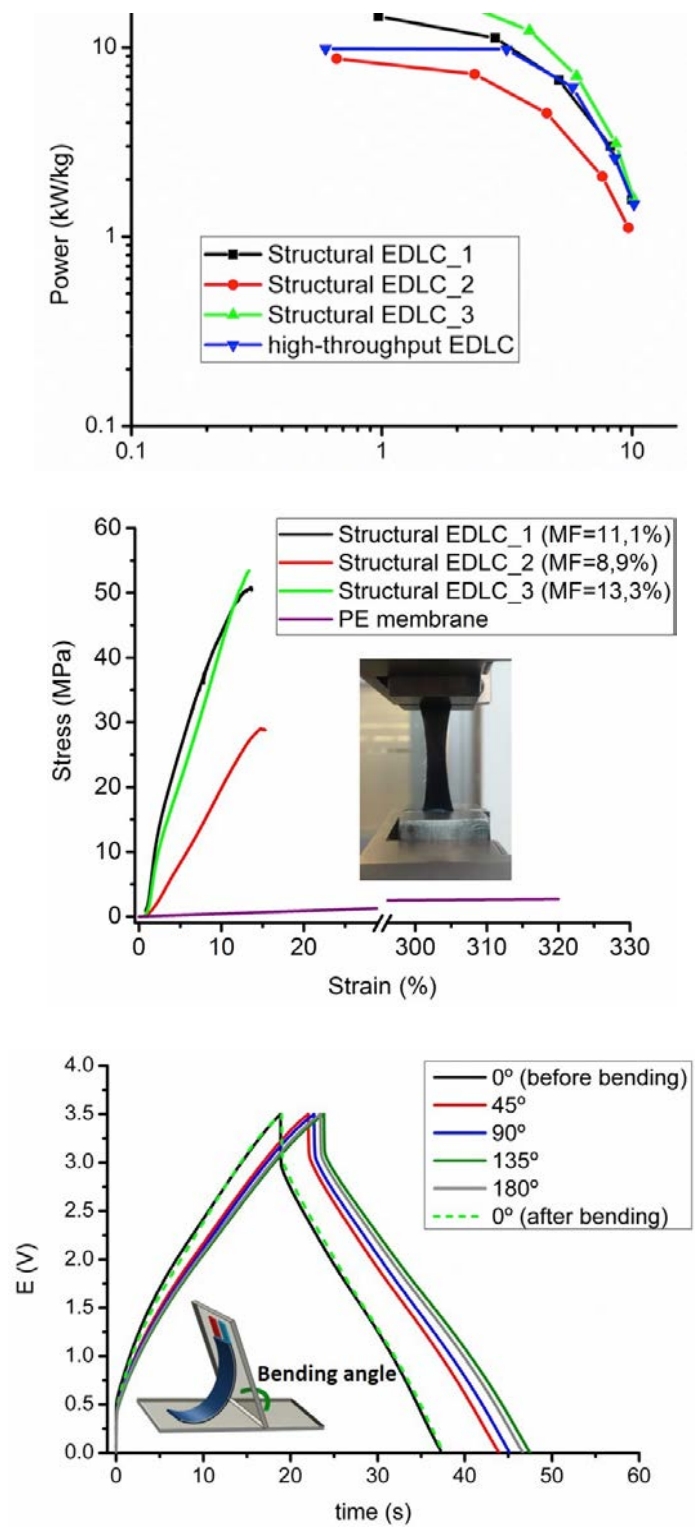

**Figure 39.** All-solid supercapacitors based on CNT fibre and polyelectrolytes. a) Photograph of a large-area device and Ragone plot with values normalised by mass of active material. b) Tensile stress-strain curves for three composites of CNT fibre and electrolyte previously tested electrochemically. c) Galvanostatic charge-discharge showing that the devices can be bent and completely folded without degradation of their electrochemical properties. [251]

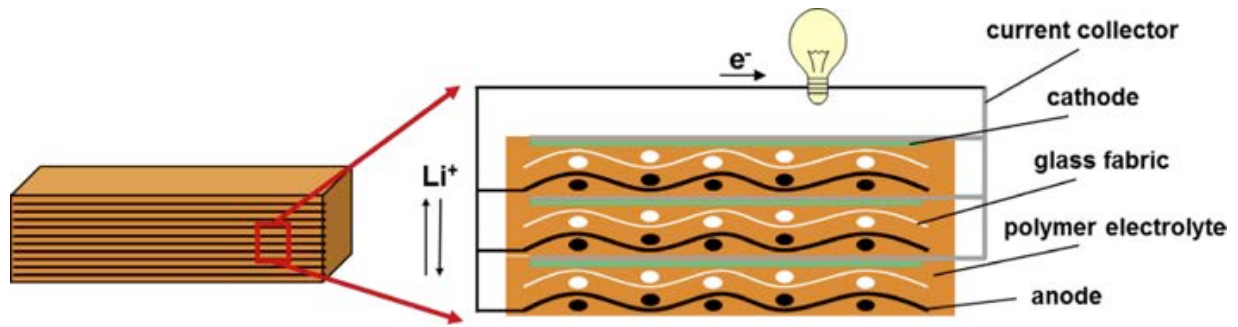

**Figure 40.** Schematic of the architecture of a structural battery with carbon fabric anode, polymer electrolyte acting also as matrix, and an oxide-based active material coated on a metal mesh as cathode [260]. Reproduced with permission from Elsevier.

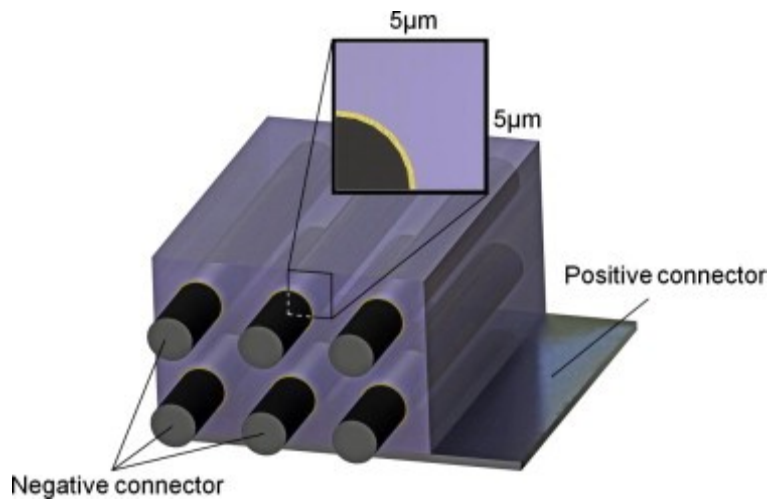

**Figure 41.** Schematic of a structural battery architecture based on carbon fiber anode with a thin coating of SPE produced by electrodeposition [244]. Reproduced with permission from Elsevier.

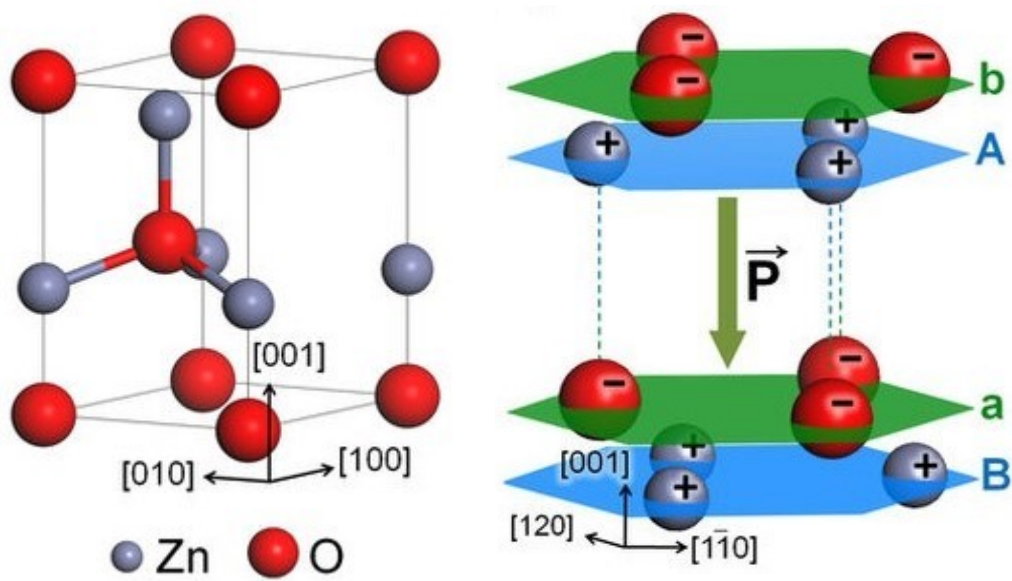

**Figure 42.** (a) Wurtzite crystal structure of ZnO. (b) Formation of dipoles under an applied stress. Modified from [272] with permission from Nature Publishing Group.

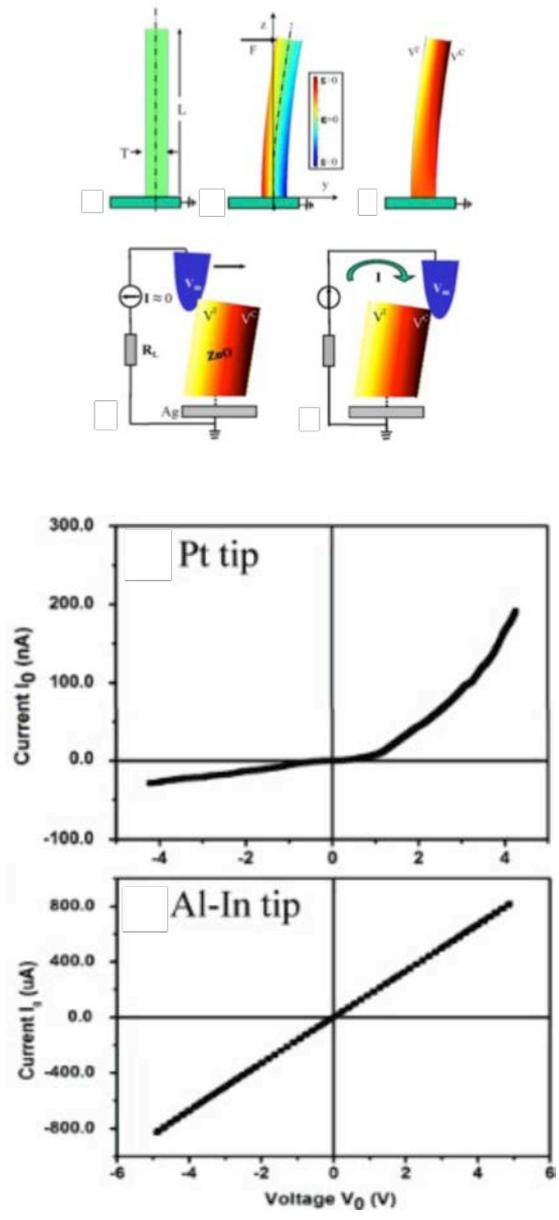

**Figure 43.** (a) Schematic showing the principle of energy harvesting from a piezoelectric. An AFM tip is used to extract current from the deflection of a piezoelectric nanowire (b) Current generated as a function of the voltage for Pt and Al tips. See text for details. Adapted with permission from [274]. Copyright (2013) American Chemical Society.

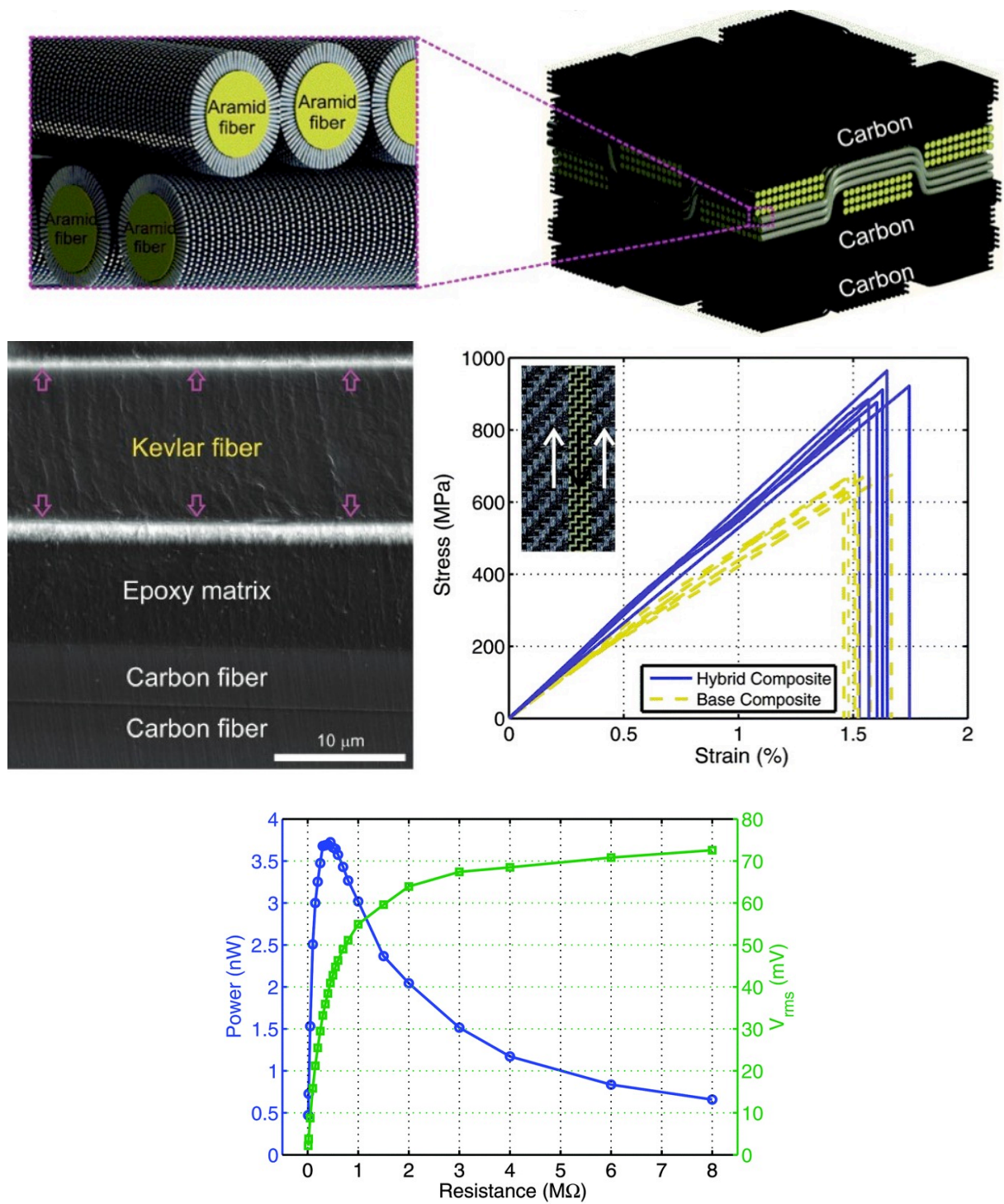

**Figure 44.** Energy harvesting laminate composite based on carbon fabric and ZnO-coated aramid fibres. a) Schematic of the laminate architecture. b) Cross section of the laminate showing the different components, with ZnO indicated by the arrows. c) Tensile properties increase as a consequence of the growth of ZnO on the aramid fibres. d) Output piezoelectric power for different resistances in series in the external circuit. Modified from [276] with permission from the Royal Society of Chemistry.

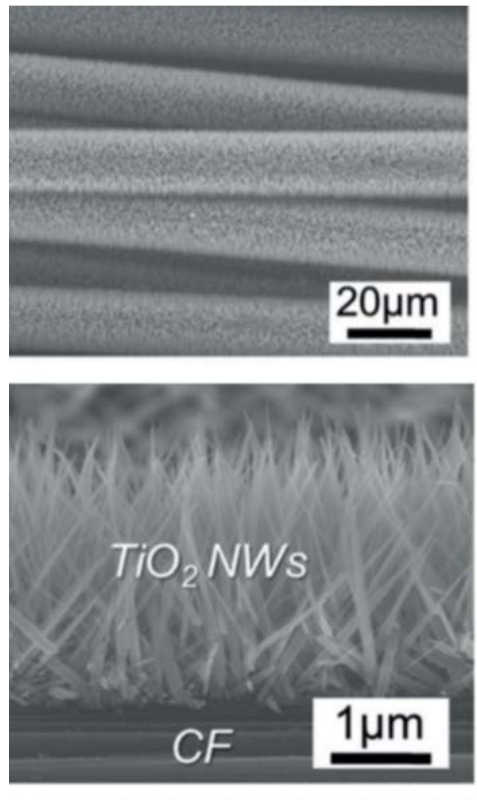

**Figure 45.** Carbon fiber coated with TiO$_2$ nanowires used in a DSSC. Electron and optical micrographs of the fibres and the CF/ZnO interface [288]. Reprinted from [288] with permission from John Wiley and Sons.

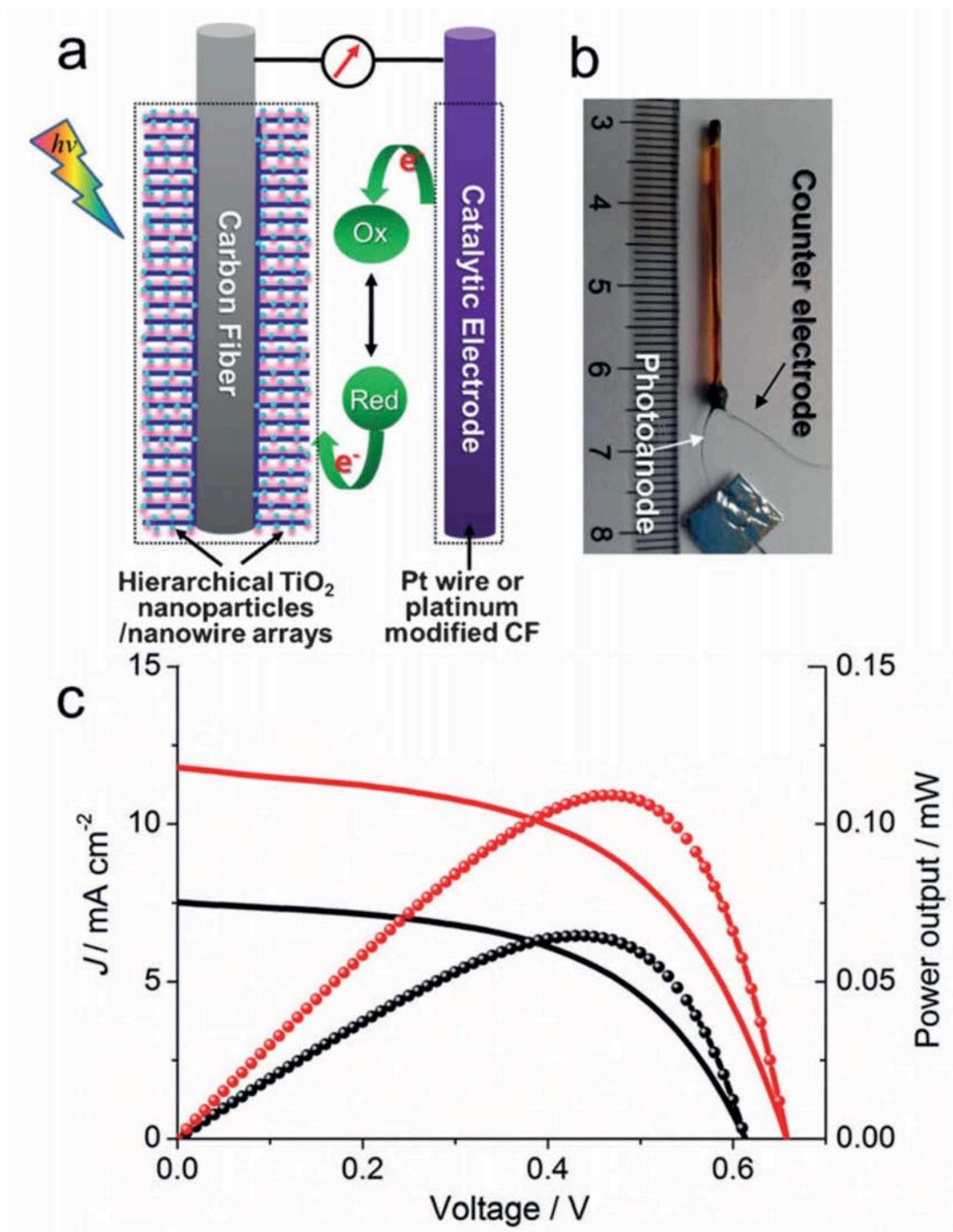

**Figure 46.** Schematic and photograph of the fibre device and PV properties: current density and output power under illumination (black – standard illumination, red – double-sided illumination) at different bias. Adapted from [288] with permission from John Wiley and Sons.

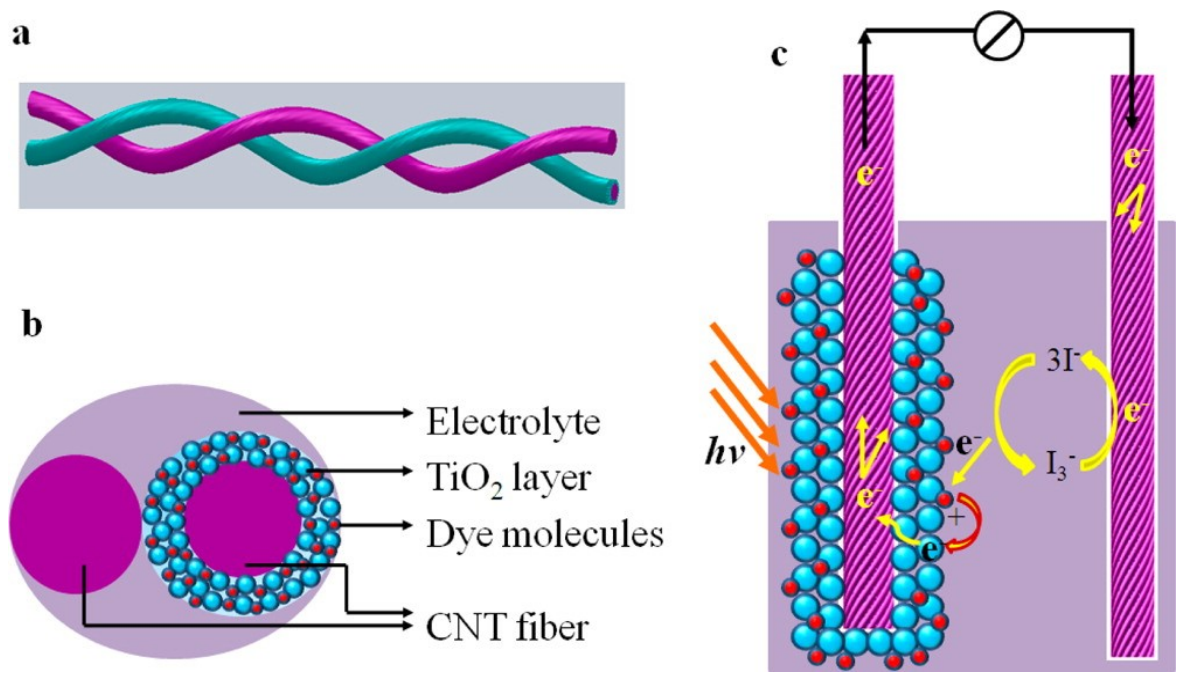

**Figure 47.** CNT fibers used as electrodes in DSSC. A CNT fibres coated with dye-loaded $TiO_2$ is intertwined with a bare CNT fibres and immersed in an electrolyte to form a DSSC. Reproduced with permission from [291]. Copyright (2012) American Chemical Society.

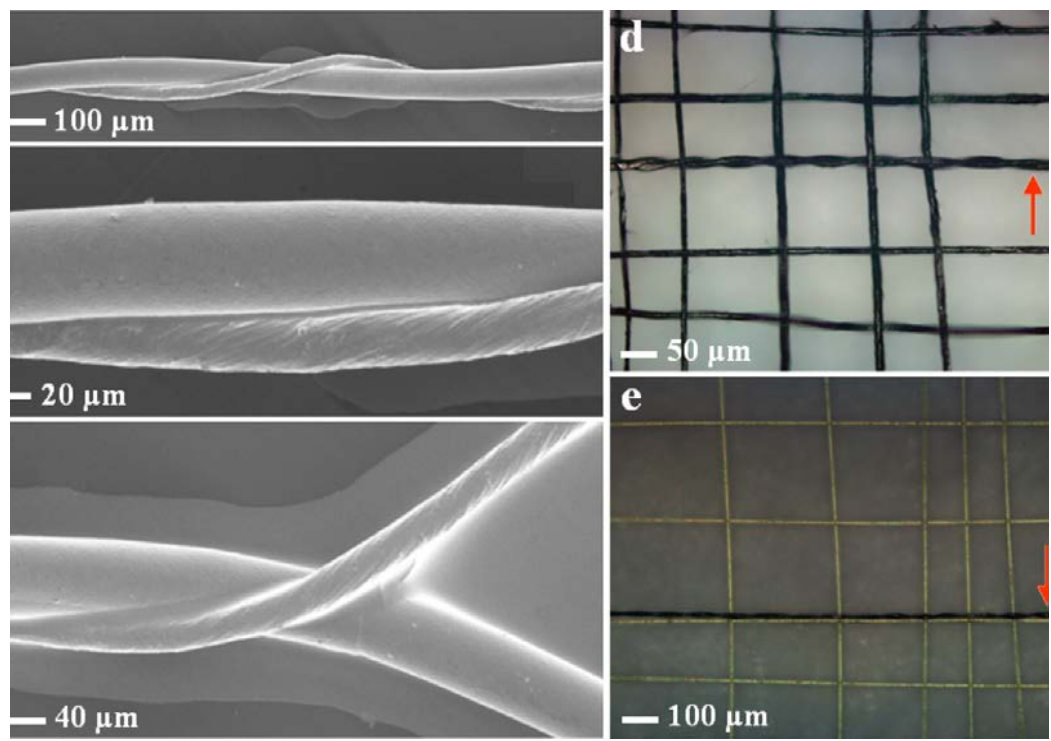

**Figure 48.** Electron micrographs of intertwined fibers and photograph of a woven array. Reproduced with permission from [291]. Copyright (2012) American Chemical Society.

# TABLES

**Table 1.** Literature data for through-thickness conductivity of hierarchical composites.

| Method | Fiber | CNT fraction (%) | Through-thickness composite conductivity (S/m) (without CNTs) | Ref |
|---|---|---|---|---|
| Thermoset powder impregnation | Hextow AS4C-GP | 6.1 vol.% | 53 (12.6) | [213] |
| Electrophoretic deposition | IM7 | 0.25 wt. % | 8.9 (2-6.5) | [214] |
| In-situ growth on fibre surface | Alumina fiber | ~1.5 wt. % | ~13 (insulator) | [216] |
| Prepreg of polymer with random dispersion of CNTs | T700GC | 0.4 wt. % | 0.22 (0.07) | [221] |
| Veils of randomly oriented CNTs | IM7 | 7.99 wt.% | 0.21 (0.04) | [222] |
| Veils of randomly oriented CNTs | UD T700S fabric | 1.09 wt. % | 20 (7) | [223] |

**Table 2.** Properties of carbon fibers, CNT fibers and their composites.

| Material | Specific strength (GPa/SG) | Specific modulus (GPa/SG) | Specific surface area (m$^2$/g) | Longitudinal conductivity (S/m) |
|---|---|---|---|---|
| AS4 Carbon fiber | 2.3 | 231 | 0.2 | $6 \times 10^4$ |
| AS4/8852 Hexply composite | 1.4 | 93 | - | $3.6 \times 10^{4a}$ |
| CNT fiber | 1 - 2 | 50 - 100 | 250 | $10^5 - 10^6$ |
| Composite of CNT fiber | 0.2 – 0.6 | 17 - 37 | - | $1.5 \times 10^{5b}$ |

[a] Assuming a rule of mixtures. [b] Assuming a rule ox mixtures with $V_f = 30\%$ and a CNT fiber longitudinal conductivity of $5 \times 10^5$ S/m.

**Table 3.** Mechanical and electrochemical properties of carbon-based electrodes

| Type of fibre | Specific strength (GPa/SG) | Specific modulus (GPa/SG) | Energy to break (J/g) | Compression strength (GPa/SG) | Capacitance (F/g) | Reference |
|---|---|---|---|---|---|---|
| HTA carbon fiber + carbon aerogel | - | - | - | - | 6 - 14 | [245] |
| HTA carbon fiber | 2.2 | 135 | 18.6 | - | 0.06 | [245, 250] |
| AS4 carbon fiber | 2.4 | 127 | 22.4 | 1.4[a] | 1.3 | [243] |
| CNT fiber | 0.9 | 44 | 30 | 0.6 | 42[b] | [237] |
| CNT fiber | 2 | 100 | 75 | 0.6 | 33[b] | [251] |
| Functionalised-CNT fiber | 1.5 | 50 | 18 | | 50-75[c] | [246] |
| CNT fiber | 0.15 | 10 | 8.625 | - | 23[d] | [252] |
| CNT fiber | .75 | 30 | 92 | 0.6 | 79.8[e] | [234] |

[a] Calculated from compressive properties of 60 vol.% composite according to manufacturer datasheet and assuming the rule of mixtures. [b] Three-electrode CV at 20mV/s in 1-butyl-1-methylpyrrolidinium bis(trifluoromethanesulfonyl)imide (PYR14TFSI) ionic liquid. [c] CV in 1M H2SO4 at 20mV/s. [d] Two-electrode CV in 1M H2SO4 at 10mV/s. [e] NaCl solution, no additional details available.

**Table 4.** Summary of electrochemical and mechanical properties of structural supercapacitors

| Electrode | Electrolyte | Capacitance (F/g) | Power density (W/kg) | Energy density (Wh/Kg) | Mechanical properties (MPa) | Ref |
|---|---|---|---|---|---|---|
| Carbon fiber+carbon aerogel[a] | PEGDE + 10% IL | 0.036 | 0.34 | 0.009 | $G_{12}$ = 8710<br>$\sigma_{12}$ = 895 | [245] |
| Carbon fiber+carbon aerogel | Epoxy resin+EMITFSI+LiTFSI | 0.27 | 18.7 | 0.21 | $G_{12}$ = 536<br>$\sigma_{12}$ = 2.8 | [256] |
| Carbon fiber[a] | PEGDE + IL + 0.1M LiTFSI | 0.026 | 2.45 | 0.013 | E = 18000<br>$X_C$ = 7.5 | [242] |
| CNT-grafted carbon fiber[a] | MTM57 + IL + LiFSI | 0.05 | 0.016 | 0.005 | $G_{12}$ = 450<br>$\sigma_{12}$ = 14<br>E = 10000<br>$X_C$ = 153 | [258] |
| Carbon fiber | CD552 + SR494 + 0.825M LiIm | 0.093 | 0.15 | 0.021 | E/SG[b] = 12000<br>$G_{12}$/SG = 310 | [257] |
| CNT fibre | Thermoplastic + TFSI | 6.7 | 3700[c] | 0.91[c] | E = 790[d]<br>$\sigma_{11}$ = 53[d] | [251] |

[a] Device properties calculated as half of those reported normalized by electrode mass. [b] Normalised by specific gravity (SG). [c] Electrochemical properties normalized by device weight without metallic current collector or encapsulation material. [d] Tensile test of active material and SPE membrane.